 \documentclass[preprint,longabstract]{aastex}

\shorttitle{Dynamics of GJ1214b's atmosphere}
\shortauthors{Charnay et al.}

\begin{document}


\title{3D modeling of GJ1214b's atmosphere:  \\ vertical mixing driven by an anti-Hadley circulation}


\author{B. Charnay\altaffilmark{1} and V. Meadows}
\affil{Astronomy Department, University of Washington,
    Seattle, WA 98125}
\affil{Virtual Planetary Laboratory, University of Washington,
    Seattle, WA 98125}

\author{J. Leconte}
\affil{Laboratoire de M\'et\'eorologie Dynamique, IPSL/CNRS/UPMC, Paris 75252, France }


\altaffiltext{1}{NASA Postdoctoral Program fellow\\ 
email: bcharnay@uw.edu}


\begin{abstract}

GJ1214b is a warm sub-Neptune transiting in front of a nearby M dwarf star. Recent observations indicate the presence of high and thick clouds or haze whose presence requires strong atmospheric mixing. In order to understand the transport and distribution of such clouds/haze, we study the atmospheric circulation and the vertical mixing of GJ1214b with a 3D General Circulation Model for cloud-free hydrogen-dominated atmospheres (metallicity of 1, 10 and 100 times the solar value) and for a water-dominated atmosphere. We analyze the effect of the atmospheric metallicity on the thermal structure and zonal winds. We also analyze the zonal mean meridional circulation and show that it corresponds to an anti-Hadley circulation in most of the atmosphere with upwelling at mid-latitude and downwelling at the equator in average. This circulation must be present on a large range of synchronously rotating exoplanets with strong impact on cloud formation and distribution. Using simple tracers, we show that vertical winds on GJ1214b can be strong enough to loft micrometric particles and that the anti-Hadley circulation leads to a minimum of tracers at the equator. We find that the strength of the vertical mixing increases with metallicity. We derive 1D equivalent eddy diffusion coefficients and find simple parametrizations from $K_{zz}=7\times 10^2 \times P_{bar}^{-0.4}$ m$^2$/s for solar metallicity to $K_{zz}=3\times 10^3 \times P_{bar}^{-0.4}$ m$^2$/s for the 100$\times$solar metallicity. These values should favor an efficient formation of photochemical haze in the upper atmosphere of GJ1214b.

\end{abstract}



\section{Introduction}
GJ1214b is a warm sub-Neptune orbiting a nearby M dwarf and discovered by the MEarth survey \citep{charbonneau09}. It is one of the rare low-mass planets whose atmosphere is characterizable in transit spectroscopy by current telescopes. GJ1214b is therefore considered as a privileged archetype of this new category of planets also called mini-Neptunes. It has a mass of 6.55$\pm$0.98 $M_\oplus$ and a radius of 2.68$\pm$0.13 $R_\oplus$, giving it a density of around 1.88 g cm$^{-3}$ \citep{charbonneau09}. This low density necessarily implies the presence of a thick atmosphere. Interior models suggest that the planet possesses either a dense iron/rock core surrounded by a thick hydrogen/helium-rich atmosphere or a water-rich core surrounded by a thick steam atmosphere \citep{rogers10, nettelmann11}. \cite{nettelmann11} found that a water-rich planet depleted in hydrogen would require a water-to-rock ratio larger than 6:1. They considered such a large ratio as unlikely and favored instead an intermediate case with a hydrogen/helium/water-rich atmosphere. GJ1214b has a tight orbit (i.e. semi-major axis of 0.014 AU) and is very likely tidally locked, with a permanent dayside and nightside.

Several observations in transit spectroscopy were performed to probe GJ1214b's atmosphere and to break the degeneracy of possible compositions. First observations by \cite{bean10} revealed a lack of spectral features between 0.78 and 1 $\mu$m, ruling out the hypothesis of a cloud-free solar composition. Other observations \citep{bean11, crossfield11, demooij12, desert11, berta12, fraine13, nascimbeni15} confirmed this flat spectrum between 0.6 and 4.5 $\mu$m while one group found a significant difference in the transit depth measured in J-band ($\sim$1.25 $\mu$m) and in Ks-band ($\sim$2.15 $\mu$m)  \citep{croll11}. Finally, precise measurements with the Wide Field Camera 3 (WFC3) instrument on the Hubble Space Telescope (HST) revealed a very flat spectrum between 1.15 and 1.65 $\mu$m \citep{kreidberg14a}. That implied the presence of high cloud/haze diffusing or absorbing the stellar radiation at low pressure. The presence of clouds is required even for a high mean-molecular-mass atmosphere with a composition dominated by water, methane, carbon monoxide, nitrogen or carbon dioxide. The cloud-top pressure in transit spectroscopy should be less than 10$^{-2}$ mbar for a solar-like composition and 10$^{-1}$ mbar for a water-dominated atmosphere \citep{kreidberg14a}.

For the conditions on GJ1214b, these clouds/haze could be either condensate clouds of potassium chloride (KCl) and zinc sulfide (ZnS) or photochemical haze produced by the photolysis of methane in the upper atmosphere \citep{miller-ricci12, morley13}. KCl and ZnS clouds are supposed to form between 0.1 and 1 bar \citep{miller-ricci12}. Their presence at low pressure would thus require a strong atmospheric circulation lofting cloud particles over several scale heights. The persistence of photochemical haze at low pressure would also require a sufficiently strong vertical mixing, counteracting particle sedimentation and bringing methane to altitudes where photolysis occurs. Therefore, the formation of high and thick cloud/haze on GJ1214b is strongly linked to the atmospheric circulation. The latter has been investigated in previous studies using General Circulation Models (GCMs) for cloud-free atmospheres \citep{zalucha12, menou12, kataria14}. \cite{menou12} used gray opacities and showed that the superrotating zonal winds and the thermal phase curves would vary greatly with the atmospheric metallicity. \cite{kataria14} analyzed the atmospheric dynamics and thermal phase curves with a more accurate GCM using non-gray opacities for different atmospheric compositions, but they did not analyze the vertical winds and the vertical mixing. 

In this paper, we analyze the circulation and the vertical mixing of GJ1214b for a cloud-free atmosphere with different compositions using the Generic LMDZ GCM, a very versatile 3D model developed to simulate any kind of planetary atmosphere.
In the next section, we describe the model and the parameters used for GJ1214b. 
In section 3, we analyze the thermal structure and atmospheric circulation (zonal, meridional and vertical winds).
In section 4, we analyze the vertical mixing of tracers and discuss its implications for cloud/haze formation in section 5.
The 3D modeling of realistic clouds with our GCM will be described in a dedicated paper, analyzing their formation, dynamics and impacts on spectra.

\section{Model}
\subsection{The Generic LMDZ GCM}
We simulated the atmosphere of GJ1214b using the Generic LMDZ GCM. This model has been specifically developed for exoplanet and paleoclimate studies. It has been used for studying the early atmospheres of the Earth, Mars and Titan \citep{charnay13, wordsworth13a, forget13, charnay14} and for studying the habitable zone for terrestrial exoplanets \citep{wordsworth11,leconte13b, leconte13a}. 
The model is derived from the LMDZ Earth GCM \citep{hourdin06}, which solves the primitive hydrostatic equations of meteorology using a finite difference dynamical core on an Arakawa C grid. This dynamical core has been used in GCMs dedicated to the present Earth \citep{hourdin06}, Mars \citep{forget99}, Venus \citep{lebonnois10}, and Titan \citep{lebonnois12}. 

In this paper, simulations were performed with a horizontal resolution of 64$\times$48 (corresponding to resolutions of 3.75$^\circ$ latitude by 5.625$^\circ$ longitude). We also did a few tests for GJ1214b with a 128$\times$96 resolution but we did not notice significant differences. For the vertical discretization, the model uses pressure coordinates. In this work, we used 45 layers equally spaced in log pressure, with the first level at 80 bars and the top level at 3 Pa. 

The radiative scheme is based on the correlated-k model, with k-coefficients calculated from high resolution absorption spectra, computed by kspectrum, a line-by-line model using the HITRAN 2012 database \citep{rothman13}. kspectrum is a tool developped by Vincent Eymet and available online at \url{http://www.meso-star.com/en\_Products.html}.
For water opacities, we used the high-temperature database HITEMP 2010 \citep{rothman10}. For methane, the HITRAN database does not include absorption lines between 9000 and 11000 cm$^{-1}$ and between 11500 and 12000 cm$^{-1}$. We added the missing lines using EXOMOL \citep{tennyson12, yurchenko14}. At a given pressure and temperatures, correlated-k coefficients in the GCM are interpolated from a matrix of coefficients stored in a 12 $\times$ 9 temperature and log-pressure grid: T = 100, 200, ...., 1000, 1200, 1500, 2000 K, P = 10$^{-1}$ , 10$^0$ , 10$^1$ , ...., 10$^7$ Pa. 
To facilitate comparison with the work of \cite{kataria13}, we computed k-coefficients over the same 11 spectral bands. The H$_2$-H$_2$ and H$_2$-He CIA from the HITRAN database were included. Rayleigh scattering by H$_2$,  H$_2$O and He was included, based on the method described in \cite{hansen74} and using the \cite{toon89} scheme to compute the radiative transfer. 

The model also includes a scheme for adiabatic temperature adjustment triggered in the regions where the atmosphere is convectively unstable. However, this condition almost never occurs in our simulations of GJ1214b.

\subsection{Atmospheric composition}

We ran simulations with four different atmospheric compositions:
\begin{itemize} 
\item H$_2$-rich atmosphere at 1$\times$solar metallicity
\item H$_2$-rich atmosphere at 10$\times$solar metallicity 
\item H$_2$-rich atmosphere at 100$\times$solar metallicity
\item pure H$_2$O atmosphere
\end{itemize}

For the H$_2$-rich atmospheric composition, we used the solar nebula atomic abundances from \cite{lodders03}.
The 10$\times$solar metallicity is similar to Saturn's metallicity, and the 100$\times$solar is quite similar to Uranus and Neptune's metallicity \citep{kreidberg14b}. Population synthesis models generally predict high metallicities between 100-400$\times$solar for low-mass planets  but lower metallicities can also arise \citep{fortney13}.

For the H$_2$-rich atmospheres, we computed the compositions at thermochemical equilibrium, assuming only H$_2$, He, H$_2$O, CH$_4$, N$_2$, NH$_3$, CO and CO$_2$.
We computed the mixing ratios analytically following the method described in the appendix of \cite{burrows99},  using molecular Gibbs energies from \cite{sharp90}. We computed  abundances from the reactions:
\begin{equation} 
\rm CO + 3H_2 = H_2O + CH_4
\end{equation} 
\begin{equation} 
\rm N_2 + 3H_2 = 2NH_3
\end{equation} 
\begin{equation} 
\rm CO_2 + H_2 = CO + H_2O 
\end{equation} 
The computation requires solving a cubic equation. 

Figure \ref{figure_1} shows the vertical molecular abundances computed for 1, 10 and 100 times the solar composition cases using temperature profiles from the 1D model (see figure \ref{figure_2}).

The opacities for a H$_2$-rich atmosphere are dominated by molecular absorption (mainly H$_2$O). However, at high pressure and for temperatures higher than around 1000 K, the pressure broadening of atomic alkalin (Na and K) spectral lines has a strong impact on the opacity in the visible. For a solar composition, atomic potassium should be the dominant K-bearing gas for temperatures higher than around 980K at 1 bar \citep{lodders06}. 
We did not include Na opacities in our model, as Na absorbs at wavelengths that are shorter than the bulk of the radiation emitted by a later-type M dwarf like GJ1214.  Also, at the high metallicities considered here, Na is likely to condense as Na$_2$S clouds in the deeper atmosphere \citep{morley13}, removing it as an atmospheric opacity source. 
For K, we simply added a continuum between 1.18$\times 10^4$ - 1.43$\times 10^4$cm$^{-1}$, assuming pressure broadening and a reference cross section for atomic K of 10$^{-19}$cm$^2$/atom at 1 bar and 1000 K for solar metallicity. This value corresponds to the absorption from \cite{burrows03} at the edges of the spectral band we considered.

\subsection{Integration}
Model parameters are given in tables 1 and 2.
For the stellar spectrum, we used a blackbody at 3026 K. We also tested with spectra of AD Leo and Gliese 581 but these more realistic spectra did not produce significantly different results. We assumed a stellar flux of 23600 W/m$^2$ at the top of the atmosphere for a null zenith angle. We also assumed an internal luminosity of 0.73 W/m$^2$ corresponding to an intrinsic effective temperature (i.e. the blackbody temperature corresponding to the internal luminosity) of 60 K, as suggested by \cite{rogers10}.

As initial states for the 3D simulations, we used the 1D temperature profiles computed with the 1D version of the model (see figure \ref{figure_2}) and with no wind. The 1D model uses the same vertical discretization and physics (e.g. radiative transfer, convective adjustment) as the GCM, assuming a stellar zenith angle of 60$^{\circ}$ and a redistribution of energy over the entire planet.
 
For the 1D calculation, we ran the model for 1000 GJ1214b days (i.e. 1600 days) and increased the radiative heating/cooling by a factor proportional to the pressure when the pressure was higher than 0.1 bar. This enables faster convergence and is equivalent to running the model for around 10$^5$ days for the deepest level at 80 bars. This technique resulted in simulations very close to equilibrium after these 1600 days of integration, with relative differences between total emitted radiation and total absorbed radiation lower than 0.1$\%$ and with no observable variation in the temperature profile for longer simulations. The 3D simulations were also run for 1600 days but we did not use the technique to accelerate the convergence since the atmosphere is not in radiative equilibrium in 3D. The differences between total emitted radiation and total absorbed radiation for the final 3D states were lower than 1$\%$.

Simulations were performed with a dynamical timestep of 60 seconds  and a physical/radiative timestep of 300 seconds. The GCM outputs are either instantaneous values or daily averaged values. All the results we present here use daily averaged values. However, the 3D simulations show small time variability with, for instance, standard deviations for temperature generally lower than 10 K.

\section{Atmospheric dynamics}
\subsection{Thermal structure}
\begin{figure}[!h] 
\begin{center} 
	\includegraphics[width=7cm]{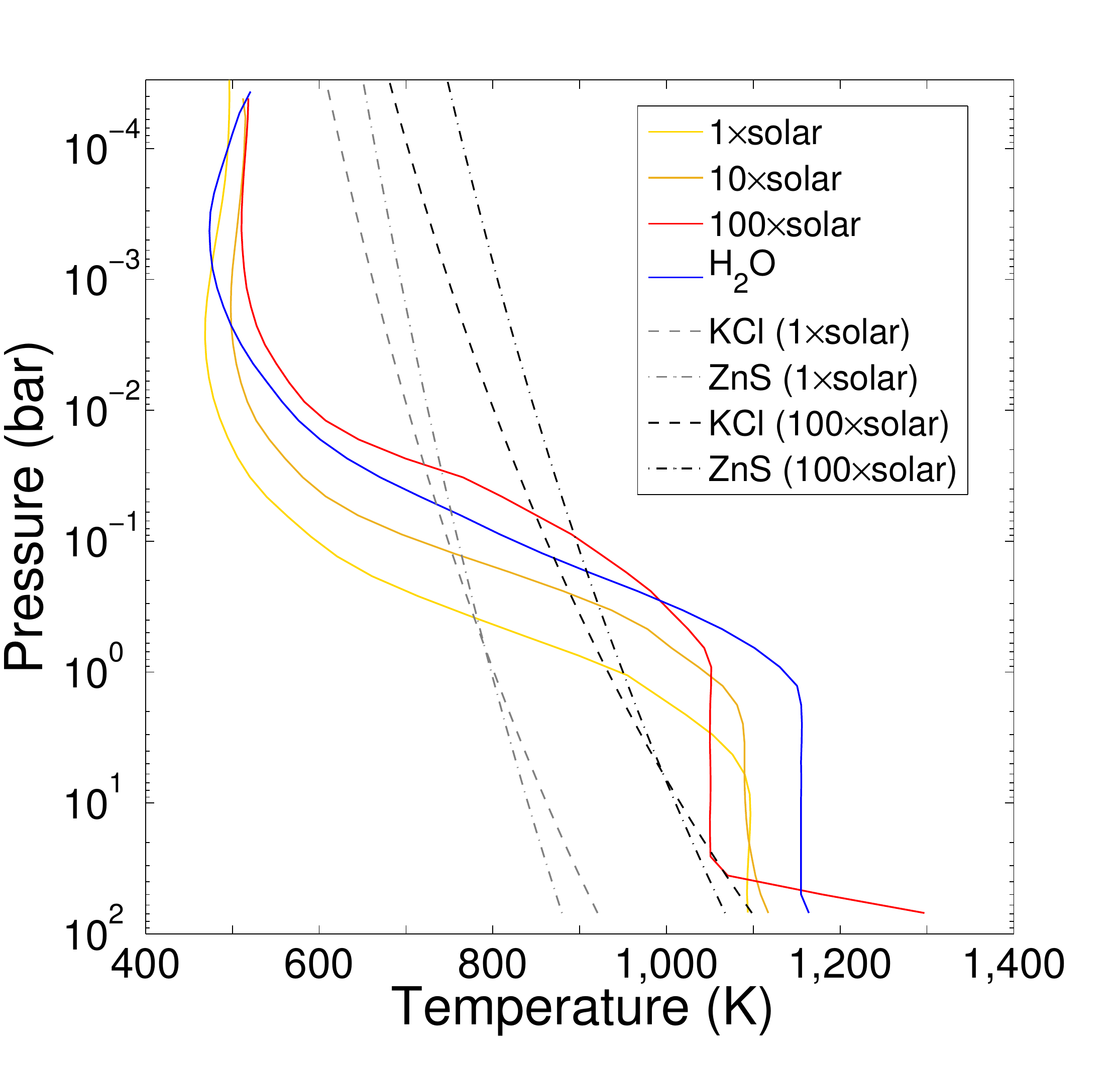}
	\includegraphics[width=7cm]{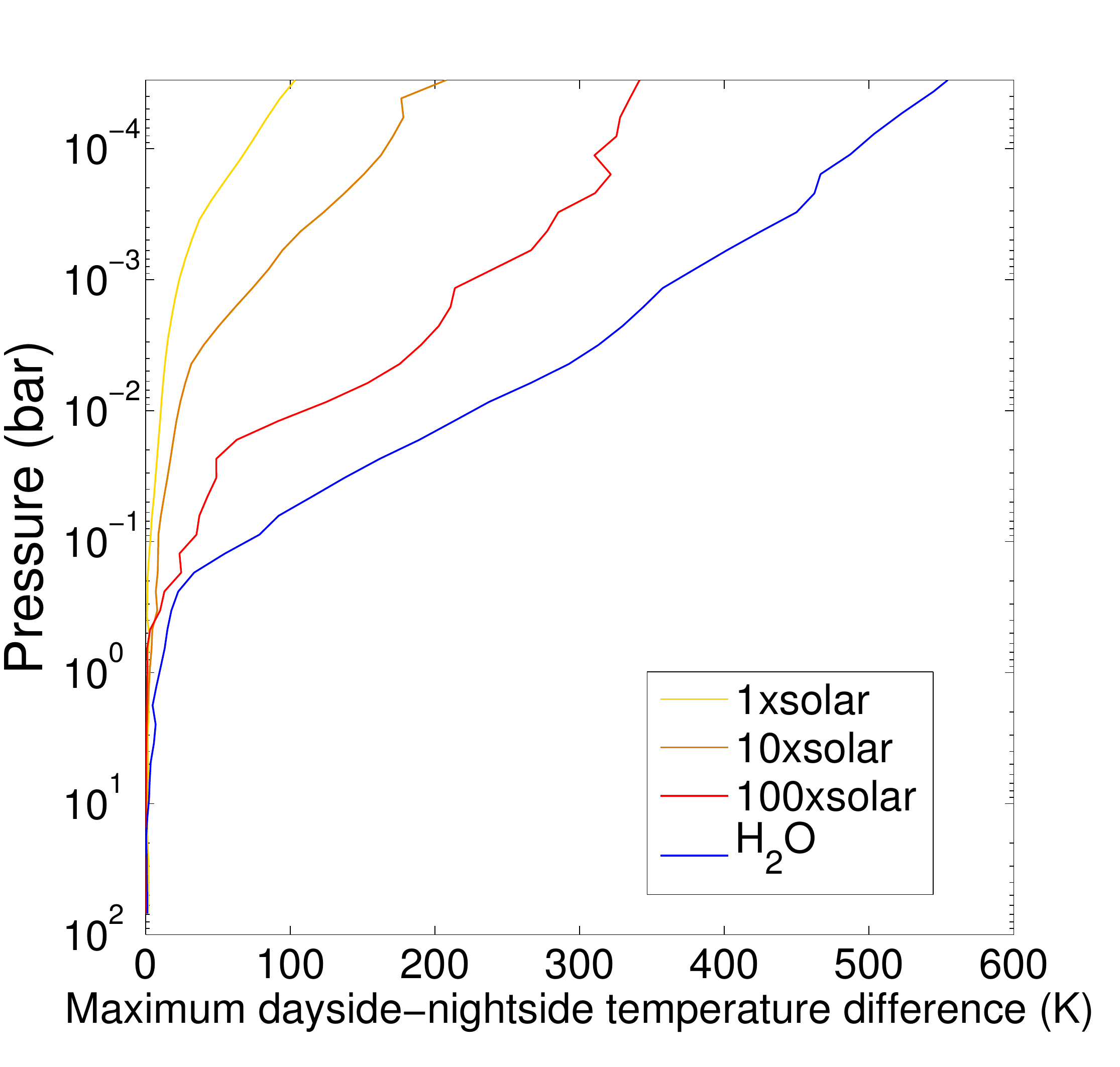}
\end{center} 
\caption{Left panel: temperature profiles from the 1D model for the different atmospheric compositions. Dashed and dotted lines correspond to the saturation vapor pressure curves for KCl and ZnS, for the solar (gray) and the 100$\times$solar (black) metallicity.
Right panel: maximum dayside-nightside temperature difference as a function of pressure. The differences are computed at each latitude and pressure level. The figure shows their maximum value at a given pressure.} 
\label{figure_2}
\end{figure}

The left panel in figure \ref{figure_2} shows the temperature profiles obtained from the 1D model to initialize the 3D runs. Dashed and dotted lines correspond to the saturation vapor pressure curves for KCl and ZnS for the solar and 100$\times$solar metallicity from \cite{morley12}. The temperature profile is only adiabatic below the isothermal region (i.e. below 40 bars for the 100$\times$solar composition and below 100 bars for the other cases) where it is controlled by the internal heat flux. For these 1D temperature profiles, KCl and ZnS should condense at around 0.4 bar for the solar metallicity and at around 60 mbar for the 100$\times$solar metallicity. These altitudes are similar to those obtained by previous 1D models \citep{miller-ricci12, morley13} and are fairly strongly dependent on the metallicity.
Our 3D simulations globally give the same condensation altitudes, however there are latitudinal variations because of the equator-pole temperature gradient. For instance, for the 100$\times$solar metallicity (the case with the strongest variations), KCl condensation would start at around 40 mbar at the equator and at around 200 mbar at the poles.

\begin{figure}[!h] 
\begin{center} 
	\includegraphics[width=7cm]{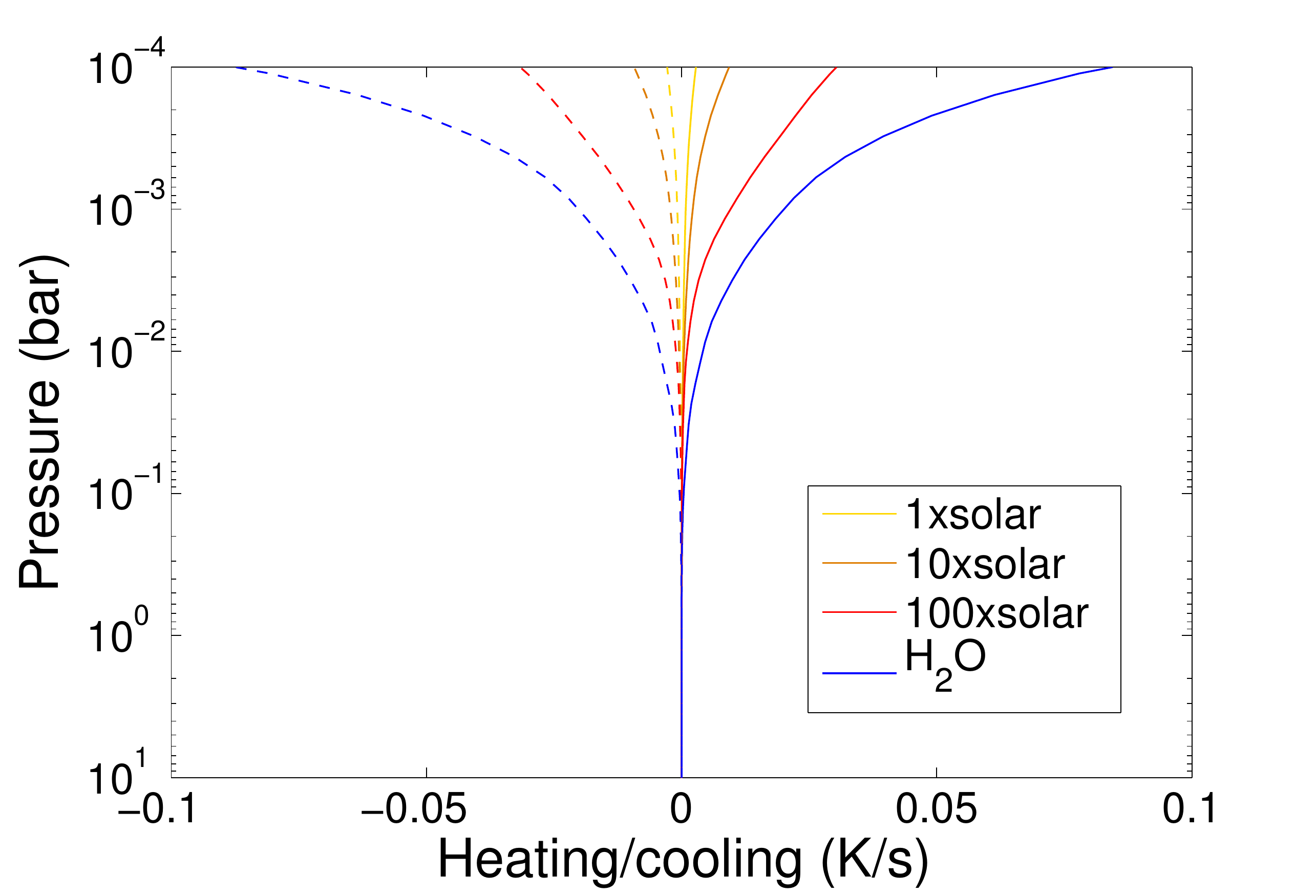}
	\includegraphics[width=7cm]{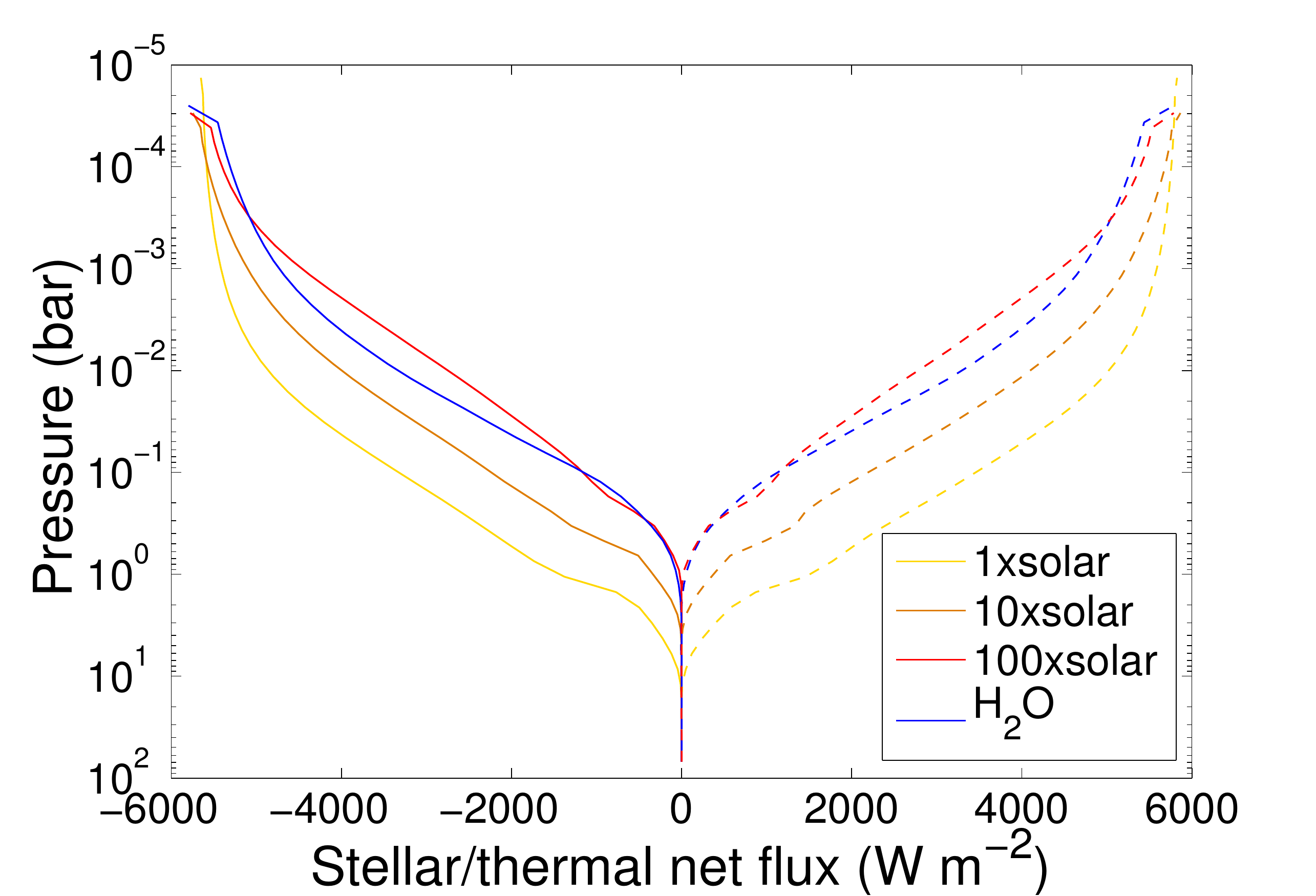}
\end{center}  
\caption{Global mean heating/cooling in K/s (left) and stellar/thermal net flux in W/m$^2$ (right) as a function of pressure. Solid lines correspond to the heating rate and the stellar net flux. Dashed lines correspond to the cooling rate and the thermal net flux.
}
\label{figure_3}
\end{figure}

In the 3D simulations, there are strong day-night temperature variations that become more pronounced with increasing atmospheric metallicity (see right panel in figure \ref{figure_2}). At 0.1 mbar the day-night temperature difference reaches 500 K for the water atmosphere but is lower than 100 K for the solar metallicity. The atmospheric opacity increases with metallicity while the specific heat capacity decreases. Both effects lead to larger heating/cooling rates above 0.1 bar for higher atmospheric metallicity (see figure \ref{figure_3} left). The altitude where stellar energy is deposited increases with metallicity (see figure \ref{figure_3} right). In our model the pure water and the 100$\times$solar atmospheres have similar net stellar flux. The lower water vapor absorption in the 100$\times$solar atmosphere compared to the pure water atmosphere is compensated by absorption by other gases, in particular methane.
With our model, we obtained net stellar and thermal fluxes similar to those in \cite{kataria14} for our four different cases. However, the day-night temperature difference for the solar metallicity is a factor of two weaker than in their model. This may be primarily due to the stronger equatorial jet we obtained for this case (see next paragraph).

\subsection{Zonal winds}
\begin{figure}[!h] 
\begin{center} 
	\includegraphics[width=7cm]{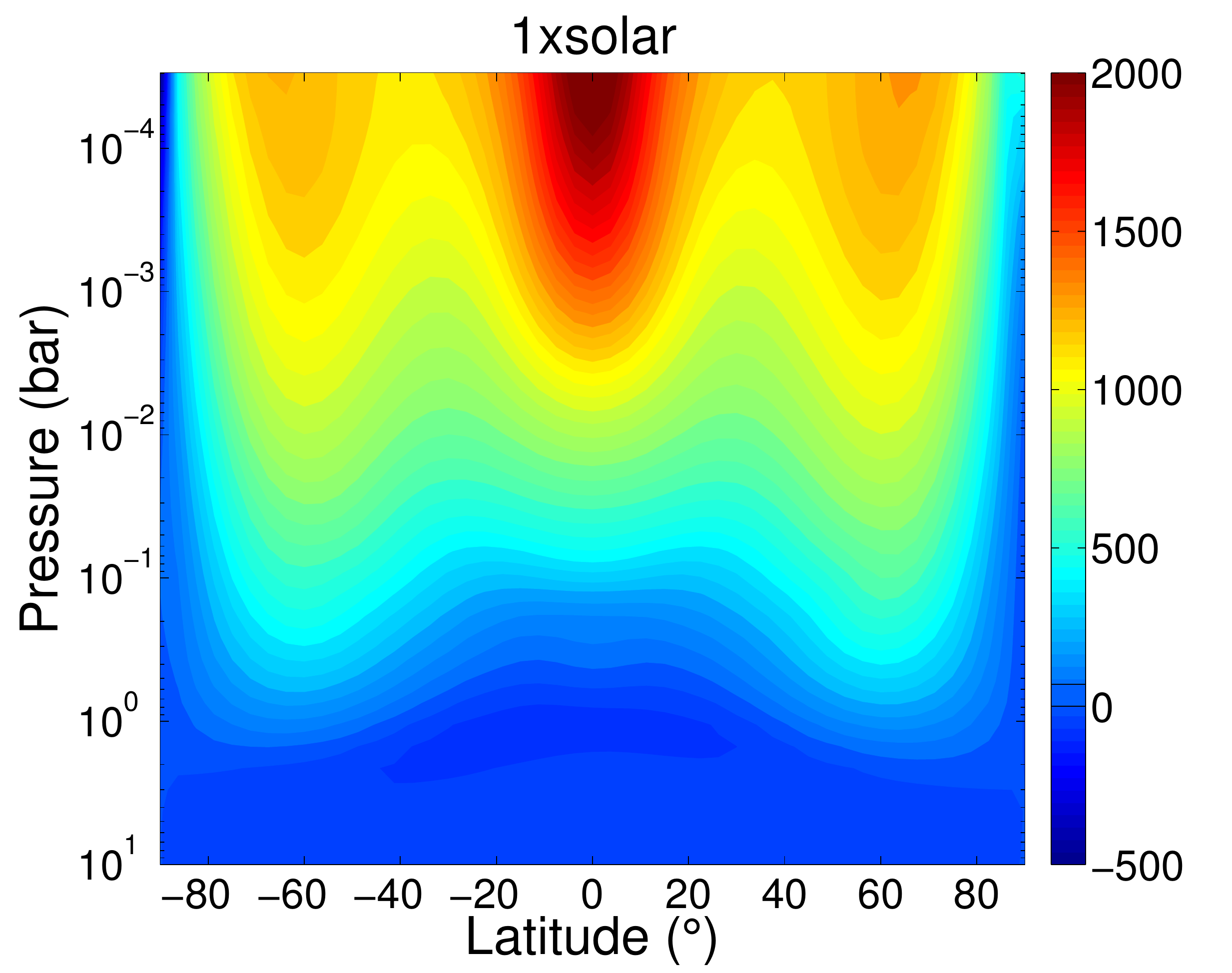}
	\includegraphics[width=7cm]{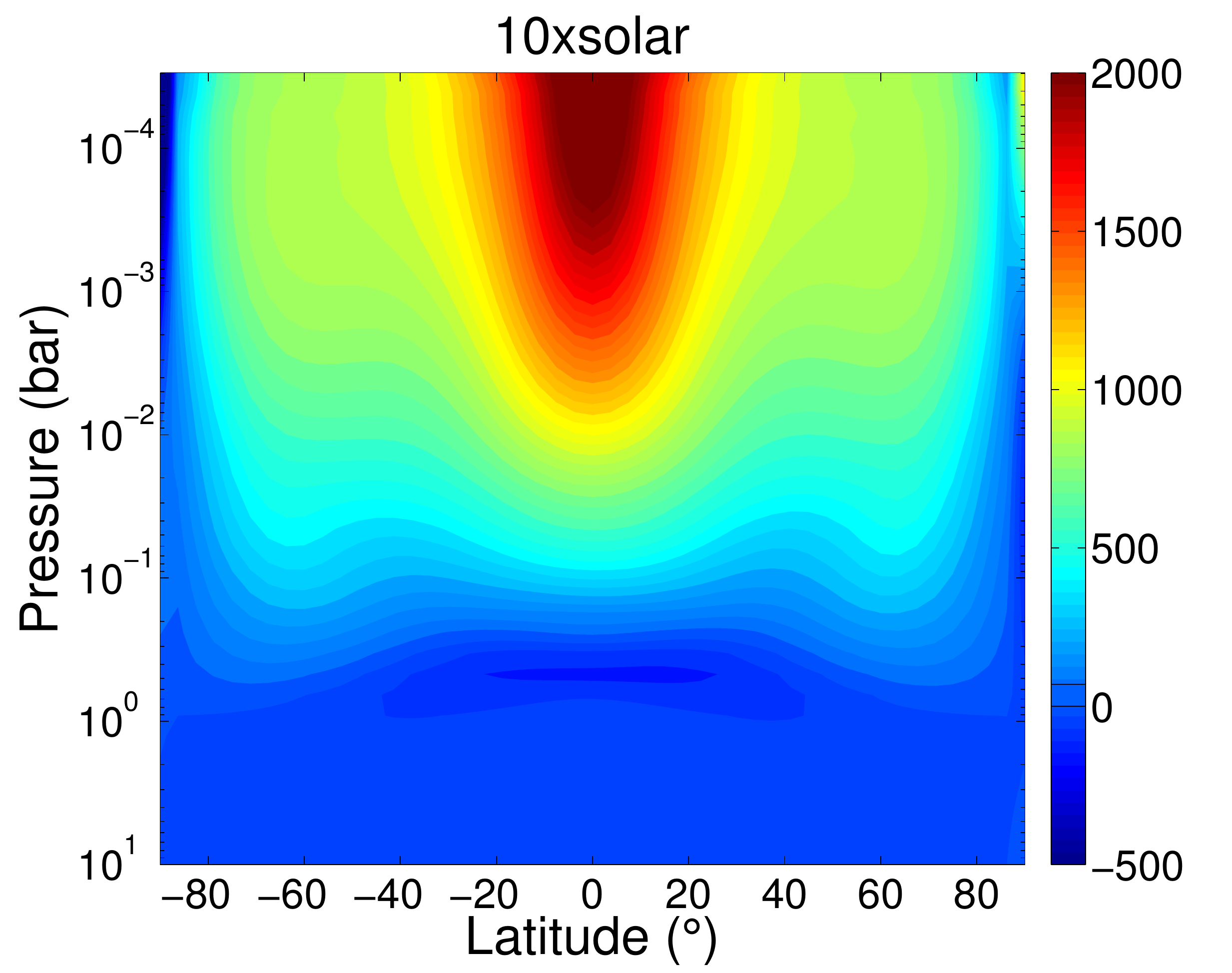}
	\includegraphics[width=7cm]{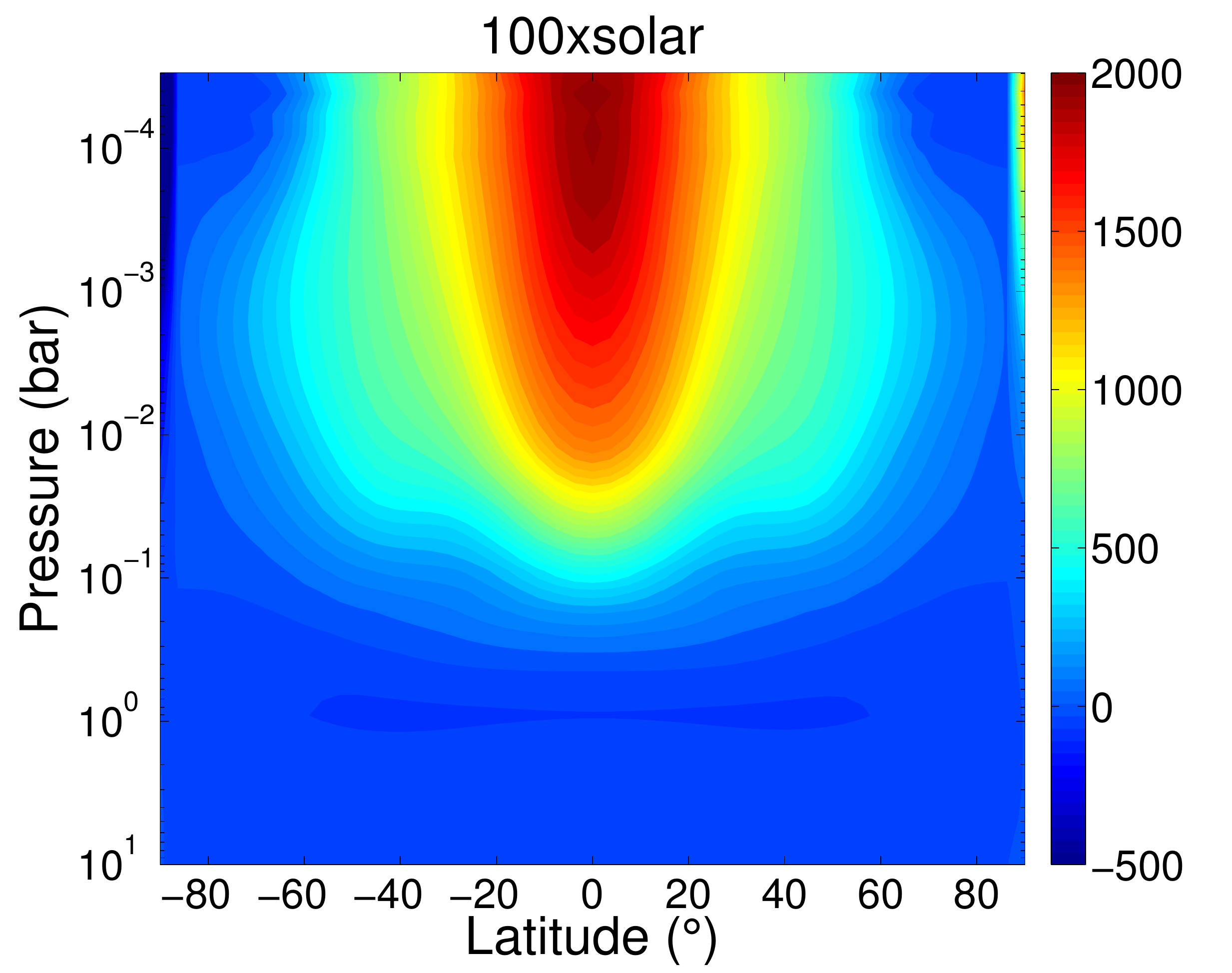}
	\includegraphics[width=7cm]{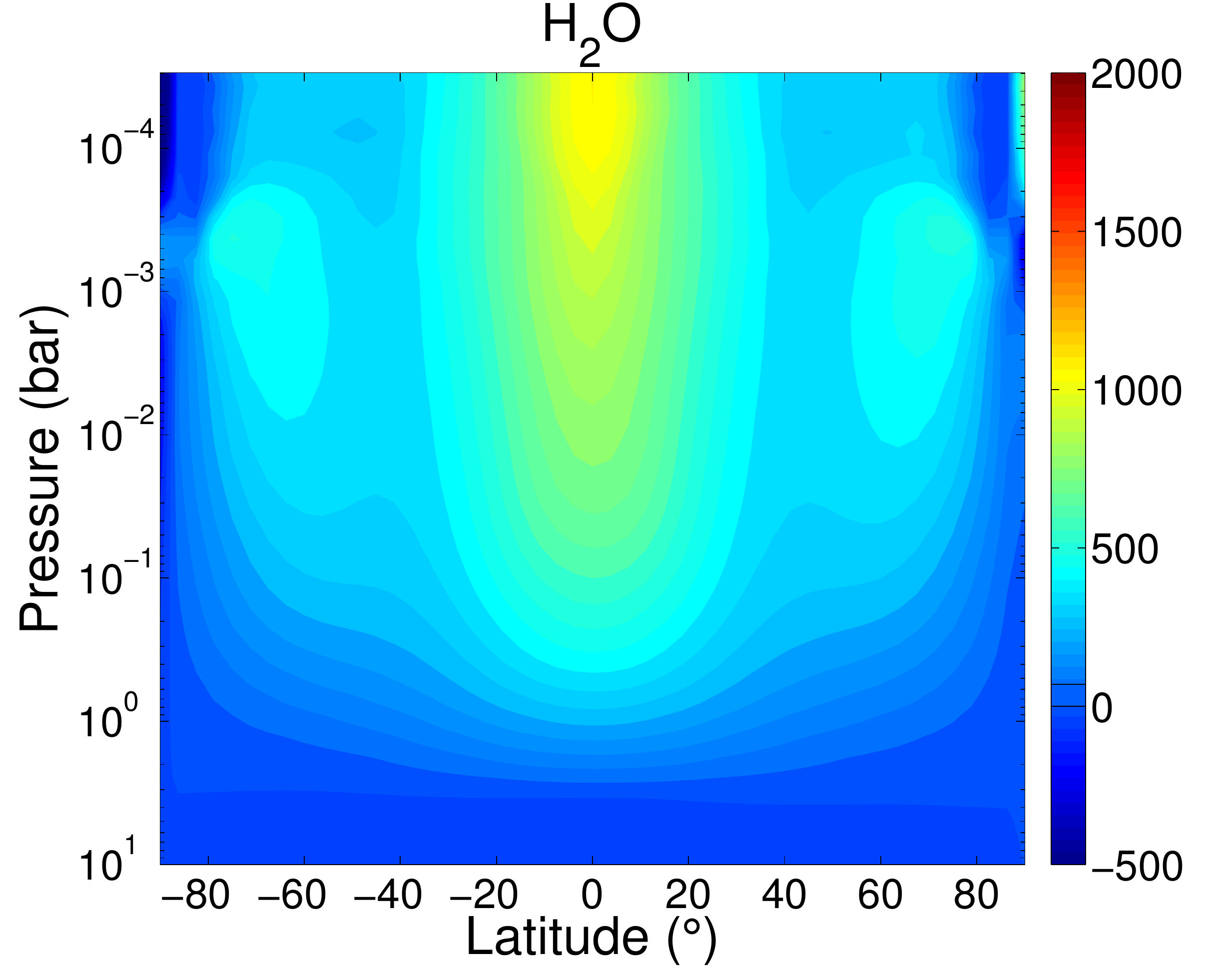}
\end{center}  
\caption{Zonally averaged zonal wind in m/s for the four atmospheric compositions.}
\label{figure_4}
\end{figure}

Concerning the zonal winds, all simulations lead to the development of an equatorial superrotating jet as expected for strongly irradiated tidally locked exoplanets \citep{showman11}. Figure \ref{figure_4} shows the zonally averaged zonal wind for our four atmospheric compositions. For the 1$\times$solar metallicity, there are also two high-latitudes jets. These jets are strongly reduced for the 10$\times$solar metallicity and vanish for the 100$\times$solar metallicity. Their development is controlled by latitudinal temperature gradients through the thermal wind equation \citep{lewis10}. With higher metallicity, stellar energy is absorbed higher in the atmosphere leading to weaker latitudinal temperature gradients in the deep atmosphere and so to less significant and shallower jets.
In contrast, the depth of the equatorial jet tends to increase with the metallicity leading to stronger dayside-nightside temperature gradients. This behavior (increase of the equatorial jet and decrease of the high-latitude jets with metallicity) was also obtained in the simulations of \cite{kataria14}.

Compared to the results from \cite{kataria14}, our GCM produces shallower high-latitude jets for the solar and 10$\times$solar metallicity cases and no jet for the 100$\times$solar metallicity. In their model, the high-latitude jets reach a maximum speed between 10$^{-1}$ and 10$^{-2}$ bar, while maximum speed occurs at the top of our model (at 0.03 mbar) for the solar and 10$\times$solar metallicity case. Therefore, high-latitude jets are stronger at low pressures in our model but weaker at high pressures.
For instance, for the solar metallicity, the high-latitude jet speed is around 1100 m/s and 500 m/s at respectively 10$^{-3}$ and 10$^{-1}$ bar in our model whereas it is around 900 m/s and 600 m/s at the same pressures in \cite{kataria14}. 
Concerning the equatorial jet, our model produces a stronger equatorial jet with a wind speed of 
1500 m/s at 10$^{-3}$ bar compared to 900 m/s in \cite{kataria14}. However, the equatorial jets are very similar between both models for higher metallicity.
These differences between both models for H-rich atmospheres may be due to some variations in atmospheric opacities. Otherwise, it may be due to the different dynamical cores,
based on a standard longitude-latitude grid with a small dissipation in the Generic LMDZ and on a cubed-sphere grid with no dissipation in \cite{kataria14}. However, our results globally remain similar to theirs.

Concerning the pure water atmopshere, our model produces an equatorial jet weaker than for H-rich atmospheres, with a maximum wind speed of around 1 km/s compared to 2 km/s for the others cases. For the water atmosphere, there also are two weak high-latitude jets with a maximum speed of around 500 m/s at 70$^\circ$N/S and at 1 mbar. \cite{kataria14} obtained the same structure and speeds for the zonal wind in the pure water atmosphere.

\subsection{Meridional circulation and vertical winds}
\begin{figure}[!h] 
\begin{center} 
	\includegraphics[width=7cm]{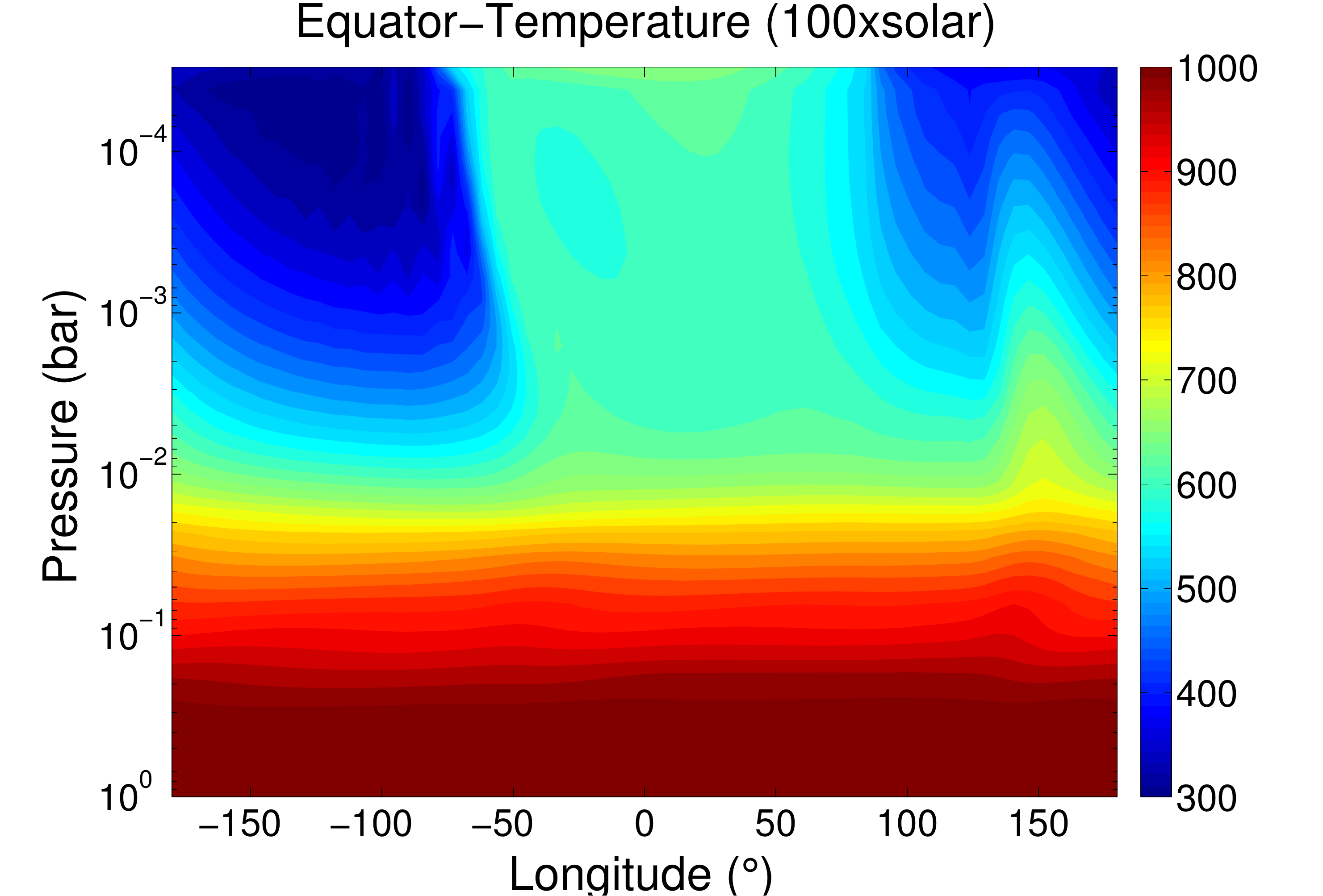}
	\includegraphics[width=7cm]{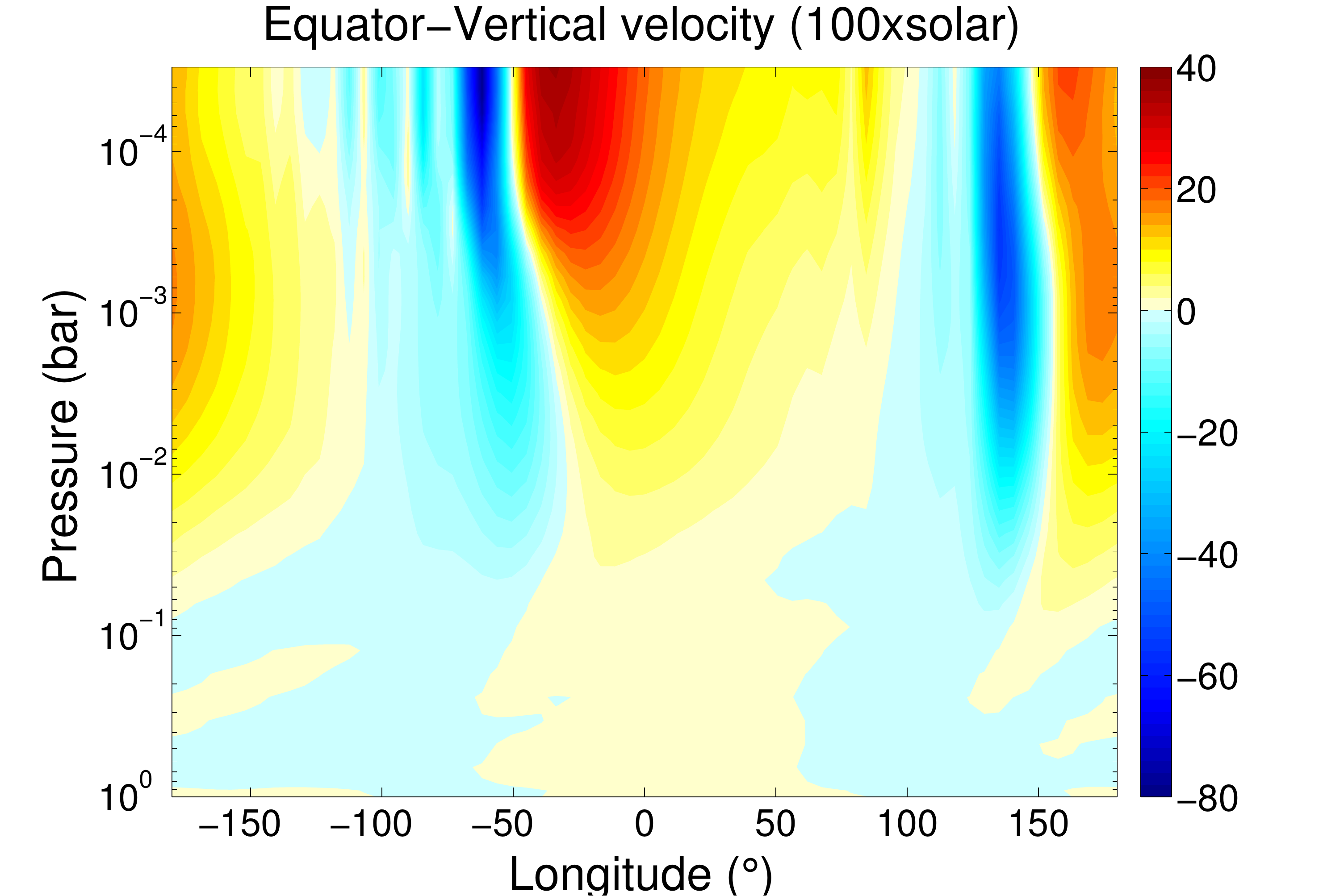}
	\includegraphics[width=7cm]{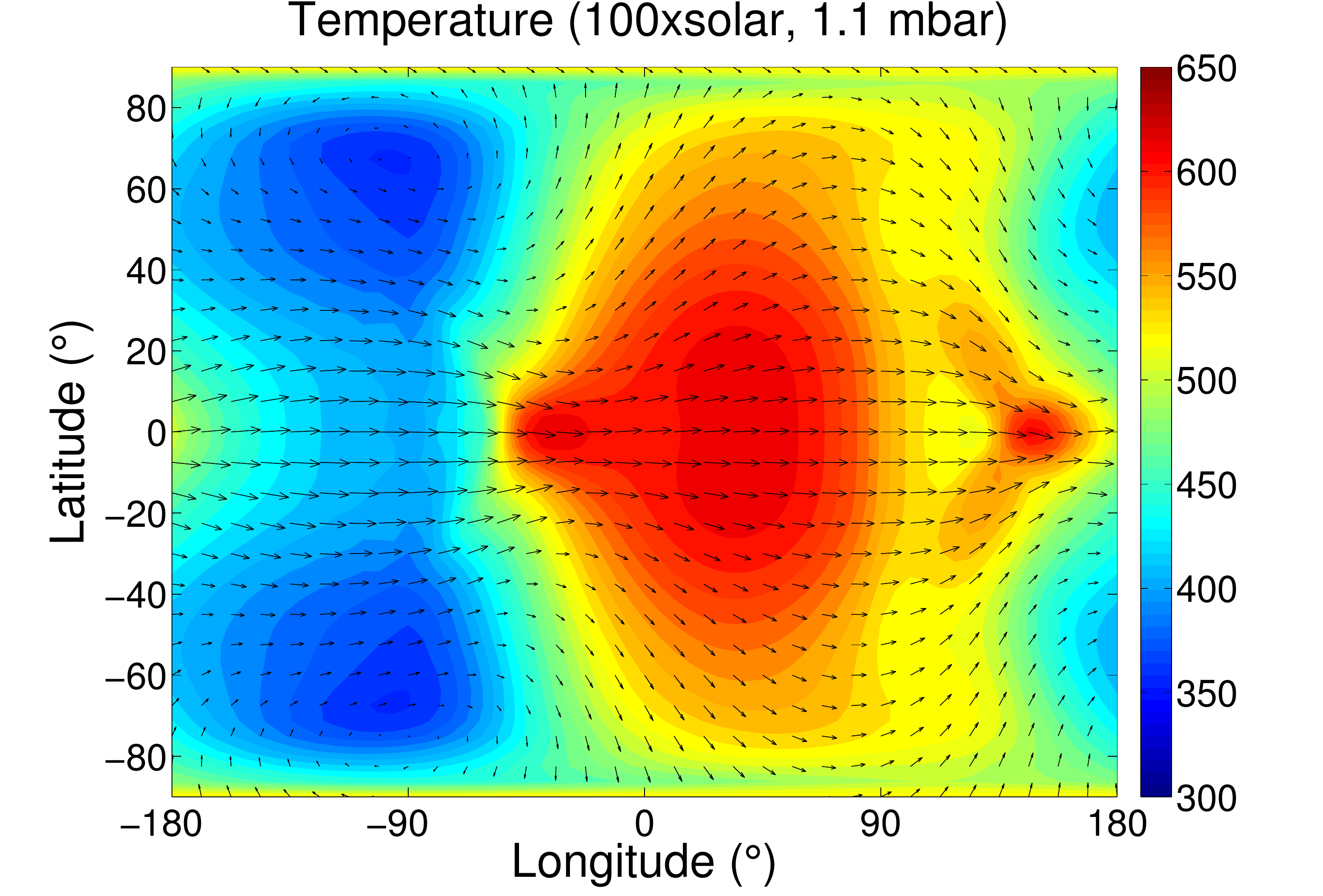}
	\includegraphics[width=7cm]{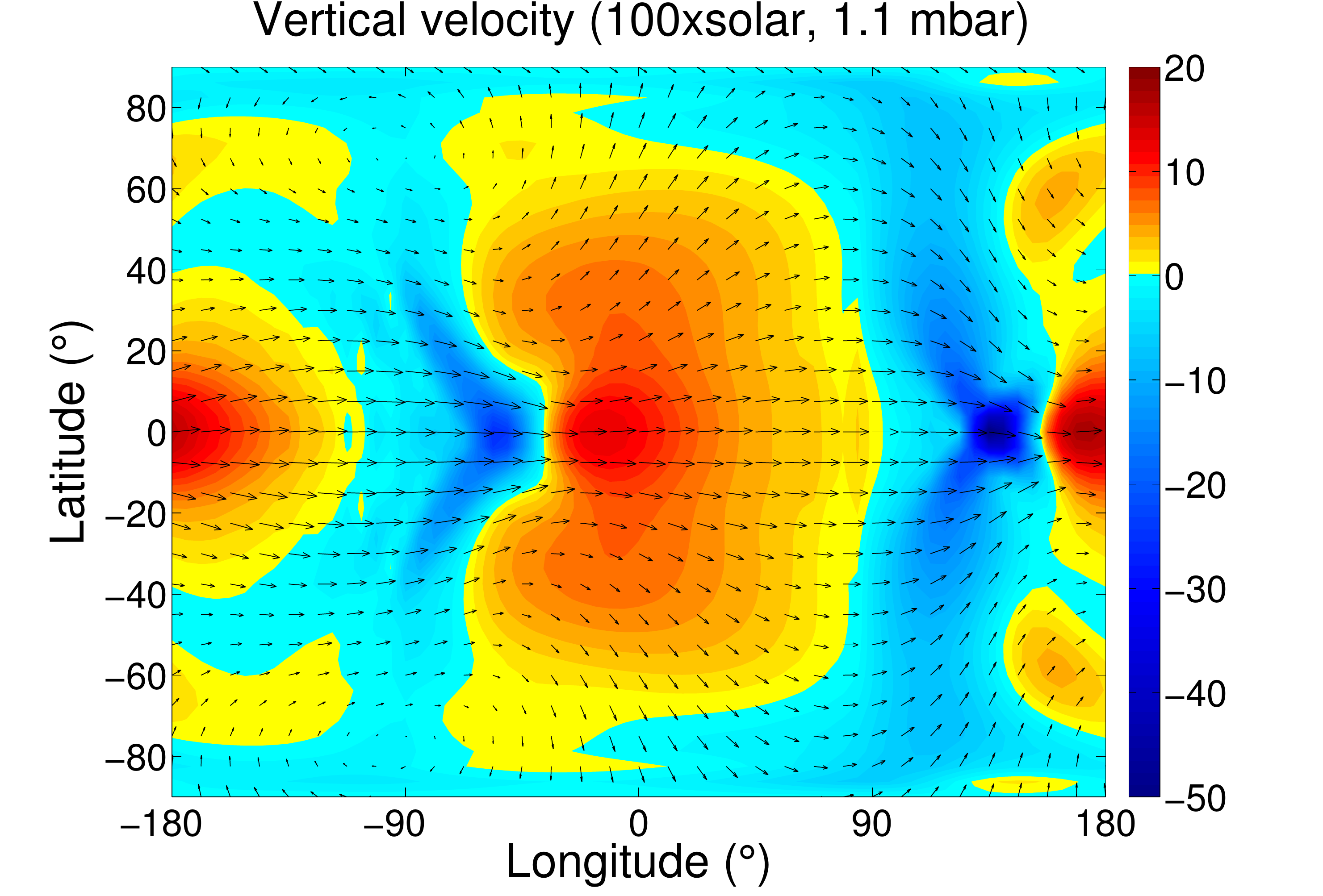}
\end{center}  
\caption{Temperature (in K) and vertical velocity (in m/s, positive value = upward wind) for the 100$\times$solar metallicity. 
The top panels show the temperature (left) and vertical velocity (right) at the equator versus longitude and pressure.
The bottom panels show the map of temperature (left) and vertical velocity (right) at 1.1 mbar. Black vectors correspond to the directions of horizontal winds.}
\label{figure_5}
\end{figure}

For a tidally locked, synchronously-rotating exoplanet, the heat redistribution from dayside to nightside is mostly done by the superrotating jets. There is also a meridional circulation redistributing heat from low to high latitudes. This is illustrated in figure \ref{figure_5}, which shows - for the 100$\times$solar metallicity -  equatorial temperature and vertical wind as a function of pressure and longitude, as well as maps of temperature and winds at 1.1 mbar (altitude close to the infrared photosphere). 
The dayside is characterized by upwelling centered near the sub-solar point with
 poleward winds transporting heat to high latitudes. The dayside poleward circulation crosses poles and is equatorward on the nightside, allowing a stronger heat redistribution. This circulation is shaped by standing Rossby and Kelvin planetary waves, responsible for the superrotation by pumping eastward momentum from high latitudes to the equator \citep{showman11}. Winds and temperatures are characterized by a chevron-shape pattern, centered at the equator and pointing to the east \citep{showman11}. Two strong downdrafts are present at the equator close to the terminators. They are associated with an adiabatic heating explaining the hot spot west of the anti-stellar point in Fig. \ref{figure_5}. They are also associated with a shock-like feature with a strong updraft at the substellar and anti-stellar points, also observed in 3D simulations of hot Jupiters \citep{showman09, rauscher10}.

\begin{figure}[!h] 
\begin{center} 
	\includegraphics[width=7cm]{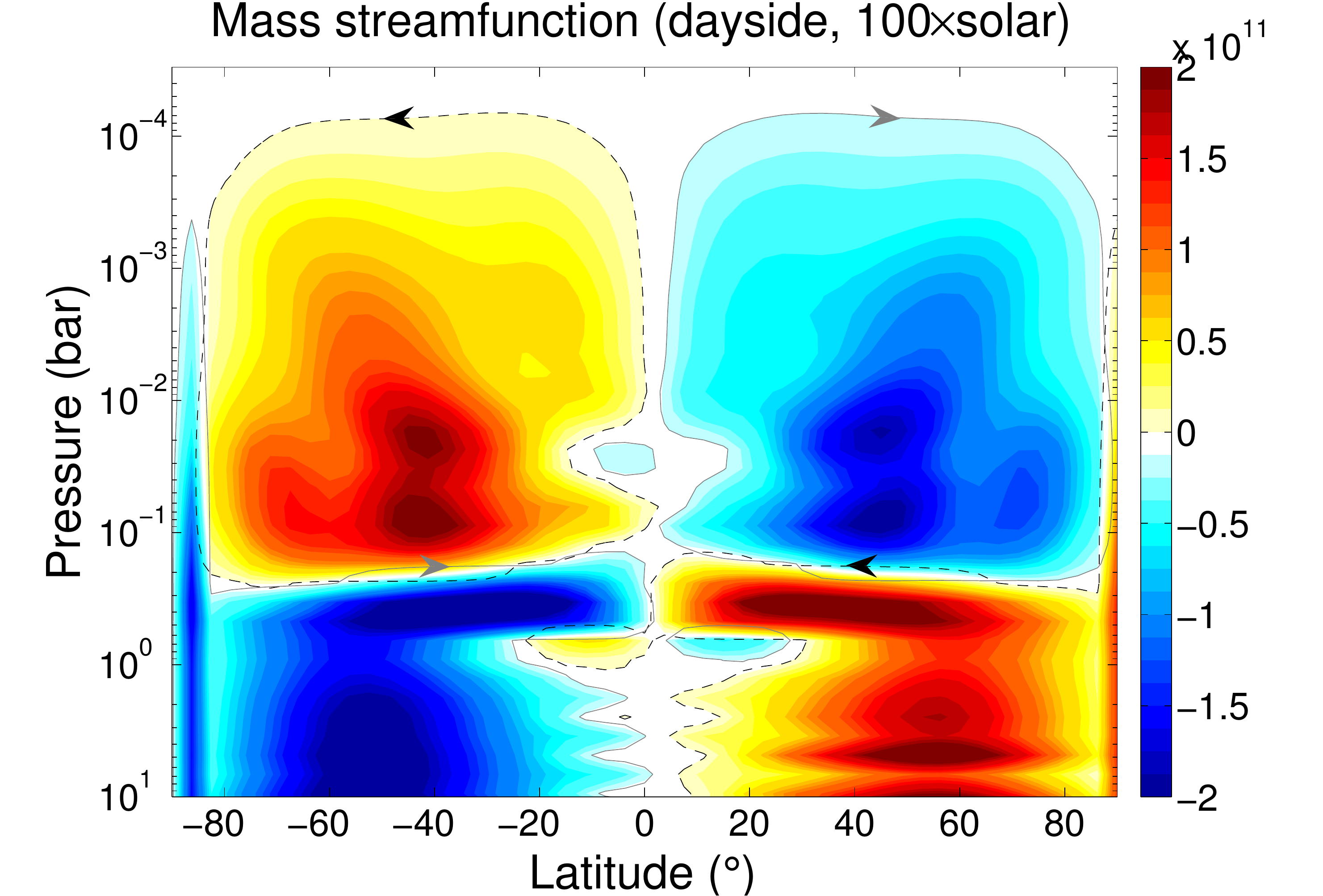}\\
	\includegraphics[width=7cm]{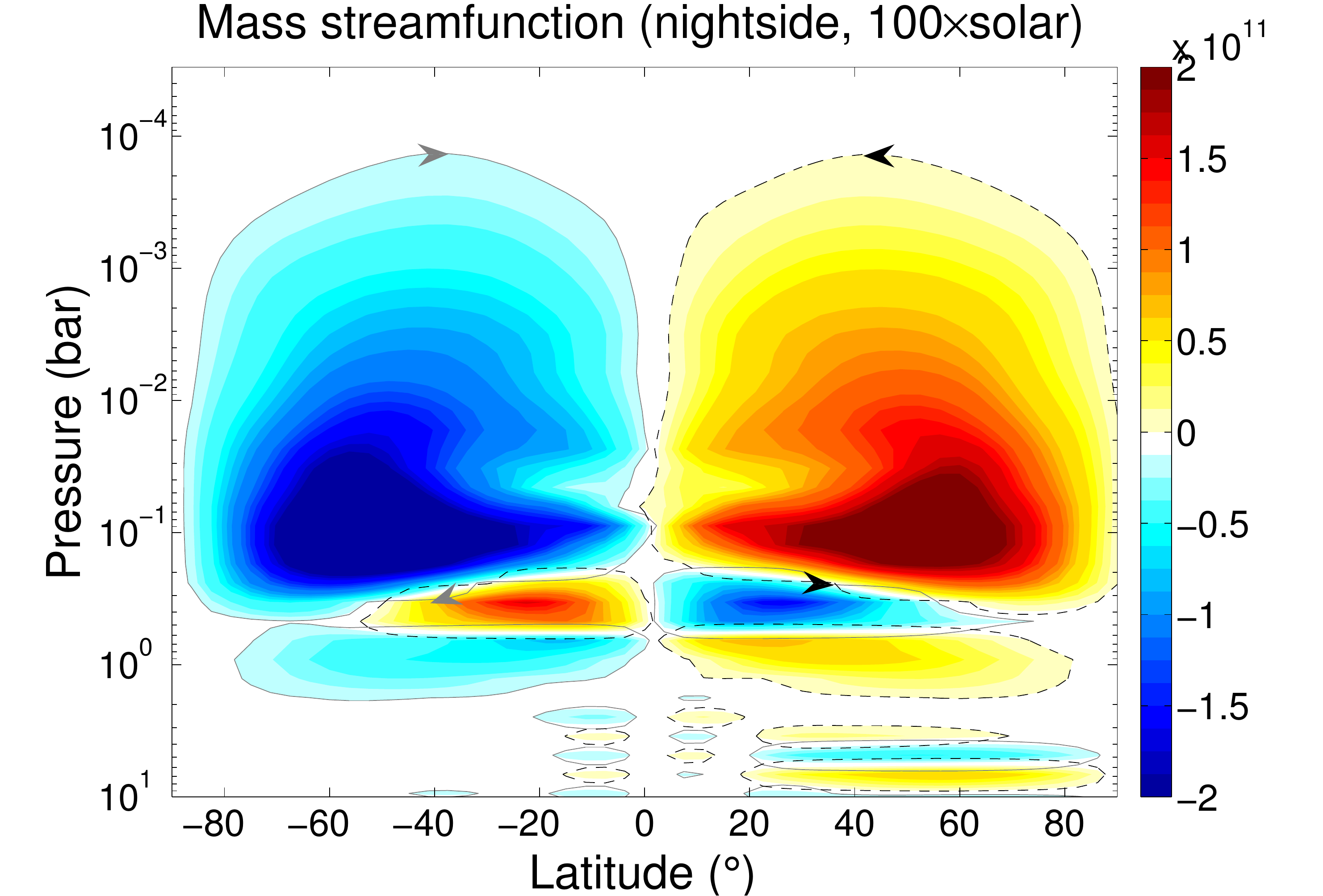}\\
        \includegraphics[width=7cm]{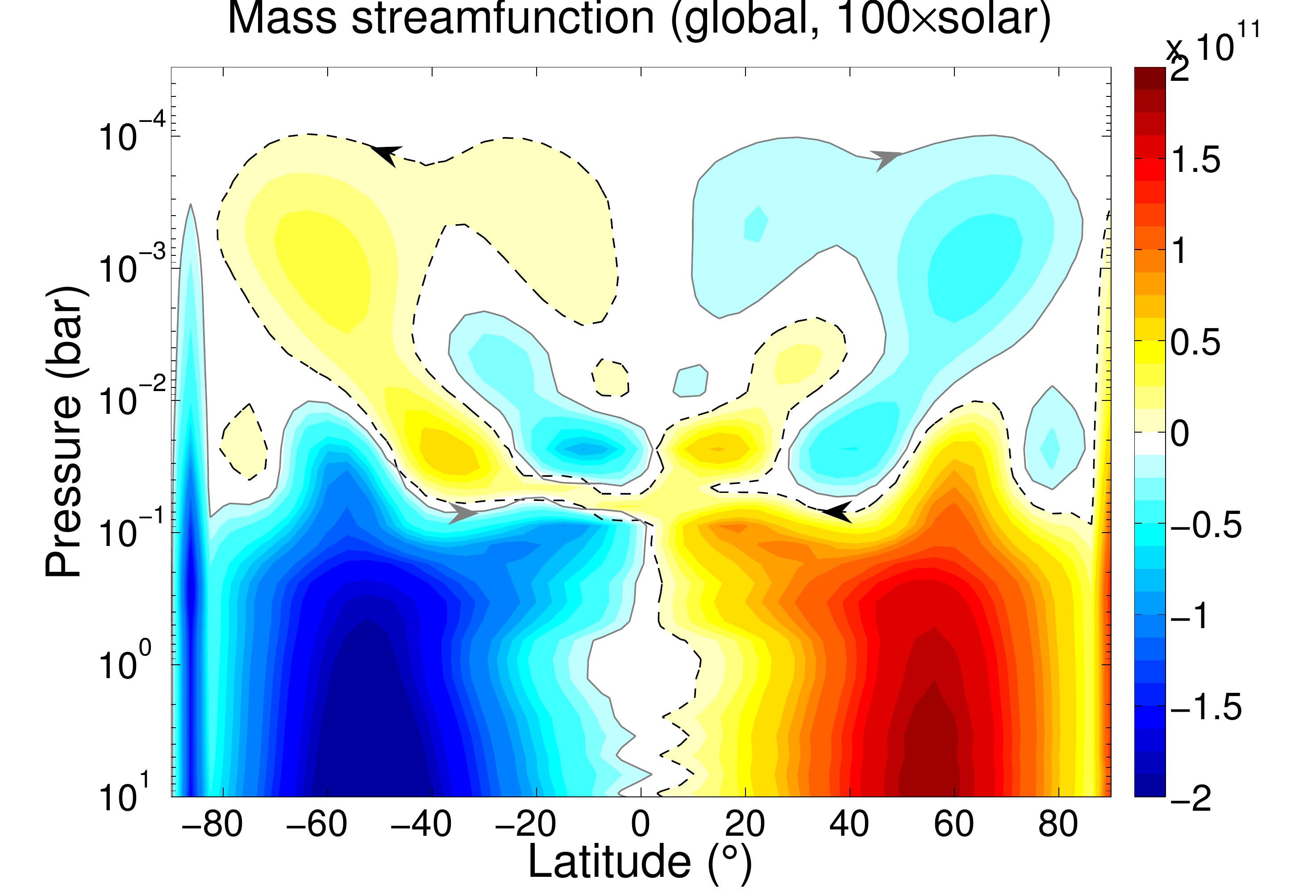}
\end{center}  
\caption{Zonal mean mass streamfunctions in kg/m/s for the 100$\times$solar metallicity case for the dayside (top), the nightside (middle) and globally (bottom). Positive (negative) values correspond to anti-clockwise (clockwise) circulation. Dashed (solid) lines are contours for the value of $\pm$10$^{10}$ kg/m/s. The dayside streamfunction was calculated by integrating longitudes between -39$^\circ$ and +90$^\circ$E. The nightside streamfunction was calculated by integrating the other longitudes (-180 $^\circ$ to -39$^\circ$E  and +90$^\circ$ to +180$^\circ$ E). The global mass streamfunction was calculated by integrating all longitudes and is equal to the sum of the dayside and nightside streamfunctions.}
\label{figure_6}
\end{figure}

Figure \ref{figure_6} shows the mass streamfunction for the 100$\times$solar metallicity case for the dayside, the nightside and globally (i.e. dayside+nightside). For the dayside, we integrated longitudes between -39$^\circ$ and +90$^\circ$ E only, to avoid the very strong downdraft located west of the substellar point. For the nightside, we integrated the mass streamfunction over all the  longitudes not included in the dayside integration.
The dayside is characterized by a large Hadley circulation for pressures below 0.1 bar with two cells extending from the equator to the poles. On the nightside, the mean meridional circulation reverses. 
This reversal is due to equatorward winds on the nightside and to the two very strong downdrafts located at the equator and close to terminators. These downdrafts are typically twice as strong as the equatorial updrafts (see figure \ref{figure_5}). With an integration between -90$^\circ$ and +90$^\circ$ E for the dayside, so including the downdraft located west of the substellar point, we obtained similar circulations but weaker at the equator.
When looking at the global mass streamfunction, it appears that the anti-Hadley circulation is stronger in the deep atmosphere (i.e. between 0.1 and 0.01 bar) than the dayside Hadley circulation, but weaker in the upper atmosphere.
In reality, the meridional circulation corresponds rather to two large cells crossing the poles and going from the dayside to the nightside but the division into a dayside Hadley circulation and a nightside anti-Hadley circulation allows an evaluation of which one dominates.
These results are for the 100$\times$solar atmospheric composition, and yet the other atmospheric compositions present the same circulations for pressure lower than around 0.1 bar.

\begin{figure}[!h] 
\begin{center} 
	\includegraphics[width=7cm]{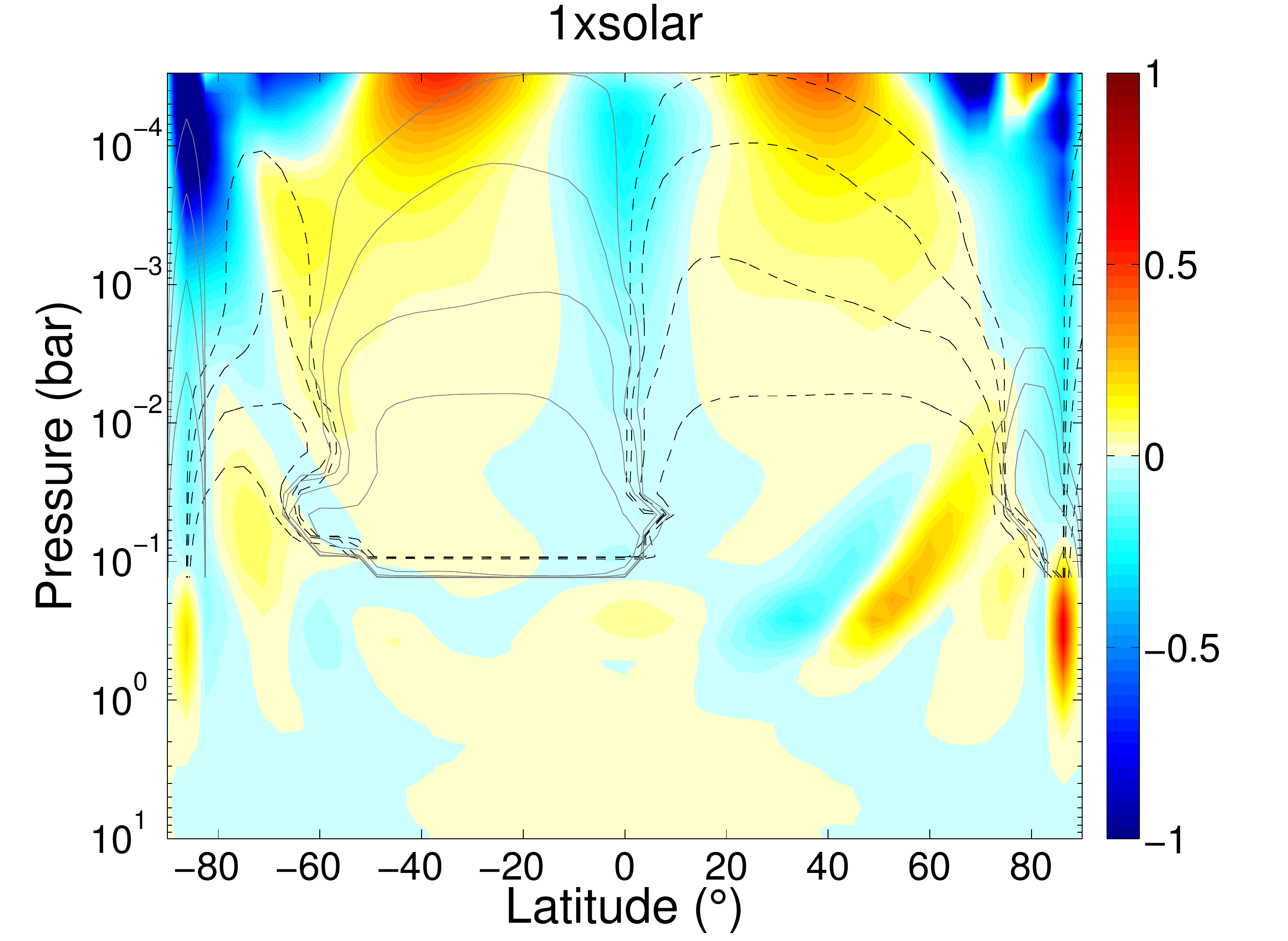}
	\includegraphics[width=7cm]{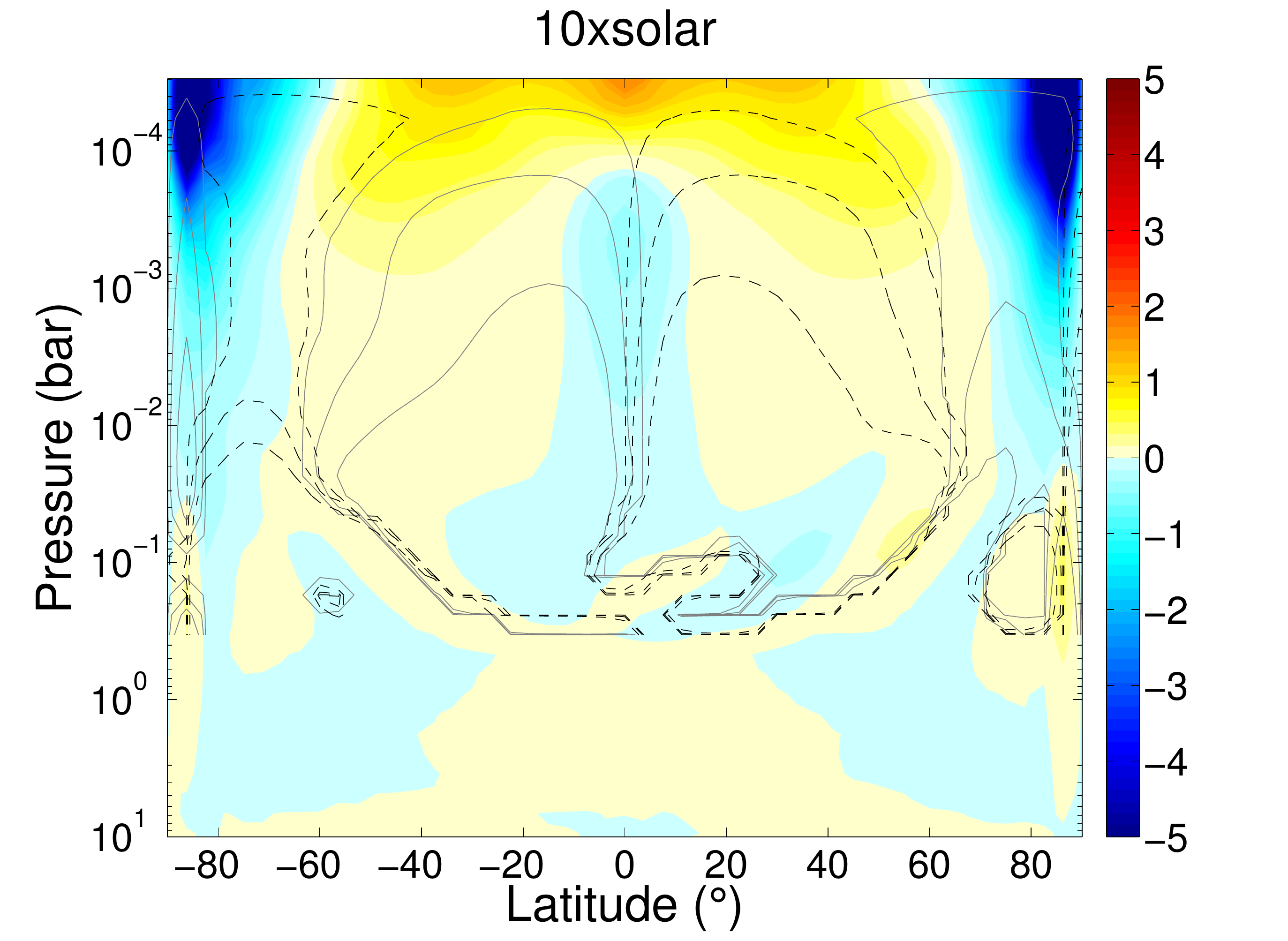}
	\includegraphics[width=7cm]{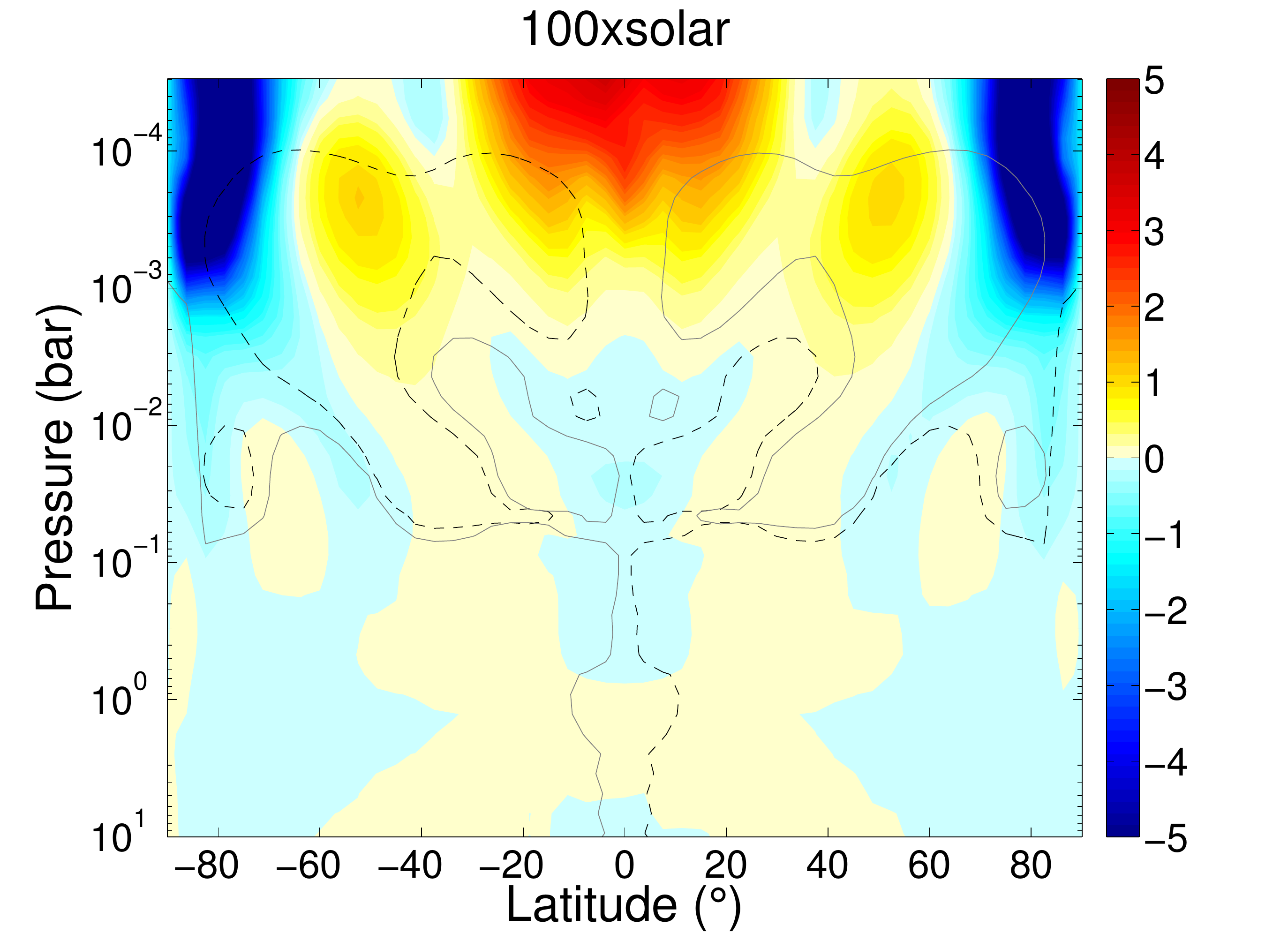}
	\includegraphics[width=7cm]{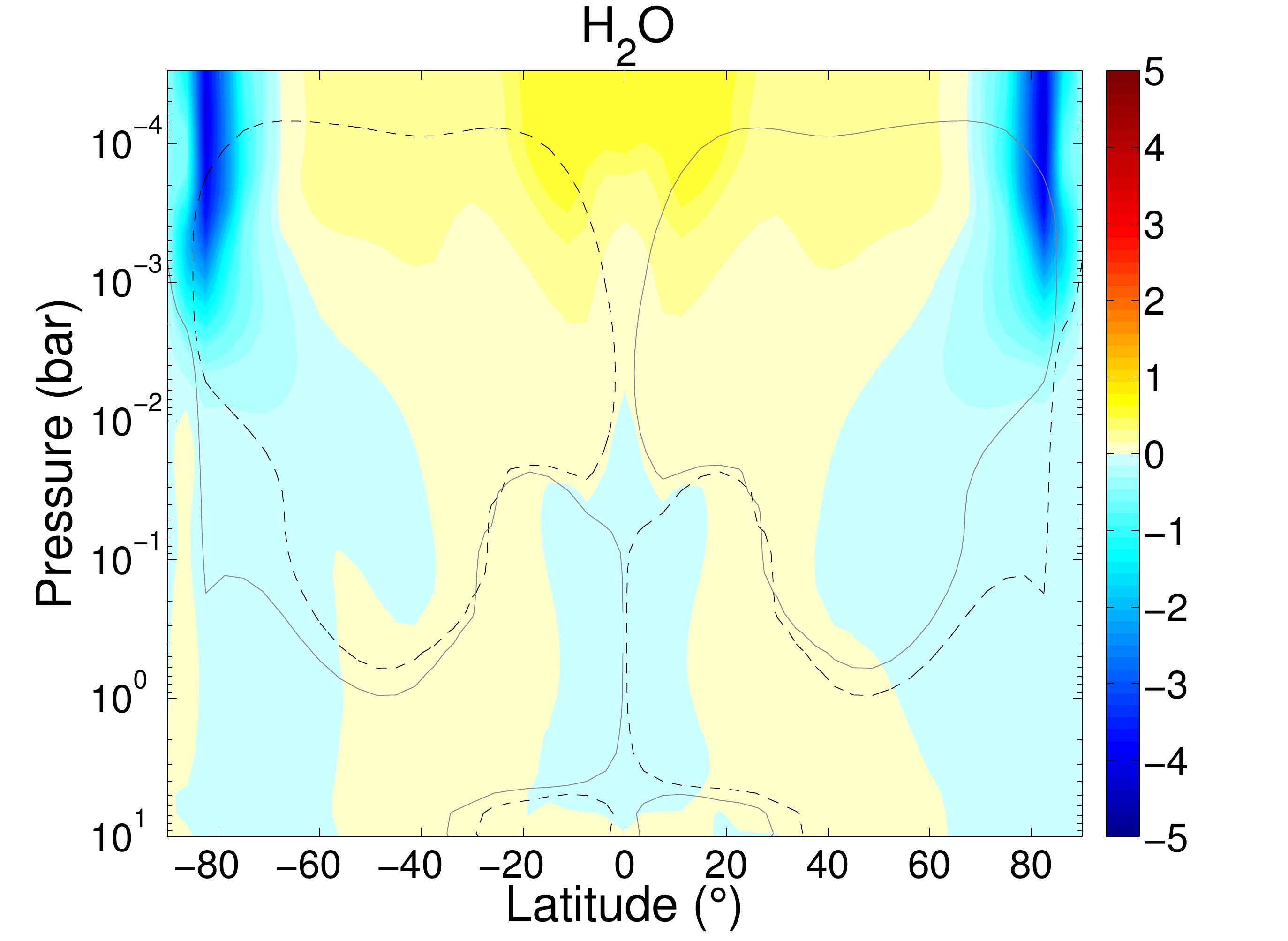}
\end{center}  
\caption{Zonally-averaged vertical wind in m/s (positive value = upward wind). Solid (dashed) lines correspond to contours of the zonally-averaged mean mass streamfunction with clockwise (anti-clockwise) rotation.}
\label{figure_7}
\end{figure}

Figure \ref{figure_7} shows a few contours of the mass streamfunctions averaged zonally over all longitudes and superposed on the zonal mean vertical winds. The mass streamfunctions follow variations of the zonal mean vertical winds.
For the 1$\times$ and 10$\times$solar metallicity at pressures lower than 0.1 bar, the zonally averaged circulation is characterized by an upwelling at mid-latitudes (between around 20$^\circ$-60$^\circ$) and a downwelling at the equator. The zonal mean circulation is thus dominated by the nightside anti-Hadley circulation.
The circulations for the 100$\times$solar metallicity and the pure water atmospheres are more complex, with the anti-Hadley circulation dominating at pressures greater than 10 mbar and the Hadley circulation above. As detailed before, the dayside Hadley circulation becomes stronger than the nightside anti-Hadley circulation in the upper atmosphere, explaining this transition. 
Therefore the zonal mean anti-Hadley circulation is present in every case but it is located deeper in the atmosphere for higher metallicity.

\begin{figure}[!h] 
\begin{center} 
	\includegraphics[width=10cm]{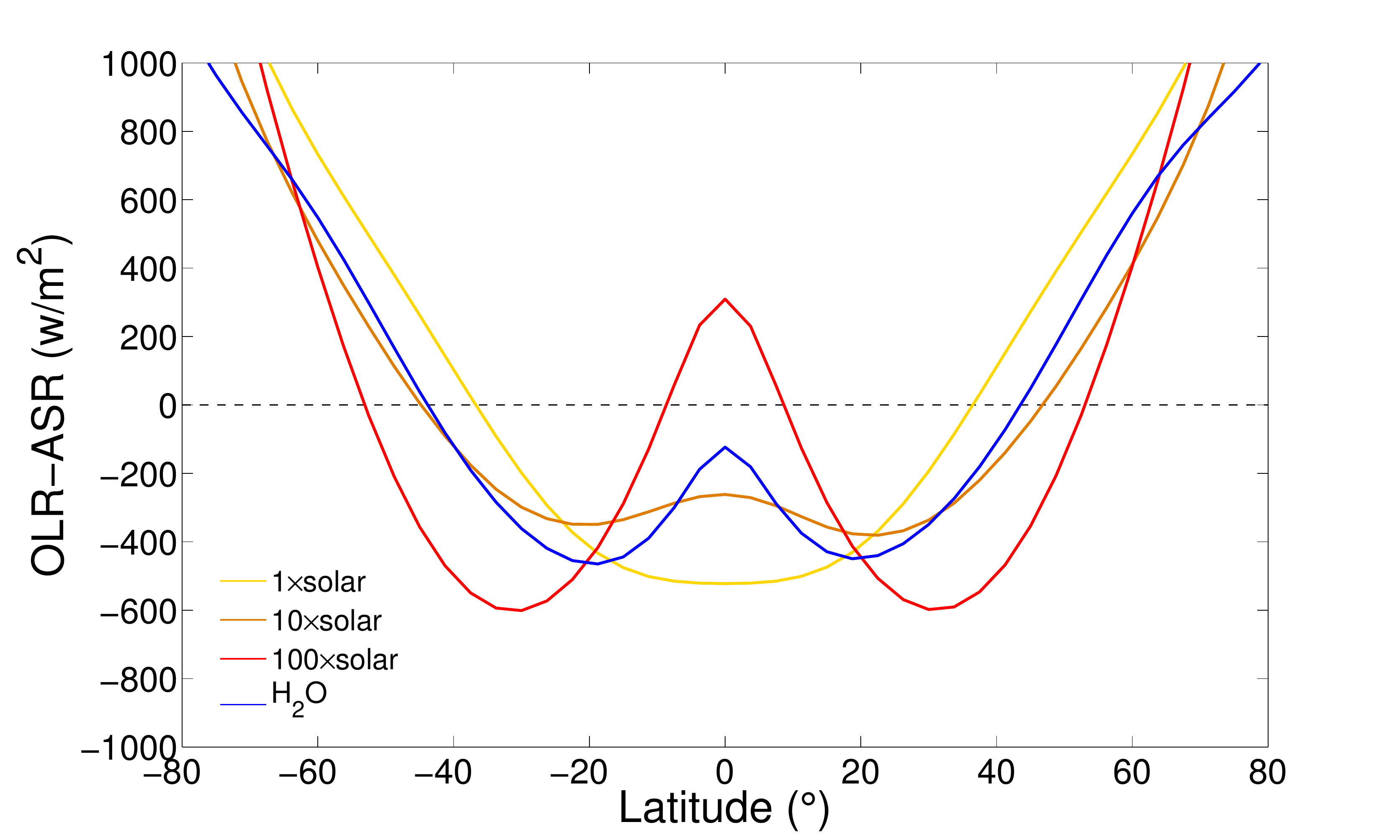}
	\includegraphics[width=10cm]{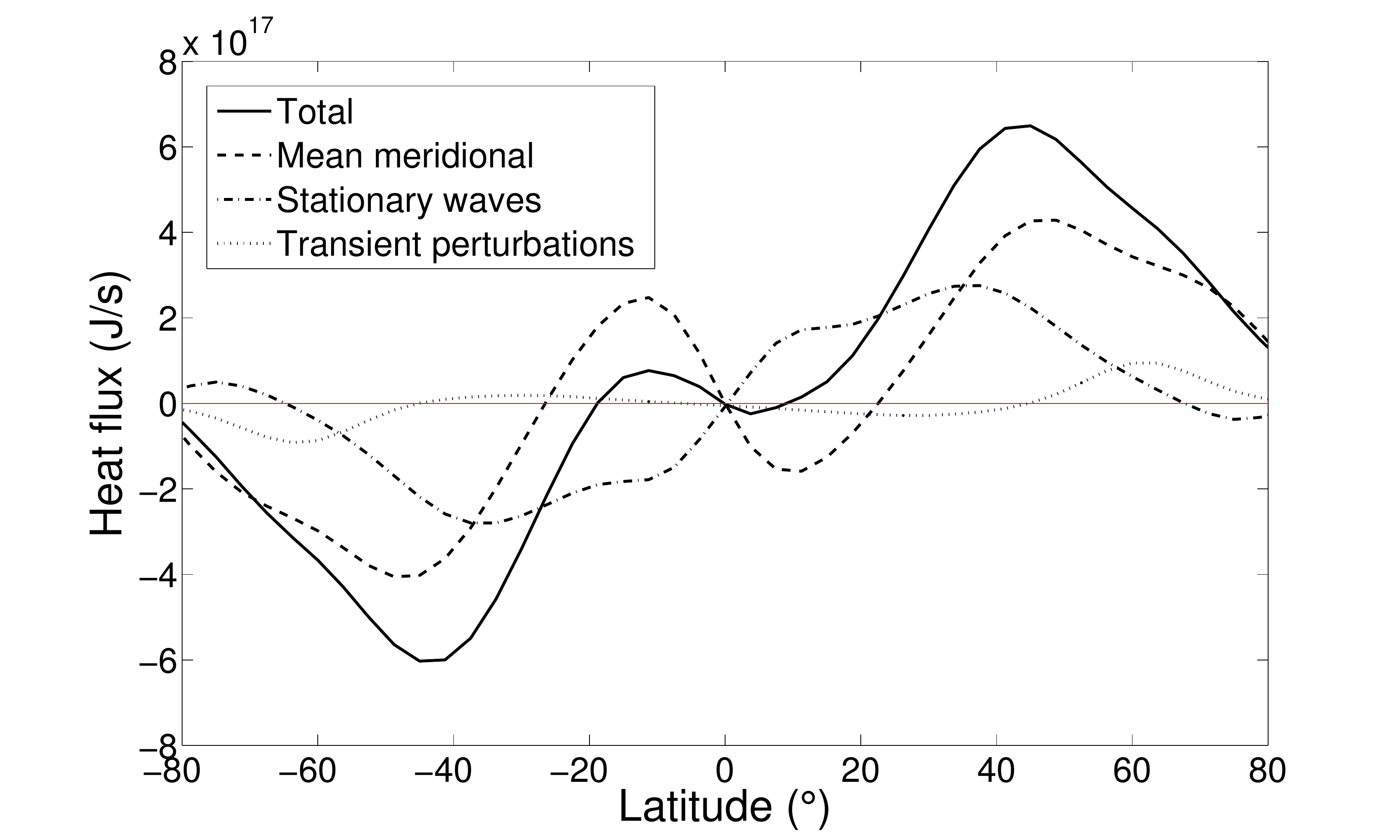}
\end{center}  
\caption{Heat transport by the meridional atmospheric circulation.
Top panel: zonal mean difference between the outgoing emitted flux and the absorbed stellar flux for the different atmospheric compositions. Positive (negative) values correspond to regions globally warmed (cooled) by the circulation.
Bottom panel: decomposition of the total heat transport (black solid line) for the 100$\times$solar metallicity case into heat transport by the mean meridional circulation (black dashed line), by stationary waves (black dashed-dotted line) and by transient perturbations (black dotted line). Positive (negative) values correspond to northward (southward) heat fluxes.}
\label{figure_7b}
\end{figure}

The top panel in figure \ref{figure_7b} shows the zonal mean difference between the outgoing longwave radiation (OLR) and the absorbed stellar radiation (ASR). Positive values correspond to regions with surplus energy (compared with the radiative equilibrium) transported by the atmospheric circulation from regions with a deficit of energy (negative values). Globally, the meridional circulation transports heat from low latitudes to high latitudes. However, the anti-Hadley circulation limits this poleward transport at low latitudes. The equatorial downdrafts produce an adiabatic heating (see for instance the equatorial temperature at 150$^\circ$E in figure \ref{figure_5}). The difference between OLR and ASR exhibits a local maximum at the equator due to this adiabatic heating. For the 100$\times$solar metallicity, the meridional circulation contributes to heating at the equator in average.

The bottom panel in figure \ref{figure_7b} shows the vertically and zonally-integrated heat flux and its decomposition into components of the heat flux transported by mean meridional circulation, stationary waves and transient perturbations. We computed the heat flux from the meridional wind $v$ and the dry static energy $e=(c_pT+gz)$, and we decomposed the heat flux as:
\begin{equation}
[\overline{ev}]=[\overline{e}][\overline{v}]+[\overline{e^*v^*}]+\overline{[e'][v']}
\end{equation}
where square brackets are time averages and asterisks are deviations from time averages, overbars are zonal averages and primes are deviations from zonal averages. The terms on the right represent, respectively, the transport by the mean meridional circulation, the transport by the stationary waves and the transport by the transient (time-dependent) perturbations.
For the 100$\times$solar metallicity case, the total heat transport is strongly impacted by the sationary waves. At a given pressure level, the dry static energy is higher (both $T$ and $z$ are higher) on the dayside with net poleward winds than on the nightside with net equatorward winds. Therefore, the stationary waves associated with the day-night contrast produce a net poleward heat transport, counterbalancing the equatorward heat transport associated with the mean anti-Hadley circulation below 10 mbar.

Finally, the anti-Hadley circulation we described in this section should be present on a large range of warm tidally locked exoplanets and has a fundamental impact on vertical mixing and cloud transport.

\section{Vertical mixing}
We performed an analysis of the strength of the vertical mixing in GJ1214b's  atmosphere using the Generic LMDZ GCM with simple tracers. The goal was to understand how the general circulation, and in particular the zonally averaged circulation described in the previous section, controls the transport and distribution of potential cloud or haze particles. 
From this 3D analysis, equivalent 1D eddy diffusion coefficients were obtained, which may be of use for 1D chemical models.

\subsection{Implementation in the GCM}
We used a similar method to the analysis of vertical mixing in the hot Jupiter HD 209458b by \cite{parmentier13}. We added particles with a fixed diameter and with a constant abundance $n_0$ below 1 bar, with this region corresponding to the source/sink of particles. We let the circulation transport these particles and allowed them to sediment at a terminal velocity described in the next section.
Contrary to \cite{parmentier13}, we do not assume that particles evaporate on the dayside, which would reduce the downward sedimentation tracer flux. Retaining the dayside particles is based on our model dayside temperatures, which are not high enough to significantly evaporate KCl or ZnS clouds even for the 100$\times$solar metallicity case.

We ran simulations for the different atmospheric compositions with particle radii from 0.1 to 10 microns and with a particle density corresponding to solid KCl (2000 kg/m$^3$), a possible composition for GJ1214b's clouds \citep{miller-ricci12}. We considered micrometric particles because they correspond to the estimated size for possible high cloud/haze particles that would produce a flat transit spectrum in visible and near-infrared \citep{miller-ricci12, morley13}.
For the initial state, we assumed a uniform particle abundance everywhere in the atmosphere and equal to $n_0$ (the constant abundance below 1 bar). We used $n_0$=10$^{-9}$ kg/kg in all simulations. This value is arbitrary and the next figures always show the relative abundance $n/n_0$.

\subsection{Sedimentation}
We computed the sedimentation of particles assuming that they fall at the terminal velocity given by
\citep{fuchs64, ackerman01}:
\begin{equation} 
V_f=\frac{2 \beta r^2 g (\rho_p-\rho)}{9\eta}
\end{equation} 
where $r$ is the particle radius, $g$ is the gravitational acceleration of GJ1214b, $\rho_p$ is the particle density (we chose $\rho_p$=2000 kg/m$^3$ corresponding to solid KCl), $\rho$ is the atmosphere density, $\eta$ is the viscosity of the atmospheric gas and $\beta$ is the Cunnigham slip factor, which describes non-continuum effects. An experimental expression of the Cunnigham slip factor is \citep{fuchs64}:

\begin{equation} 
\beta= 1+ K_n \left( 1.256 + 0.4 e^{-1.1/K_n} \right)
\label{eq_beta}
\end{equation} 

where $K_n$ is the Knudsen number, equal to the ratio of the mean free path to the size of radius of the particle:

\begin{equation} 
K_n= \frac{\lambda}{r}
\end{equation} 

The mean free path $\lambda$ is given by:

\begin{equation} 
\lambda = \frac{k_B T}{\sqrt{2} \pi d^2} \frac{1}{P} 
\end{equation} 
with $k_B$ the Boltzmann constant, $T$ and $P$ the temperature and pressure of the gas and $d$ the molecular diameter.

For the hydrogen-rich atmospheres, we considered that the dynamical viscosity was equal to that of pure H$_2$. Following \cite{ackerman01} and \cite{parmentier13}, we used the formula:

\begin{equation} 
\eta_{\rm H_2} = \frac{5}{16} \frac{\sqrt{\pi m k_B T}}{\pi d^2}  \frac{(k_B T/ \epsilon)^{0.16}}{1.22}
\end{equation} 
where m is the molecular mass, $d$=2.827$\times 10^{-10}$ is the molecular diameter and $\epsilon$=59.7k$_B$ K is the Lennard-Jones potential for H$_2$. The ratio on the right in the formula corresponds to the deviation from the hard-sphere model.

H$_2$O is a polar molecule that deviates too strongly from the hard-sphere model to be described by formula (8) \citep{crifo89}. For the water atmosphere, we used a parameterization of the dynamical viscosity of H$_2$O for dilute conditions and given by \cite{sengers84}:

\begin{equation} 
\eta_{\rm H_2O} = \frac{ \eta^{\star} \sqrt{\frac{T}{T_\star}}} { \sum\limits_{k=0}^{3} a_k \left( \frac{T}{T_\star} \right)^{-k}   }
\end{equation} 

with $\eta^{\star}$= 10$^{-6}$ Pa/s, $T_\star$= 647.27 K, $a_0$= 0.0181583, $a_1$= 0.0177624, $a_2$= 0.0105287 and $a_3$= -0.0036744.
This parameterization is valid for temperatures ranging from around 300 K to 1100 K. The dilute approximation is valid up to 10 bars for GJ1214b conditions (see \cite{sengers84} for an expression with  correction terms to the dilute approximation), and is thus compatible with the pressure range we considered for vertical mixing.
For the calculation of $K_n$ for the water-rich atmosphere, we used a molecular diameter $d_{H_2O}$  depending on temperature and given by
\citep{crifo89}:

\begin{equation} 
d_{H_2O}=  4.597\times 10^{-10} {   \rm m}  \left(\frac{ T}{300 {\rm K}}\right)^{-0.3} 
\end{equation} 

\begin{figure}[!h] 
\begin{center} 
	\includegraphics[width=7cm]{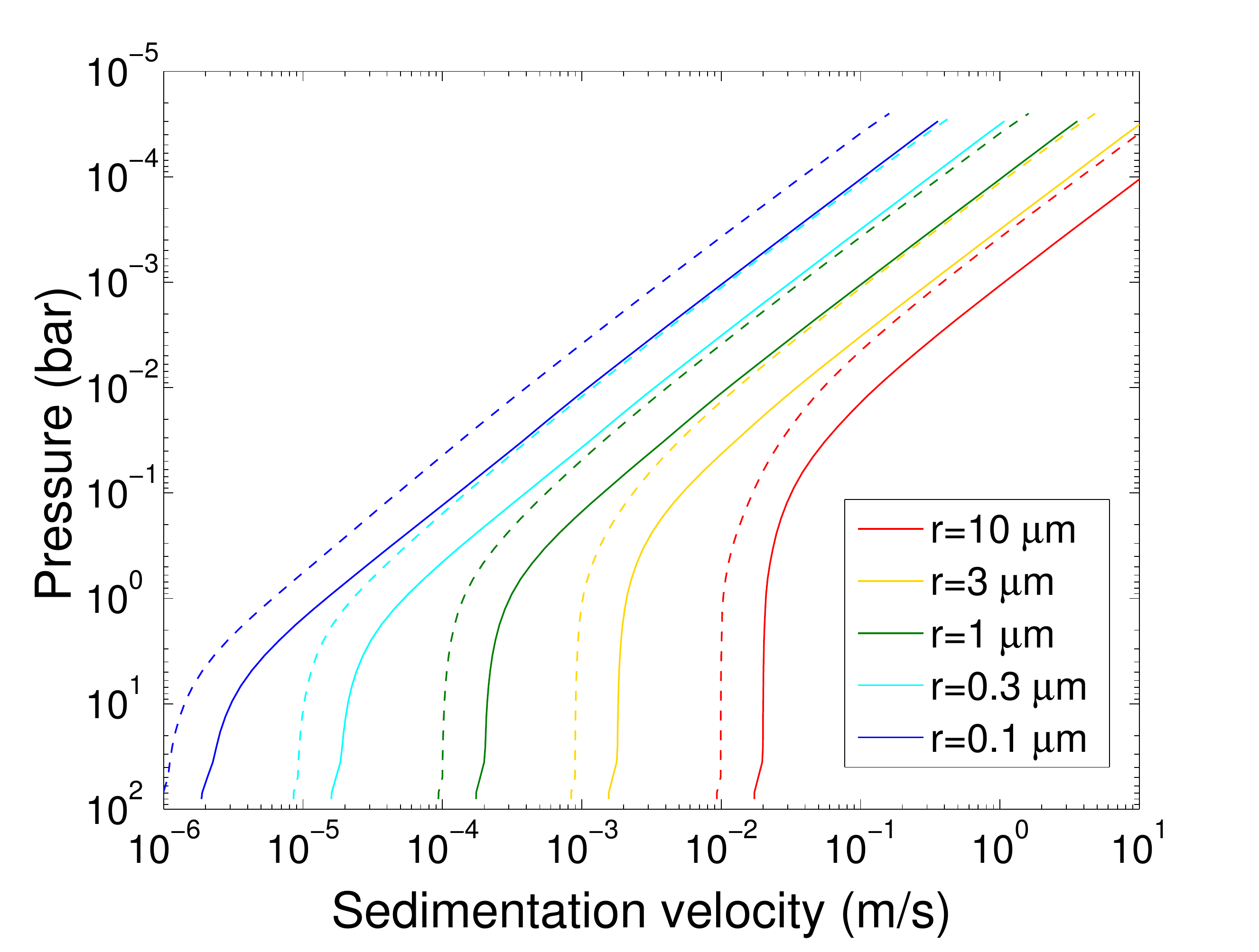}\\
	\includegraphics[width=7cm]{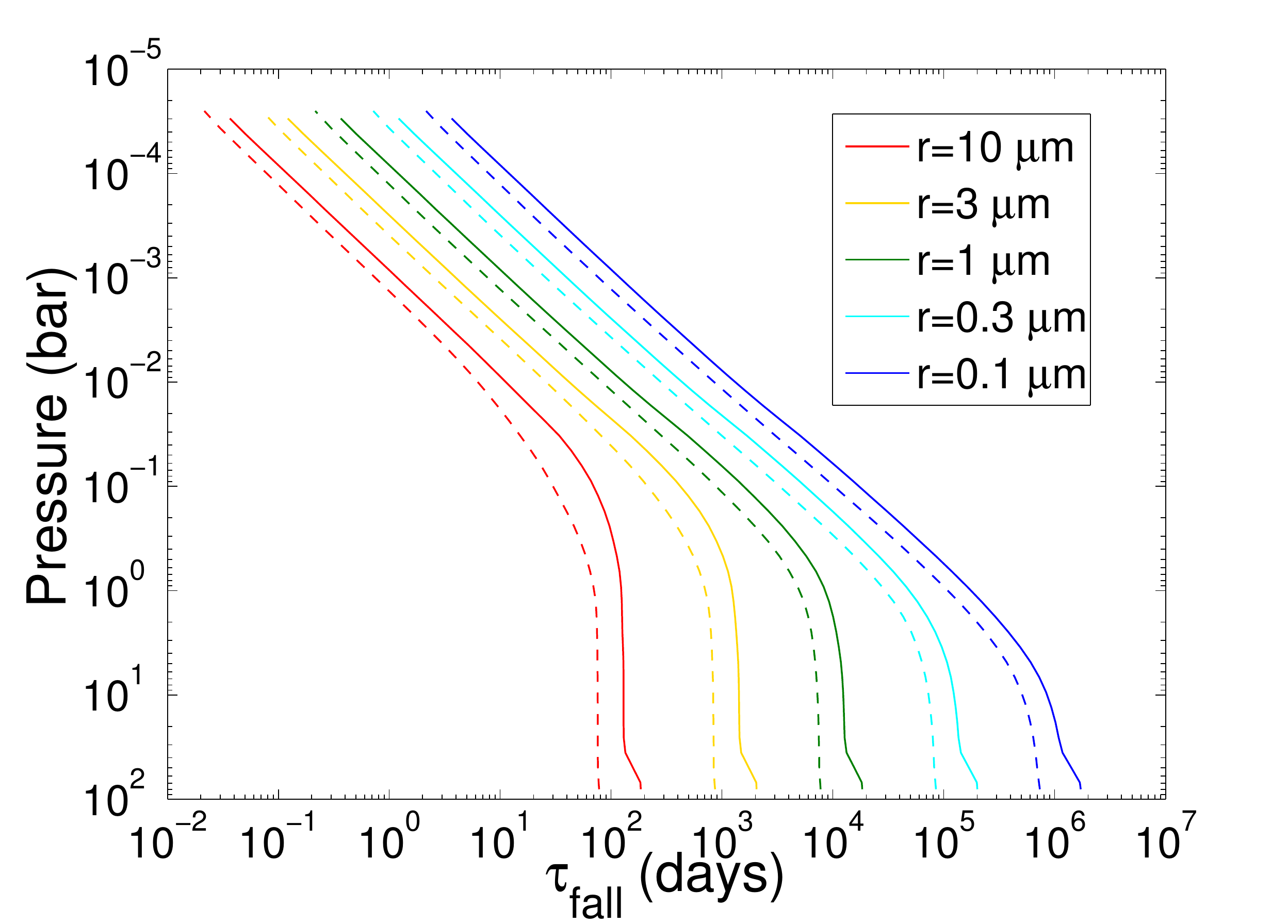}
\end{center}  
\caption{Variations with pressure of the sedimentation velocity (top) and the sedimentation rate (bottom) for particles with radii from 0.1 to 10 microns.  The solid and dashed lines use the 1D temperature profiles for the 100$\times$solar metallicity (solid) and the pure water atmosphere (dashed).}
\label{figure_8}
\end{figure}

The top panel in figure \ref{figure_8} shows the terminal velocity for particle radii from 0.1 to 10 microns, using $\rho_p$=2000 kg m$^{-3}$ (similar to the density of KCl) and for the 100$\times$solar and the pure water atmospheric composition cases (temperature profiles from the 1D model). For both compositions, two regimes are present. 
At high pressure, the Knudsen number is smaller than 1. The Cunningham slip factor and thus the terminal velocity are almost constant.
At low pressure, the Knudsen number is much larger than 1. The Cunningham slip factor and thus the terminal velocity are inversely proportional to the pressure.
For a given particle radius, the terminal velocity and the corresponding sedimentation rate are always lower in the H$_2$O atmosphere than in the H-rich atmosphere. In the upper atmosphere, the terminal velocities are lower by around a factor 3. This is due to both the larger molecular diameter and the larger viscosity of H$_2$O compared to H$_2$.

\begin{figure}[!h] 
\begin{center} 
	\includegraphics[width=7cm]{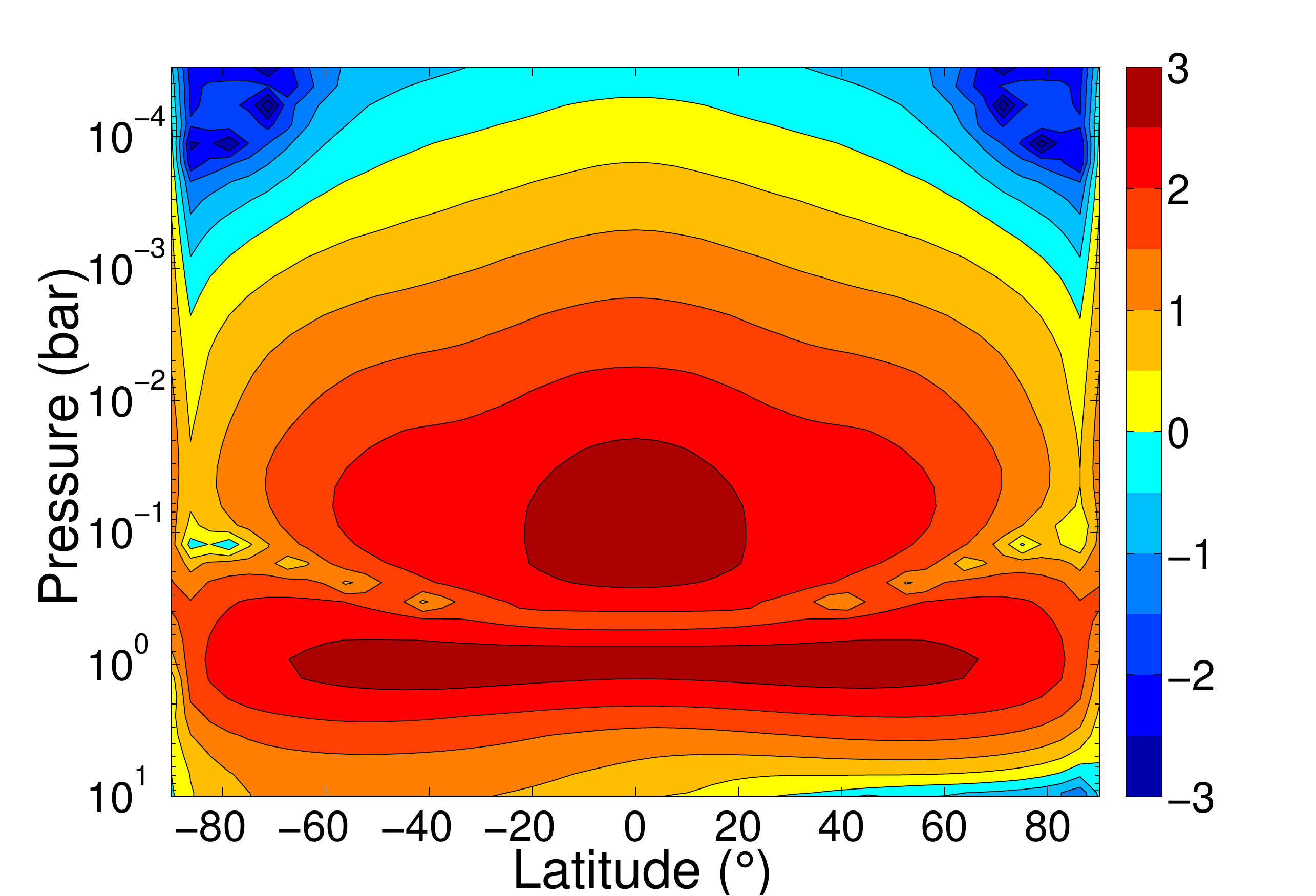}
\end{center}  
\caption{Zonally-averaged ratio of the sedimentation timescale by the advection timescale for the 100$\times$solar metallicity. Contour colors are in log scale.}
\label{figure_9}
\end{figure}

The sedimentation timescale is defined as the ratio of the scale height to the sedimentation velocity  $\tau_{s}=H/V_f$. The bottom panel in figure \ref{figure_8} shows the sedimentation timescales for the 100$\times$solar and the pure water atmosphere. The vertical mixing simulation is valid when the duration of the simulation is longer than the sedimentation timescale. Since our simulations were run for around 1600 days, the vertical mixing simulations are valid for pressures lower than 0.1 bar for a particle radius of 1 micron, and for pressure lower than 1 bar for particle radii of 3-10 micron. For evaluating the vertical mixing of the tracers, it is useful to perform simulations with different particle radii.  Large particles provide the most robust results at higher atmospheric pressures, but can be completely removed from the upper atmosphere, and so smaller particles give more accurate estimates for vertical mixing at these levels.
Figure \ref{figure_9} shows the ratio (in log scale) of the sedimentation timescale to the advection timescale ($\tau_{adv}=2\pi R/U$ with $R$ GJ1214b's radius and $U$ the mean zonal wind) for the 100$\times$solar case. This ratio is very similar for the different atmospheric compositions. When the sedimentation timescale is longer than the advection timescale (i.e. particles fall by less than a scale height during a full rotation around the planet), the particle transport is mostly driven by the zonal mean circulation and there are small longitudinal variations in the particle distribution. According to figure \ref{figure_9}, the particle distribution is mostly controlled by the zonal mean circulation and there should be almost no variations in longitude for pressures higher than 1 mbar. Moreover, longitudinal variations should be enhanced at high latitude.

\subsection{Results for the 3D mixing}
 Figure \ref{figure_10} shows the global mean particle abundance as a function of pressure for the different atmospheric compositions for a particle radius of 1 micron and for the 100$\times$solar metallicity case for particle radii from 0.1 to 10 micron. The particle abundance in the upper atmosphere increases with atmospheric metallicity. The abundance is reduced by 50$\%$ at around 5 mbar for the solar and 10$\times$solar metallicity cases, and at 0.2-0.4 mbar for the other cases. The strength of the atmospheric circulation above 0.1 bar increases with metallicity. 
For the solar metallicity case, the circulation cannot efficiently transport micrometer particles to the upper atmosphere (0.1-0.01 mbar).
For the 100$\times$solar metallicity, the vertical mixing is weak between 1 bar and 0.1 bar and very strong between 0.1 bar and 1 mbar. Particles with a radius of 0.1 micron are well mixed above the particle source layer while particles with a radius of 10 microns cannot go higher than 0.5 bar.

\begin{figure}[!h] 
\begin{center} 
	\includegraphics[width=7cm]{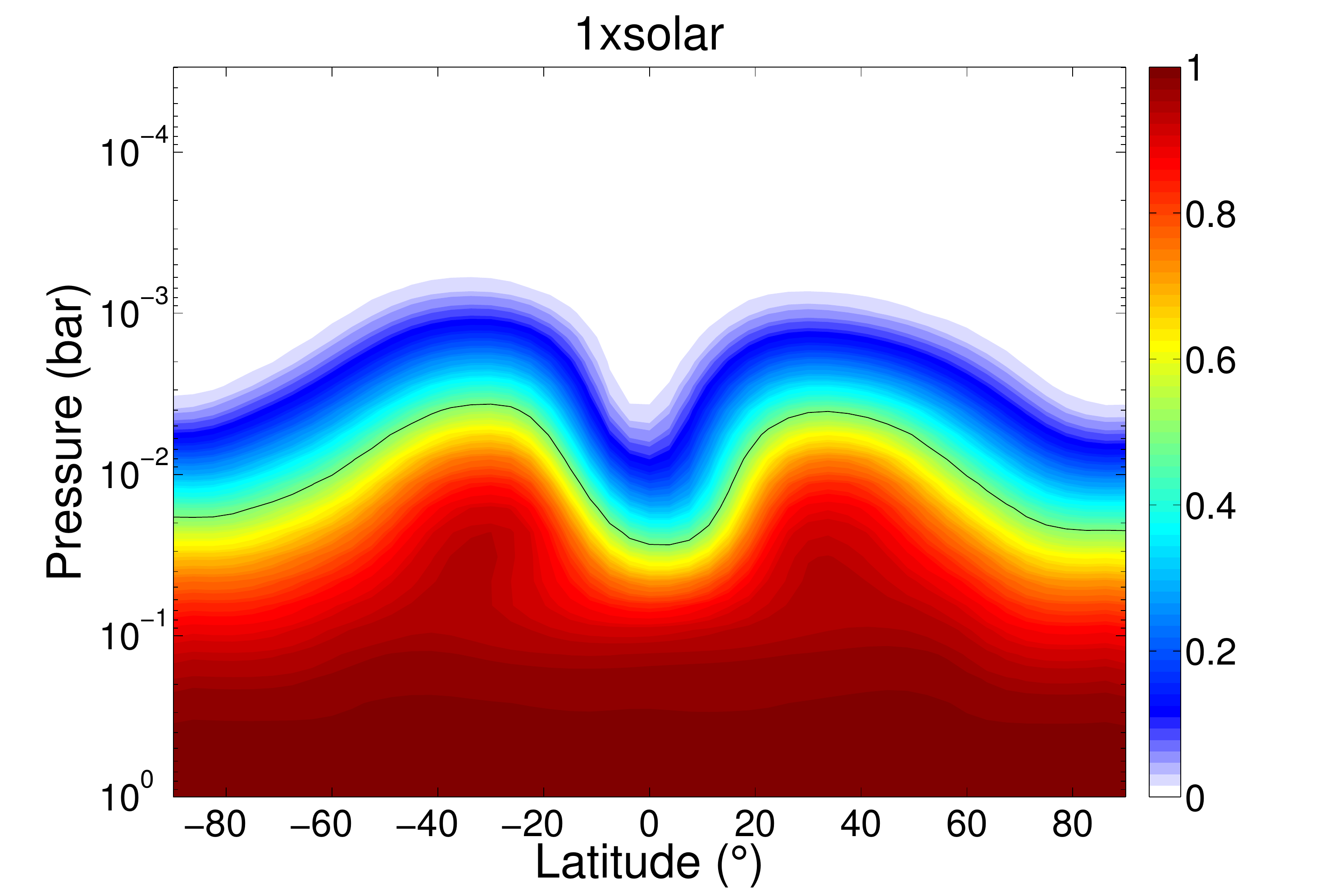}
	\includegraphics[width=7cm]{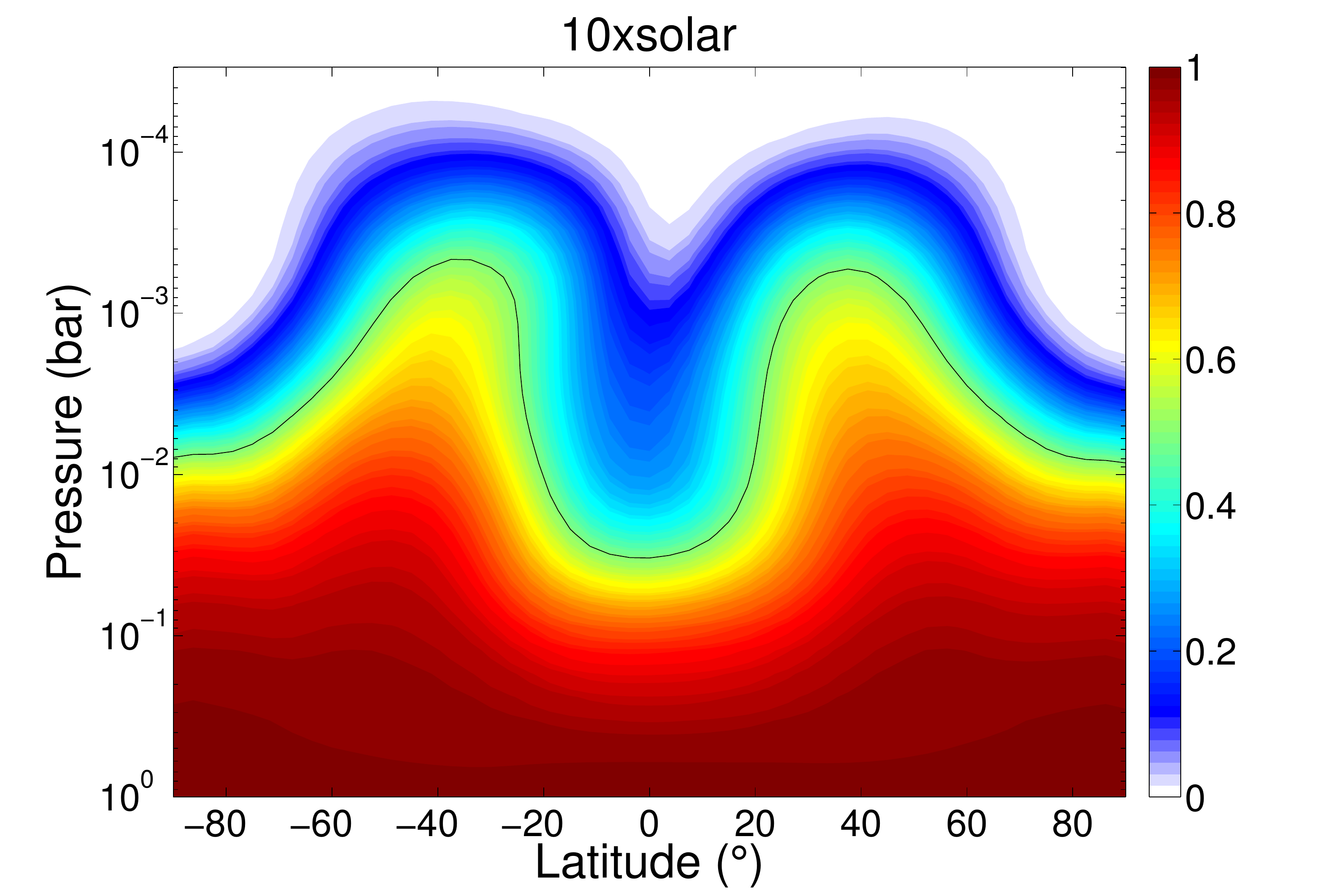}
	\includegraphics[width=7cm]{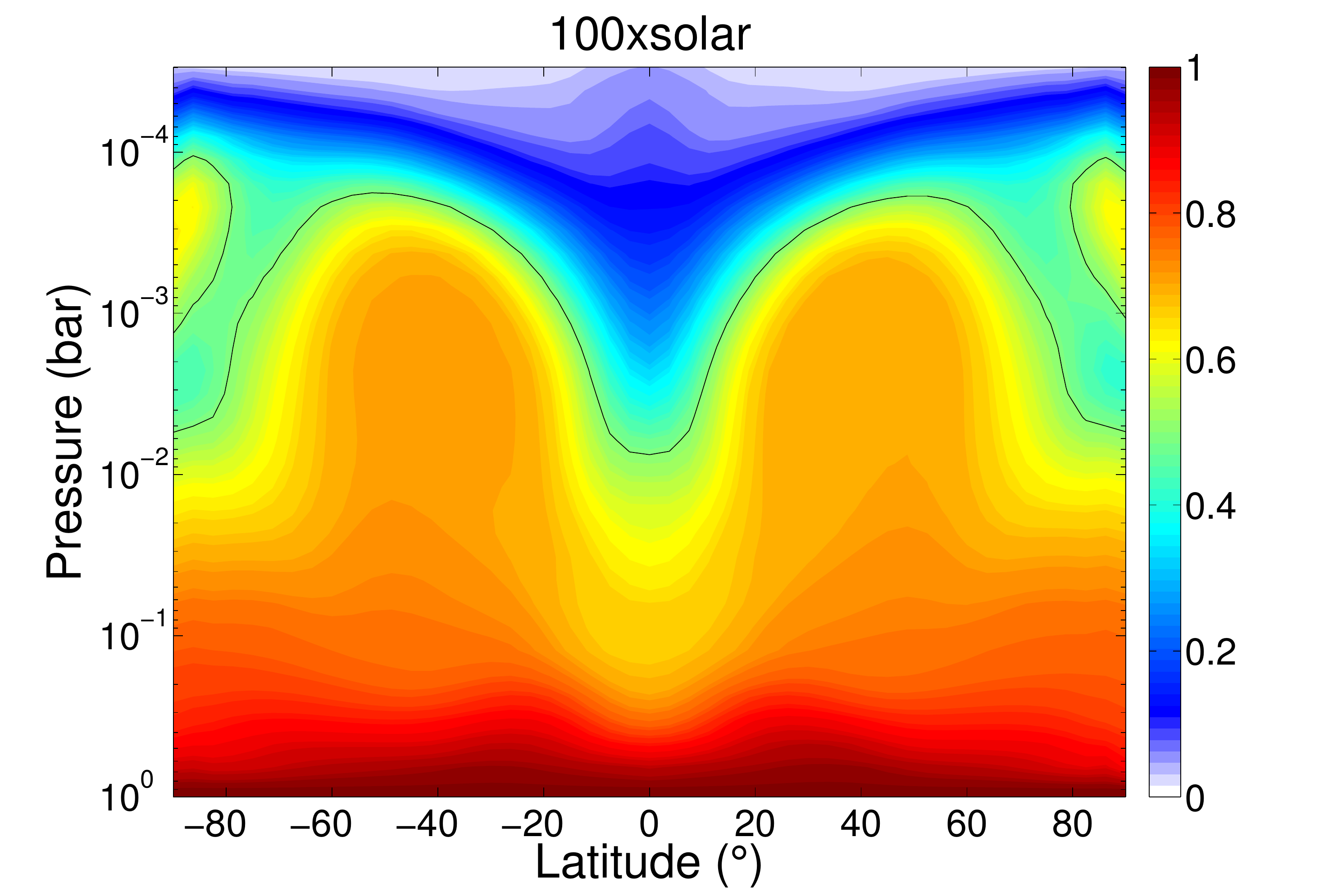}
	\includegraphics[width=7cm]{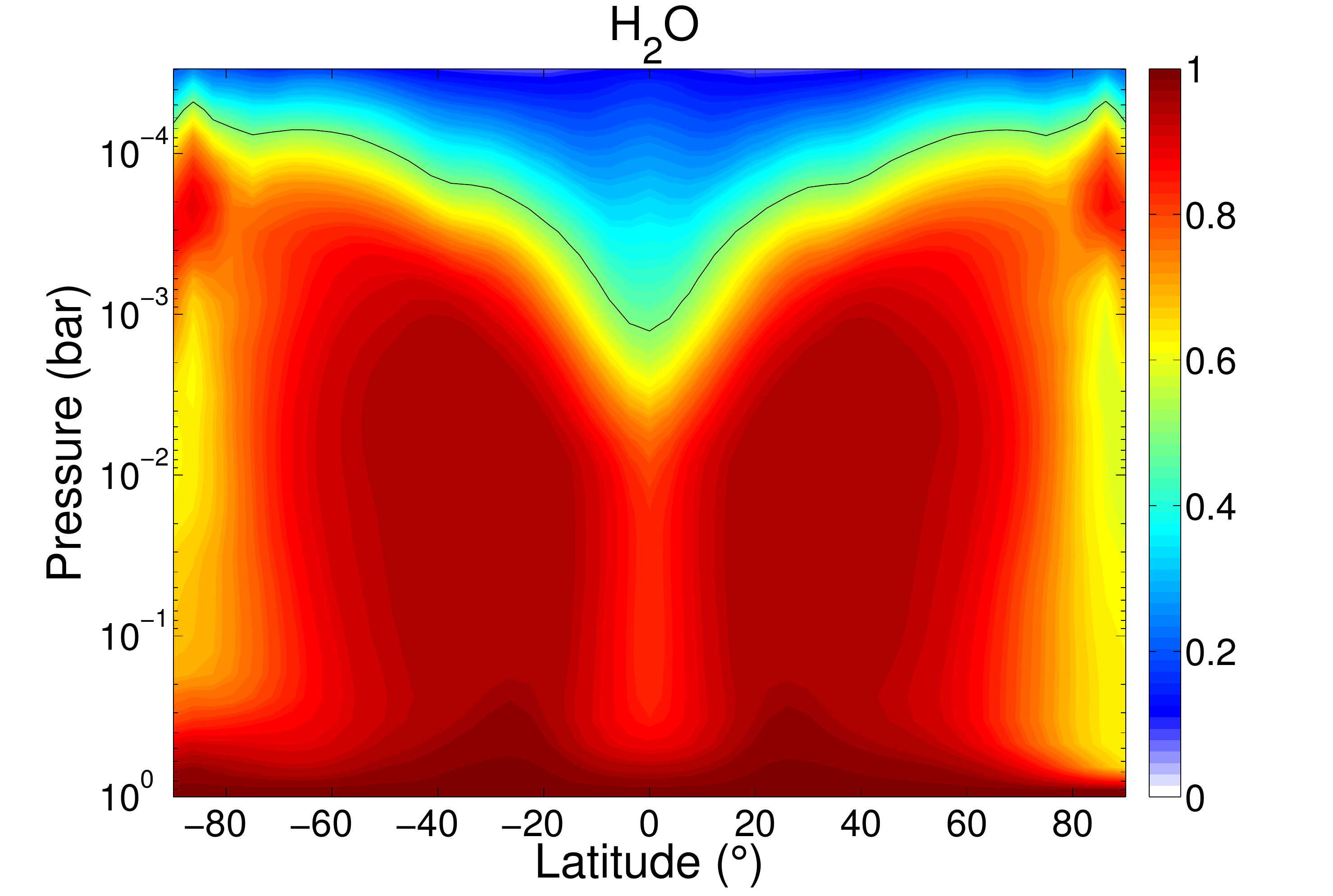}
\end{center}  
\caption{Zonally-averaged relative abundance of tracers with particle radius of 1 micron for the different atmospheric compositions. The black line corresponds to pressures where the relative abundance is 50$\%$. }
\label{figure_11}
\end{figure}

Figure \ref{figure_11} shows the zonal mean tracer abundance for a particle radius of 1 micron. The black lines correspond to the 50$\%$ abundance contours. Tracers reach the highest altitudes at mid-latitude (at around 40$^\circ$N/S). Except at low pressure for high metallicity (i.e. 100$\times$solar and the pure water atmosphere), there is a minimum of particle abundance at the equator. As explained before, the particle distribution is mostly controlled by the zonal mean circulation with in particular the anti-Hadley circulation (see figure \ref{figure_7}). On average, particles are thus lofted at mid-latitudes and sediment at the equator. This behavior on a synchronously rotating planet is opposite to that for a non-synchronously rotating planet like the Earth, for which there is an upward transport of water vapor and chemical species at the ITCZ (Inter-Tropical Convergence Zone), and thus a maximum of clouds at the equator.

\begin{figure}[!h] 
\begin{center} 
	\includegraphics[width=5cm]{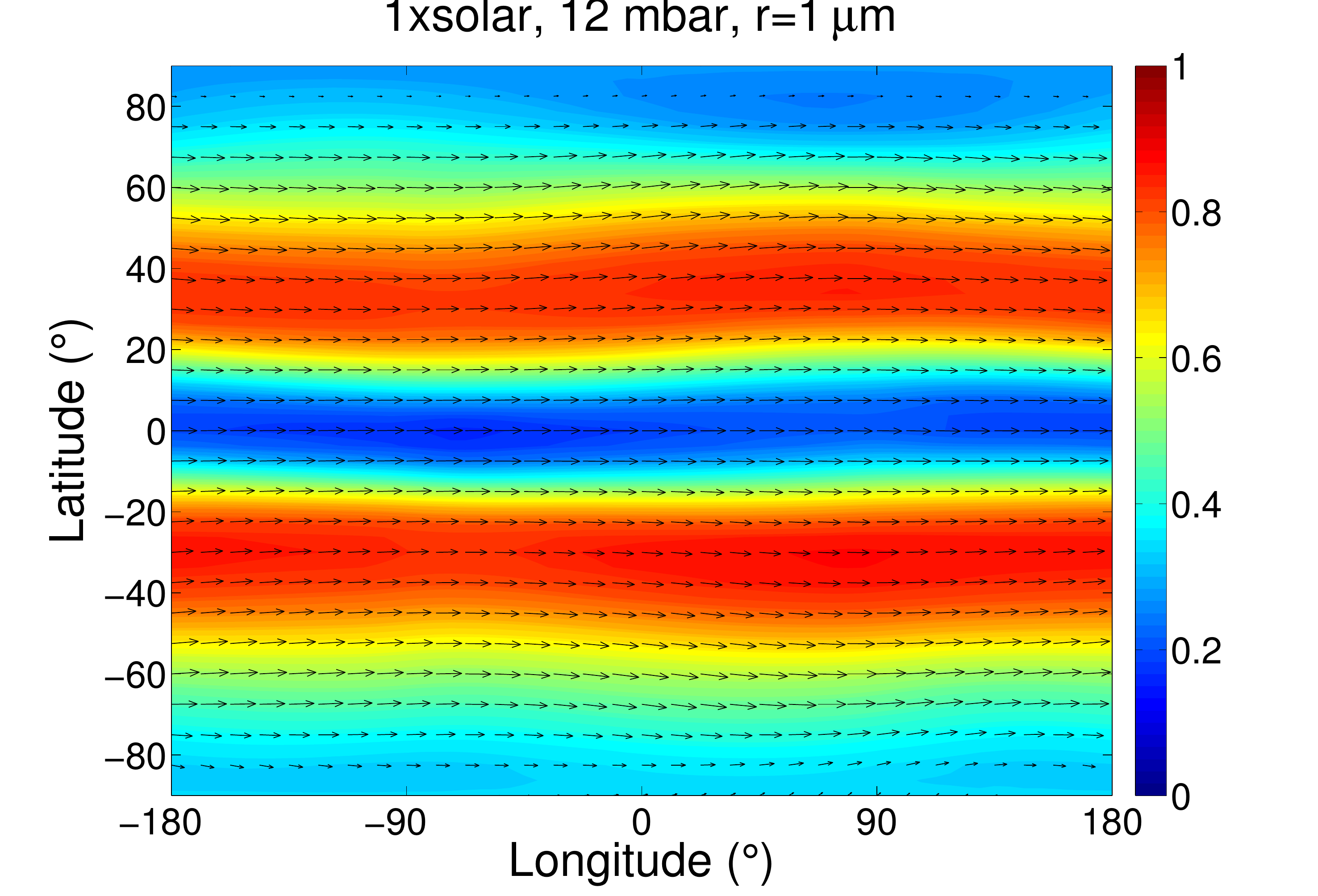}
	\includegraphics[width=5cm]{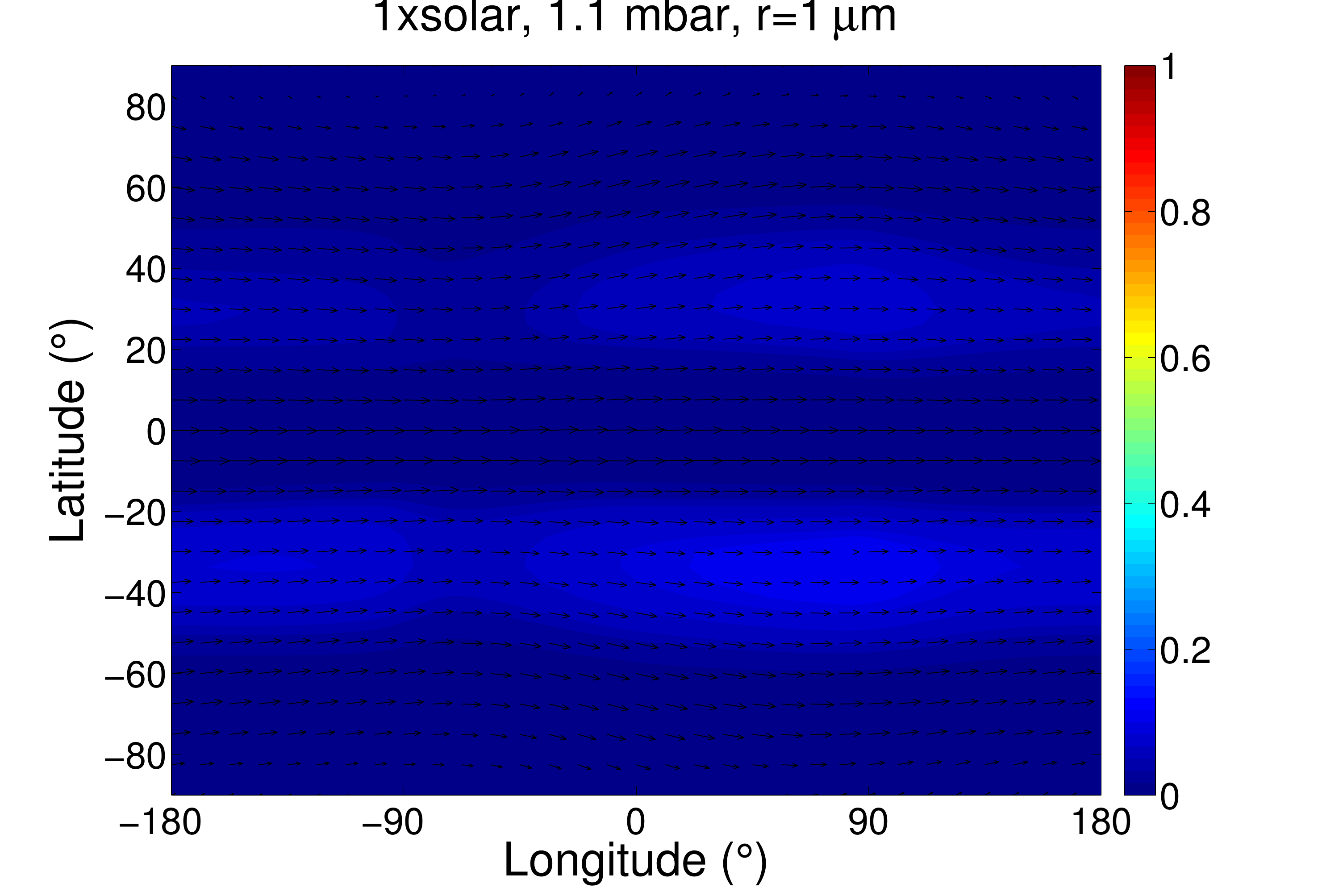}
	\includegraphics[width=5cm]{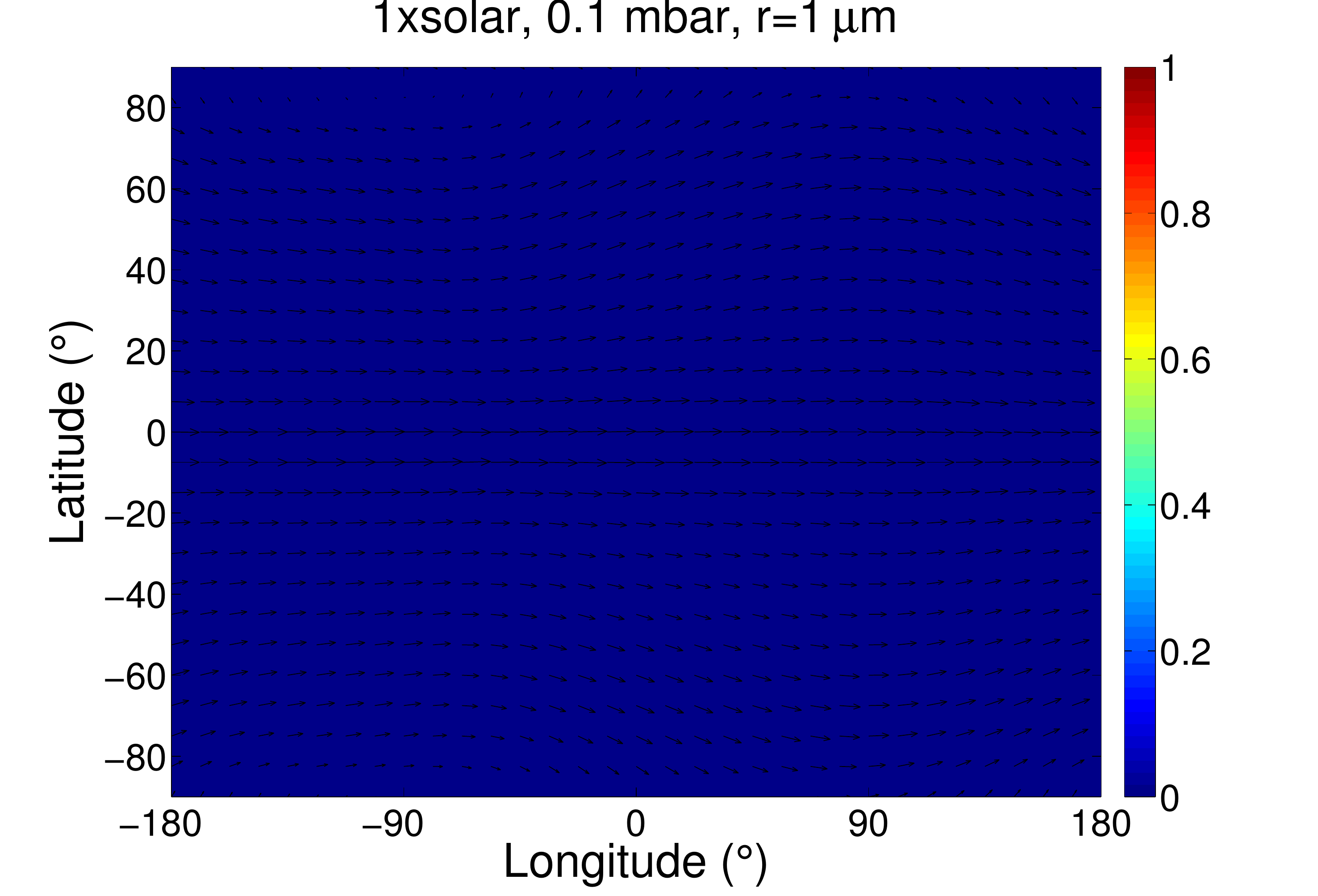}

	\includegraphics[width=5cm]{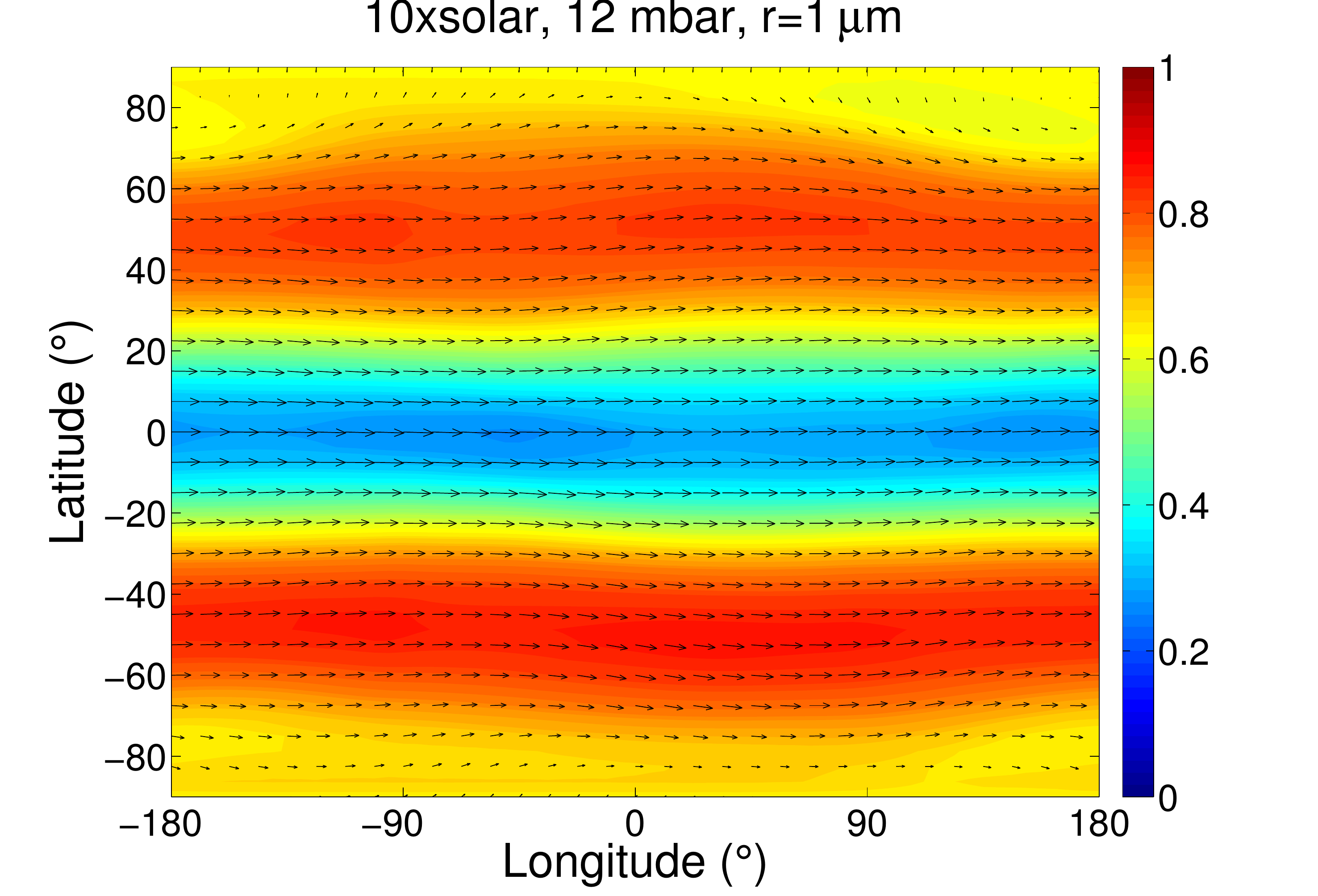}
	\includegraphics[width=5cm]{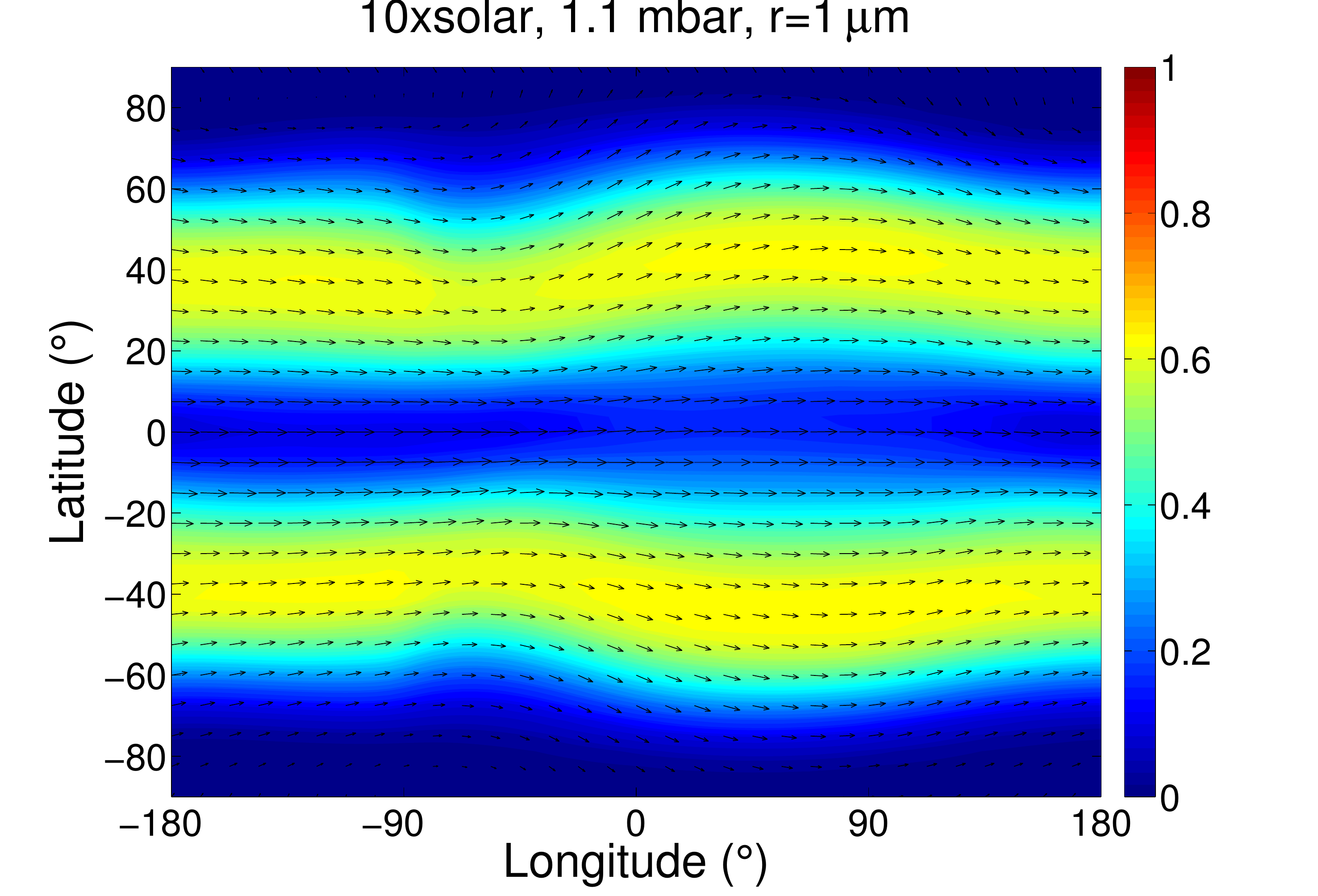}
	\includegraphics[width=5cm]{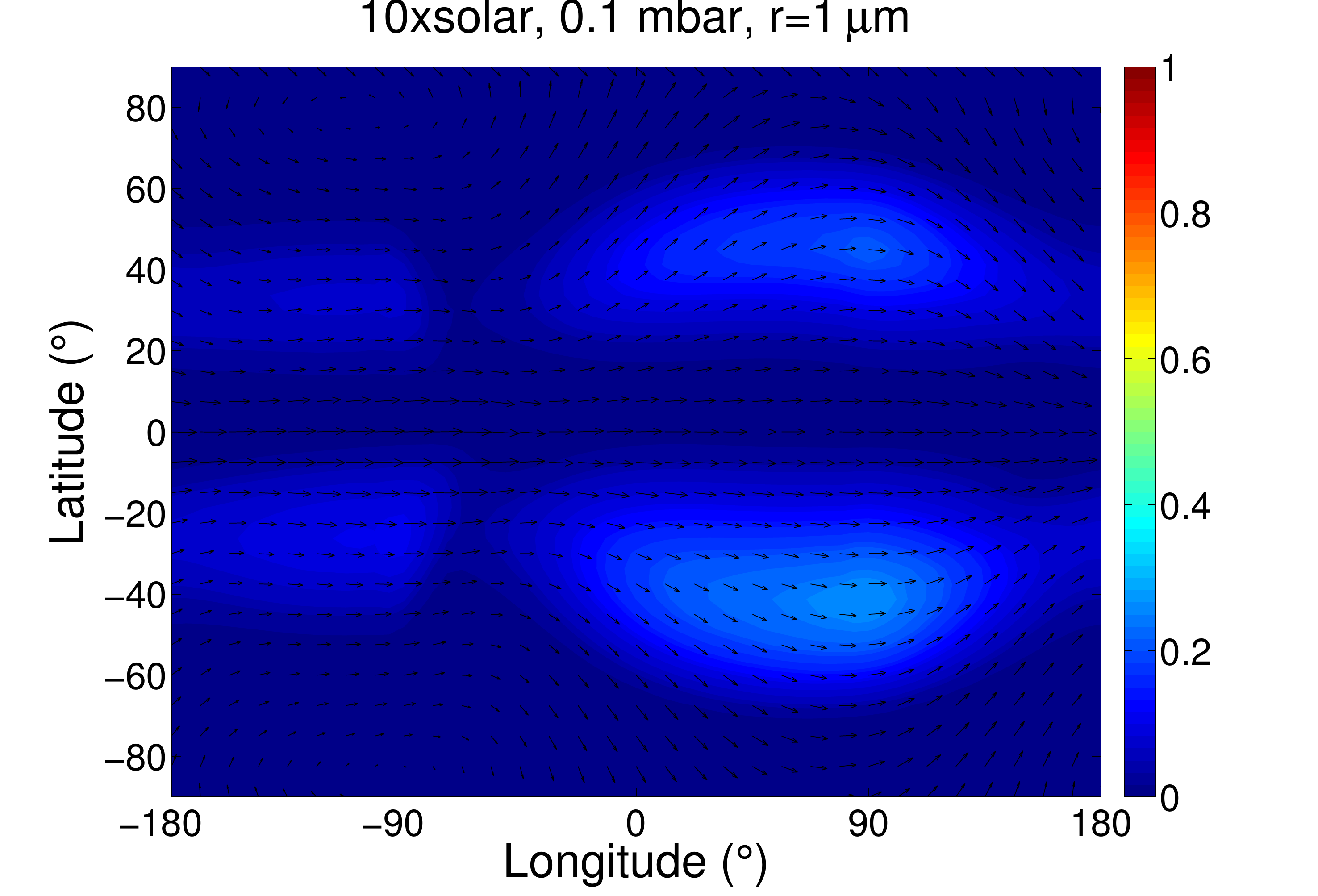}

	\includegraphics[width=5cm]{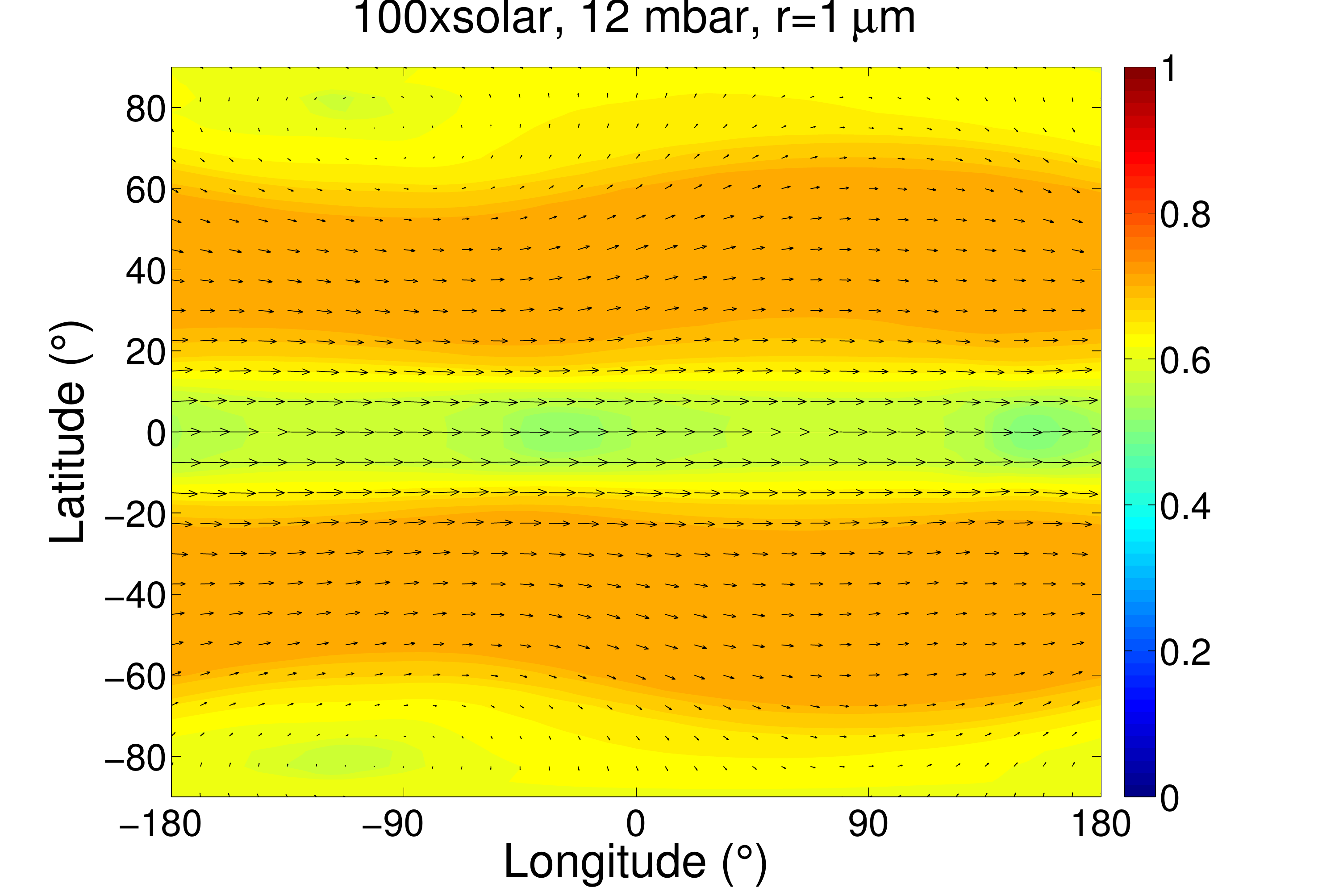}
	\includegraphics[width=5cm]{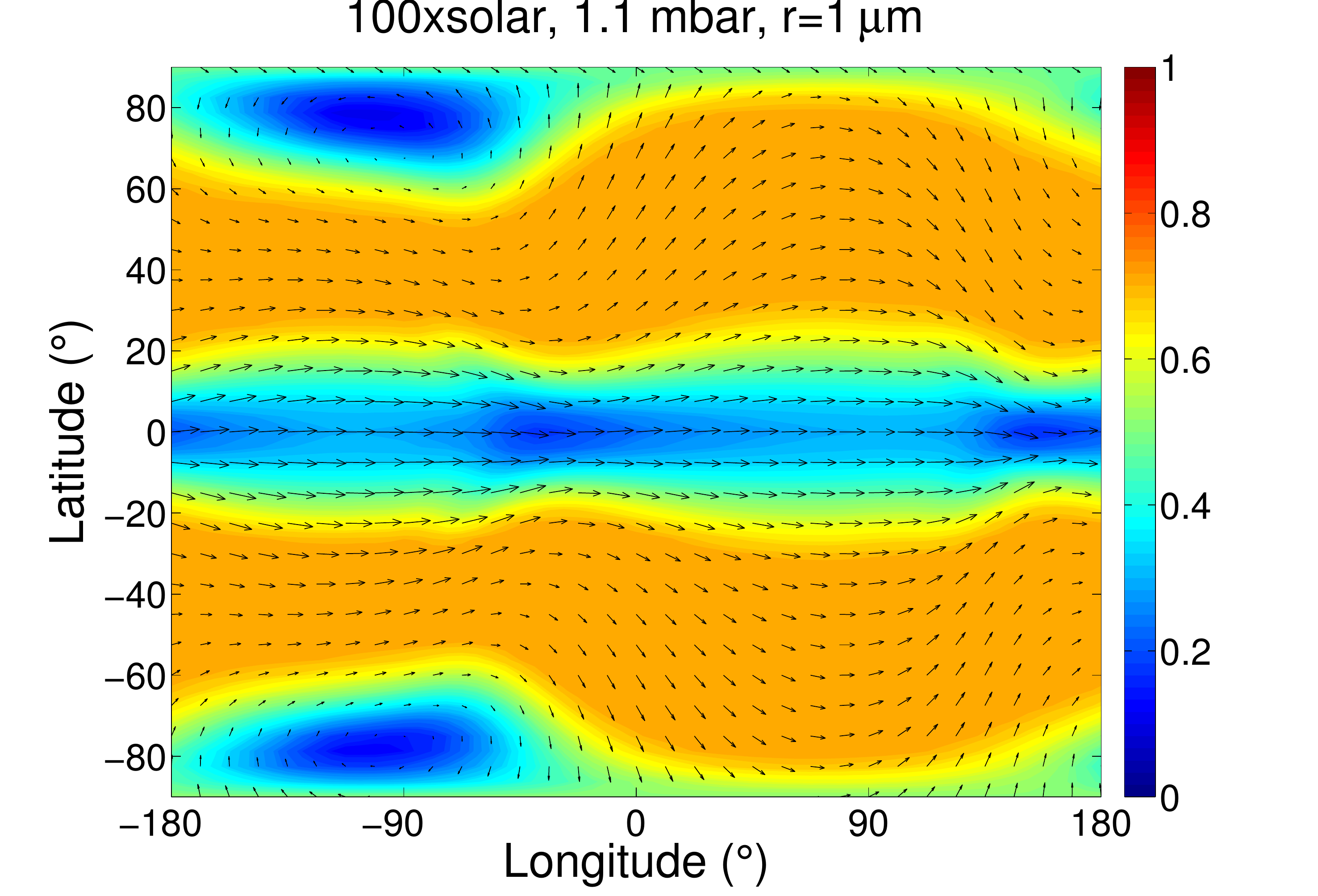}
	\includegraphics[width=5cm]{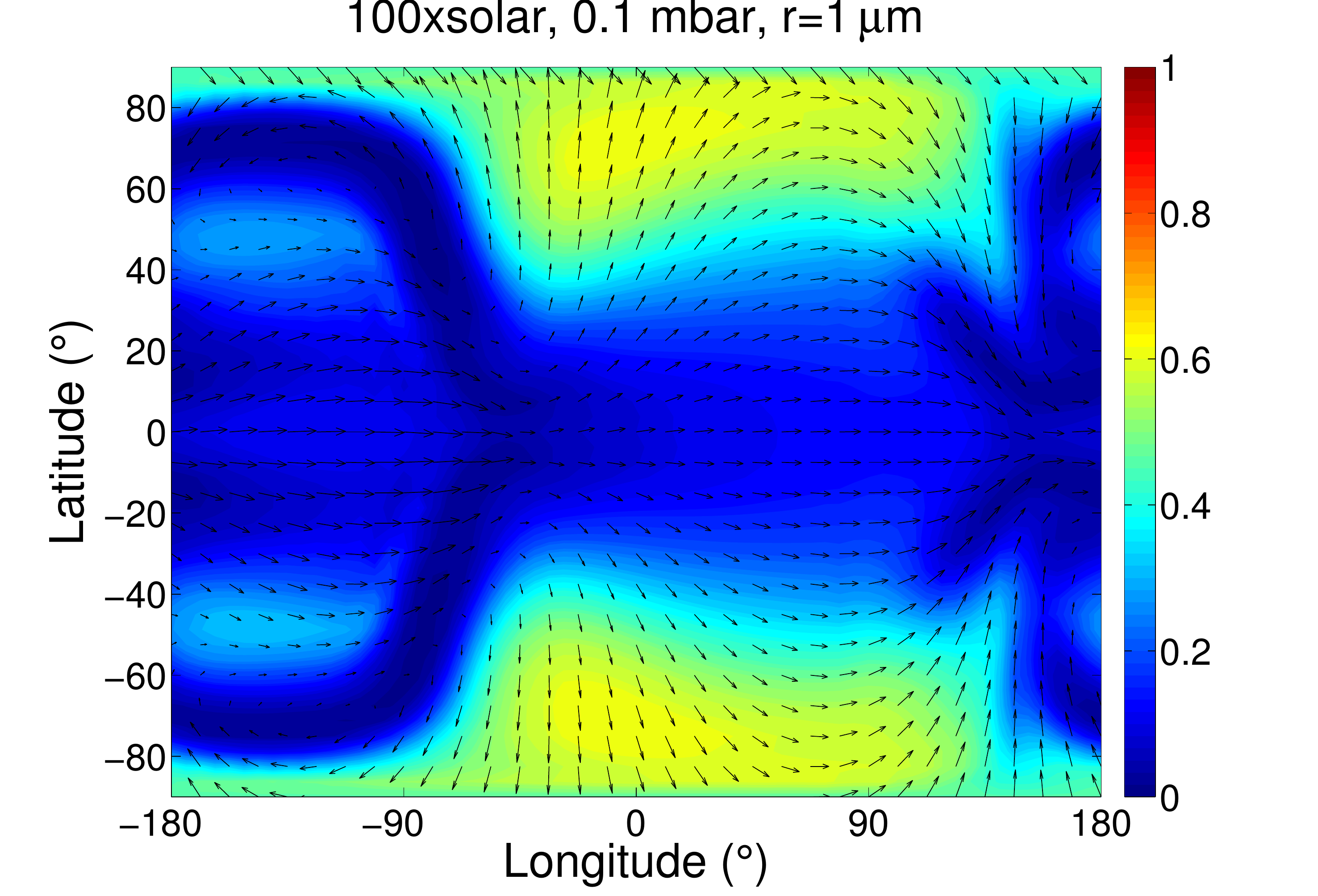}

	\includegraphics[width=5cm]{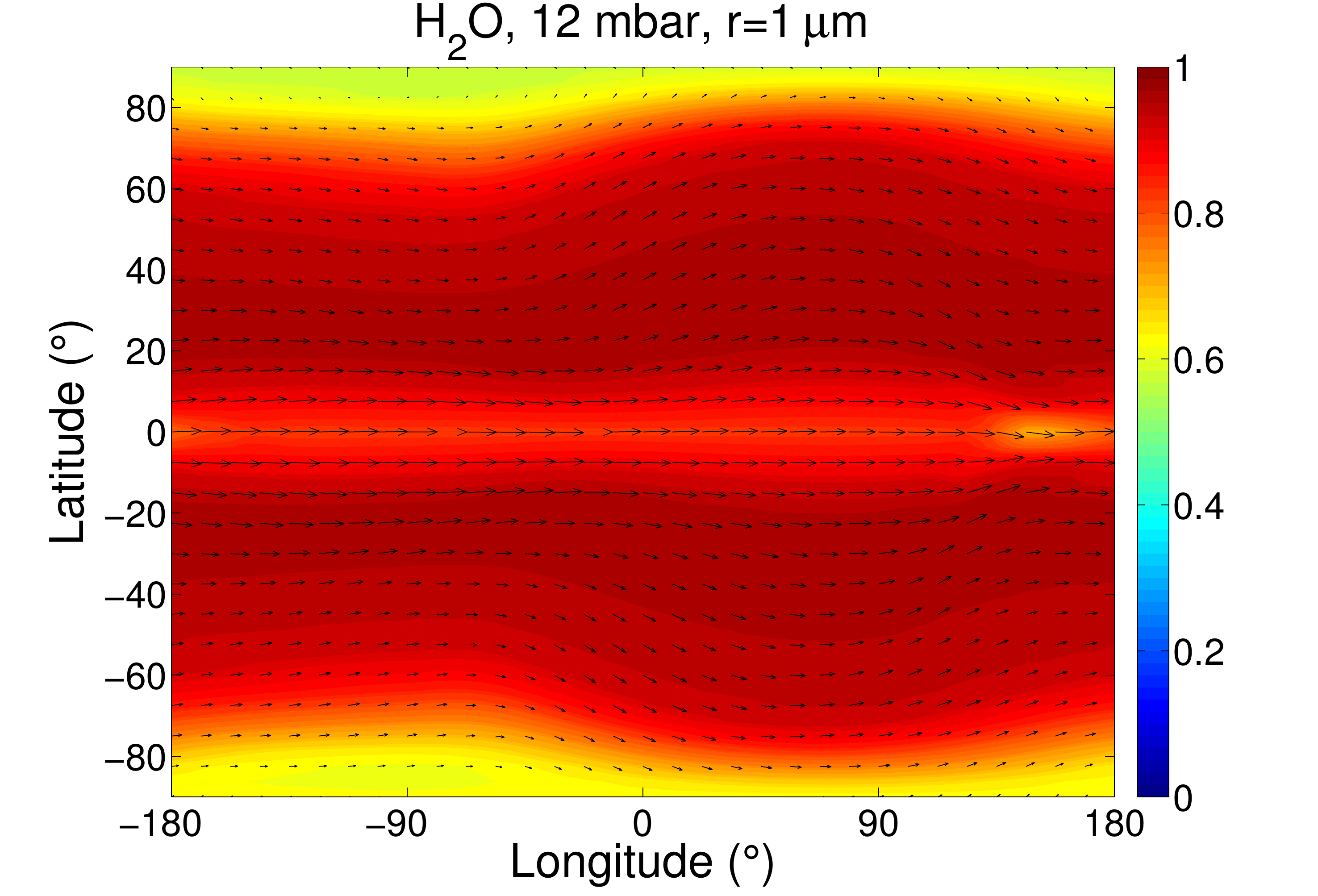}
	\includegraphics[width=5cm]{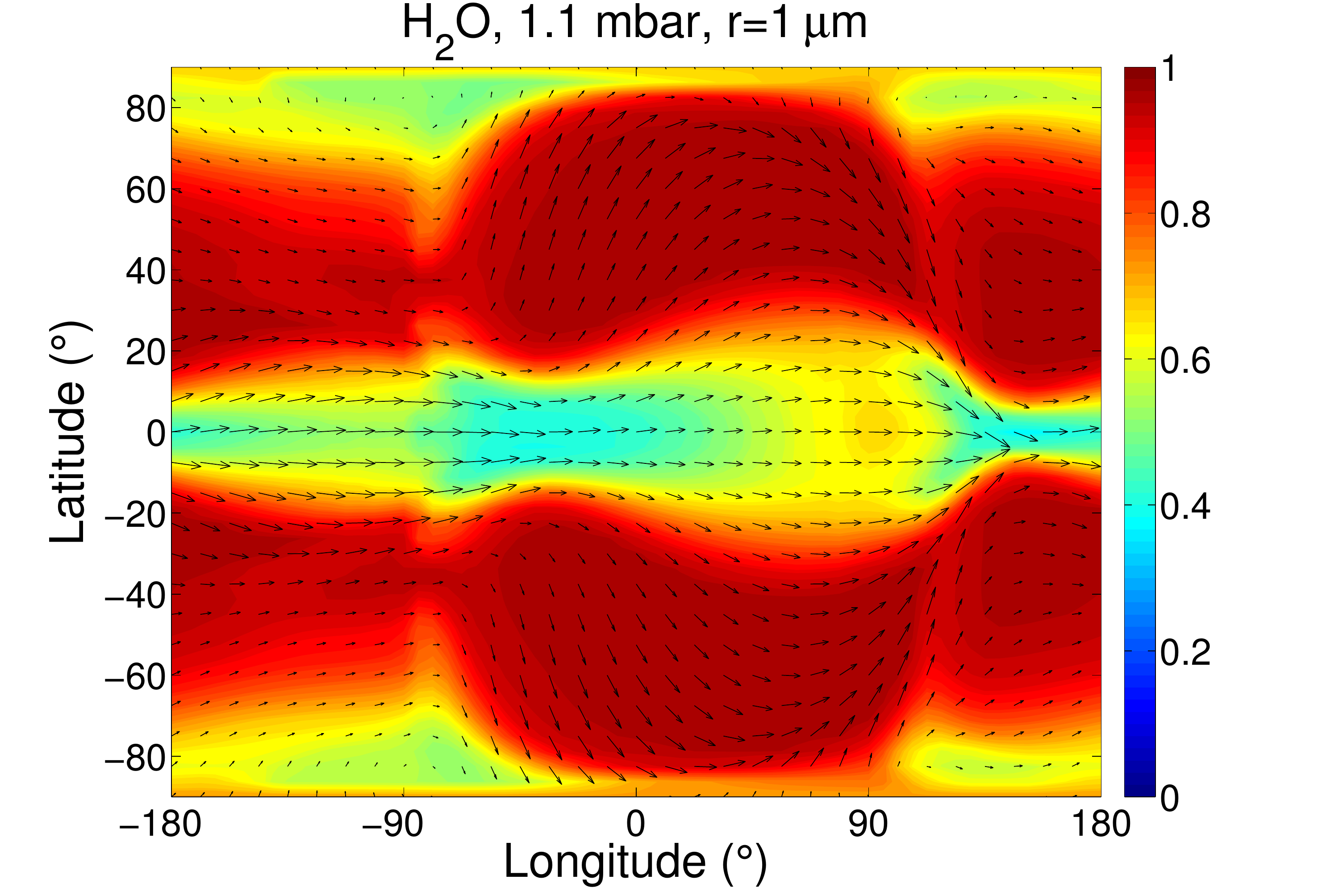}
	\includegraphics[width=5cm]{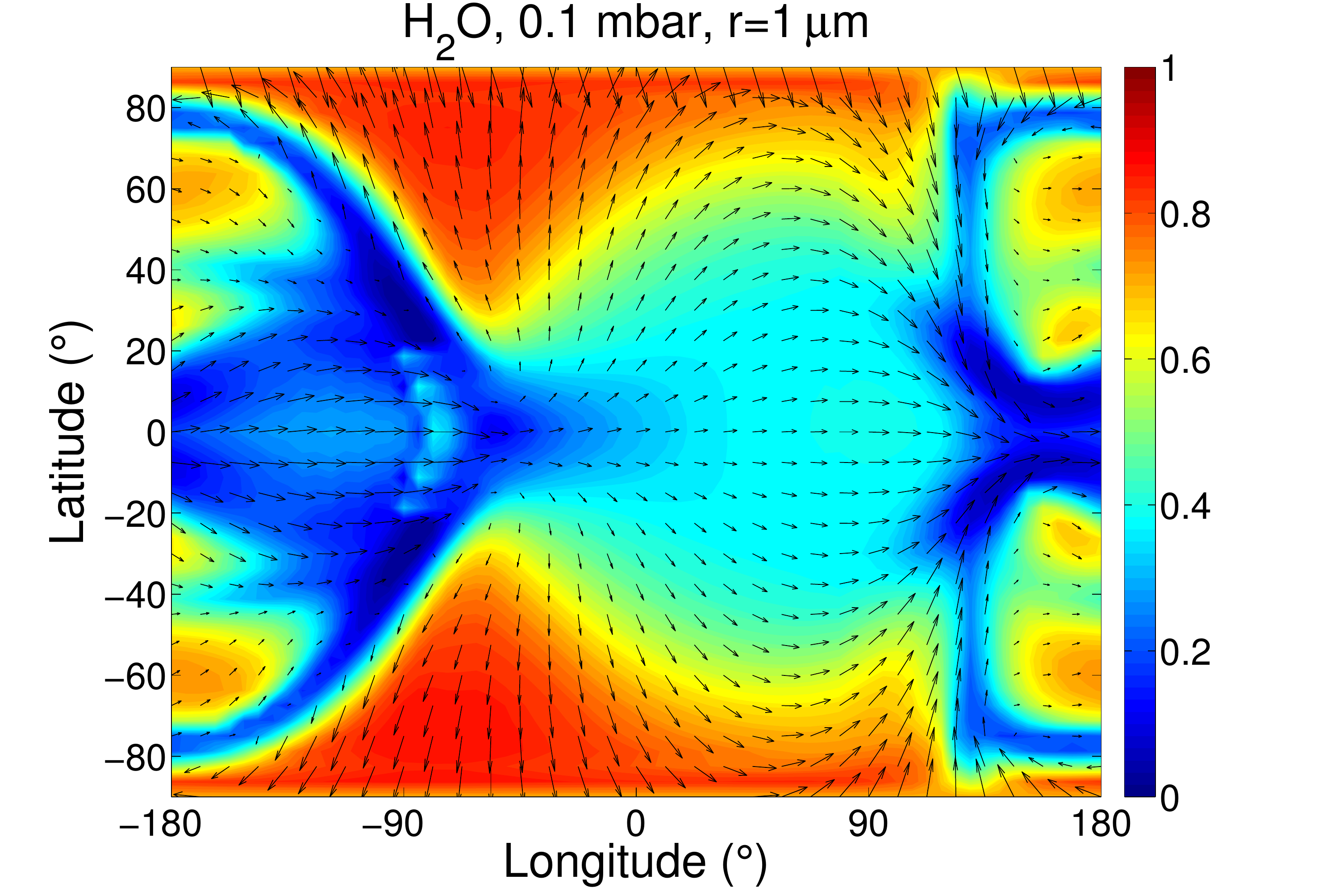}
\end{center}  
\caption{Maps of relative tracer abundance (particle radius of 1 micron) for the different atmospheric compositions (1$\times$, 10$\times$, 100$\times$solar and pure H$_2$O from top to down) and for pressure of 12 mbar (left) 1.1 mbar (middle) and 0.1 mbar (right).}
\label{figure_12}
\end{figure}

Figure \ref{figure_12} shows maps of particle abundance at 12, 1.1 and 0.1 mbar. As discussed in the section 4.1, there is almost no longitudinal variation at pressures higher than 1 mbar. The zonal wind spatially homogenizes the particle distribution. At pressures lower than 1 mbar, strong longitudinal variations appear. These variations are controlled by local upward/downward particle fluxes and also by meridional winds, producing more complex distributions. In particular, in the upper atmosphere, the meridional circulation tends to transport particles poleward. This is clear in figure \ref{figure_12} for H$_2$O and the 100$\times$solar metallicity case at 0.1-1 mbar. The poleward transport is mostly due to the stationary waves, similar to the heat transport in figure \ref{figure_7b}. Indeed, at low pressures, the dayside-nightside contrast is strong and stationnary waves dominate the transport. The particle abundance is higher in the dayside, where there is upwelling, than in the nightside, where there is downwelling. The poleward wind in the dayside and the equatorward wind in the nightside, therefore produce a net 
poleward particles flux.
This meridional transport maintains a high particle abundance at high latitudes in the upper atmosphere, and the maximum of particle abundance migrates poleward with altitude (see figure \ref{figure_11}).

\subsection{Equivalent 1D mixing}
Most studies concerning the photochemistry or the cloud formation on exoplanets are based on 1D models that take into account the vertical mixing by using a single  eddy diffusion coefficient $K_{zz}$ to represent the global 3D mixing. In these studies, the values used for the eddy diffusion coefficient are quite arbitrary, exploring a range of values and generally without altitude dependence. However, our 3D model results can be used to provide eddy diffusion coefficients for use by 1D models.

At steady state,  the downward particle flux by sedimentation is compensated by the upward vertical mixing. Thus the particle abundance $n$ (in kg/kg or mol/mol) in 1D  is given by:

\begin{equation} \label{equation_diff}
-K_{zz} \rho \frac{\partial n}{\partial z}  - \rho n V_f = 0 
\end{equation} 

With the outputs of the GCM, an equivalent 1D $K_{zz}$ for the global 3D mixing can therefore be defined as \citep{parmentier13}:

\begin{equation}
K_{zz}=-\frac{\langle\rho n V_f \rangle}  {\langle \rho \frac{\partial n}{\partial z} \rangle }
\end{equation} 

where the brackets correspond to a horizontal and time average over the entire planet.

Defining the diffusive time scale $\tau_d = \frac{H^2}{K_{zz}}$ and the sedimentation time scale 
$\tau_s = \frac{H}{V_f}$ (both depending on altitude), we can also write equation (11) in pressure coordinate as \citep{parmentier13}:

\begin{equation} 
\frac{\partial n}{\partial P}   = \frac{\tau_d}{\tau_s} \frac{n}{P}
\label{equation_diff}
\end{equation}

If we assume 1) a simple form for the dependence of $K_{zz}$ on pressure, $K_{zz}$=$K_{zz0}(P_0/P)^\alpha$ with $\alpha \neq$ 0 and $\alpha \neq$1  (in our simulation $\alpha$ $\simeq$ 0.4), 2) an isothermal atmosphere and 3) $\beta$=$1.656 K_n$ for the terminal velocity in equation \ref{eq_beta} (corresponding to a regime where the sedimentation speed is proportional to the pressure), then the particle abundance $n$ at a pressure $P$ can be expressed in  function of $n_0$ at a pressure $P_0$ as (see also \cite{parmentier13}):

\begin{equation} 
n=n_0  exp\left( \frac{\tau_{d_0}}{\tau_{s_0}}  \frac{1}{\alpha-1} \left(  \left( \frac{P}{P_0} \right)^{\alpha-1} -1      \right) \right)
\end{equation} 

\begin{figure}[!h] 
\begin{center} 
	\includegraphics[width=7cm]{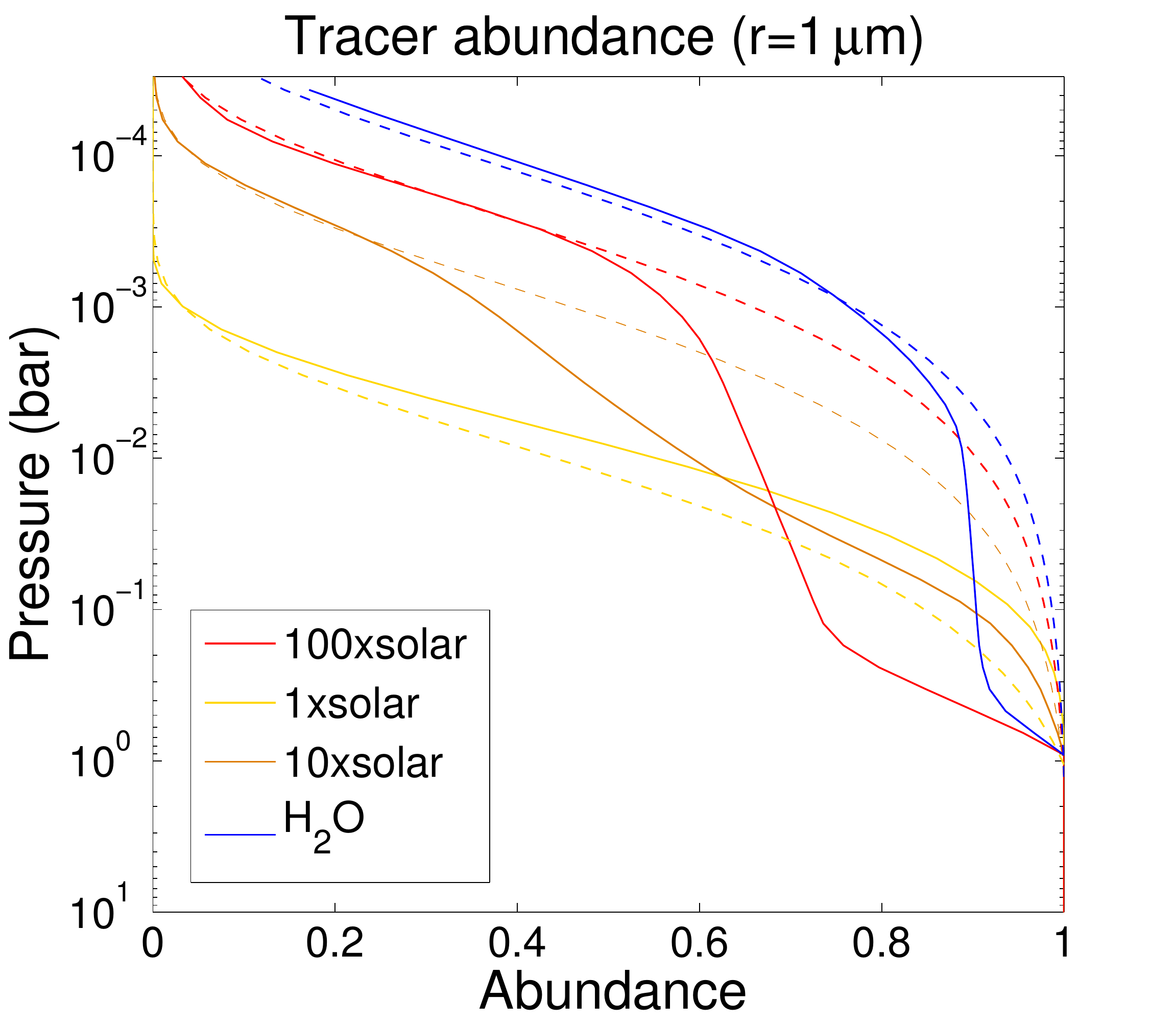}
	\includegraphics[width=7cm]{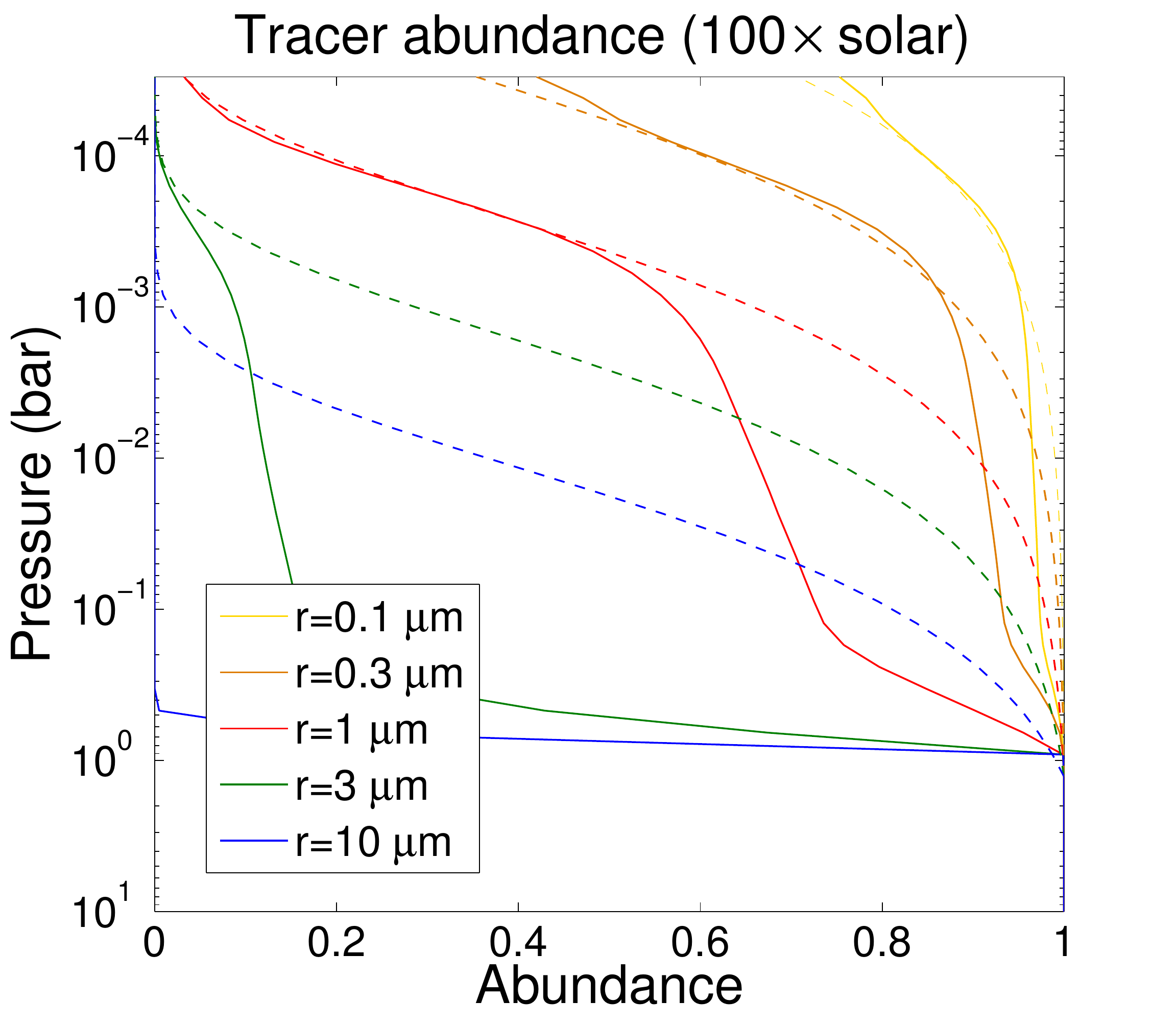}
\end{center}  
\caption{Mean profiles of the relative abundance of tracers.
Left panel: for the different atmospheric compositions with particle radius of 1 micron. 
Right panel: for the 100$\times$solar metallicity with particle radii from 0.1 to 10 microns. 
For both figures, the relative abundance is fixed to 1 below 1 bar. The dashed lines correspond to the analytical profiles using a parametrized $K_{zz}$ (see section 4.4).
}
\label{figure_10}
\end{figure}

This relation can match quite well the mean tracer abundance in figure \ref{figure_10}, for instance for the 100$\times$solar metallicity using $K_{zz0}=3\times 10^3$ m$^2$/s, $P_0$=1 bar, $\alpha$=0.4 and $n_0$=1.

\begin{figure}[!h] 
\begin{center} 
	\includegraphics[width=8cm]{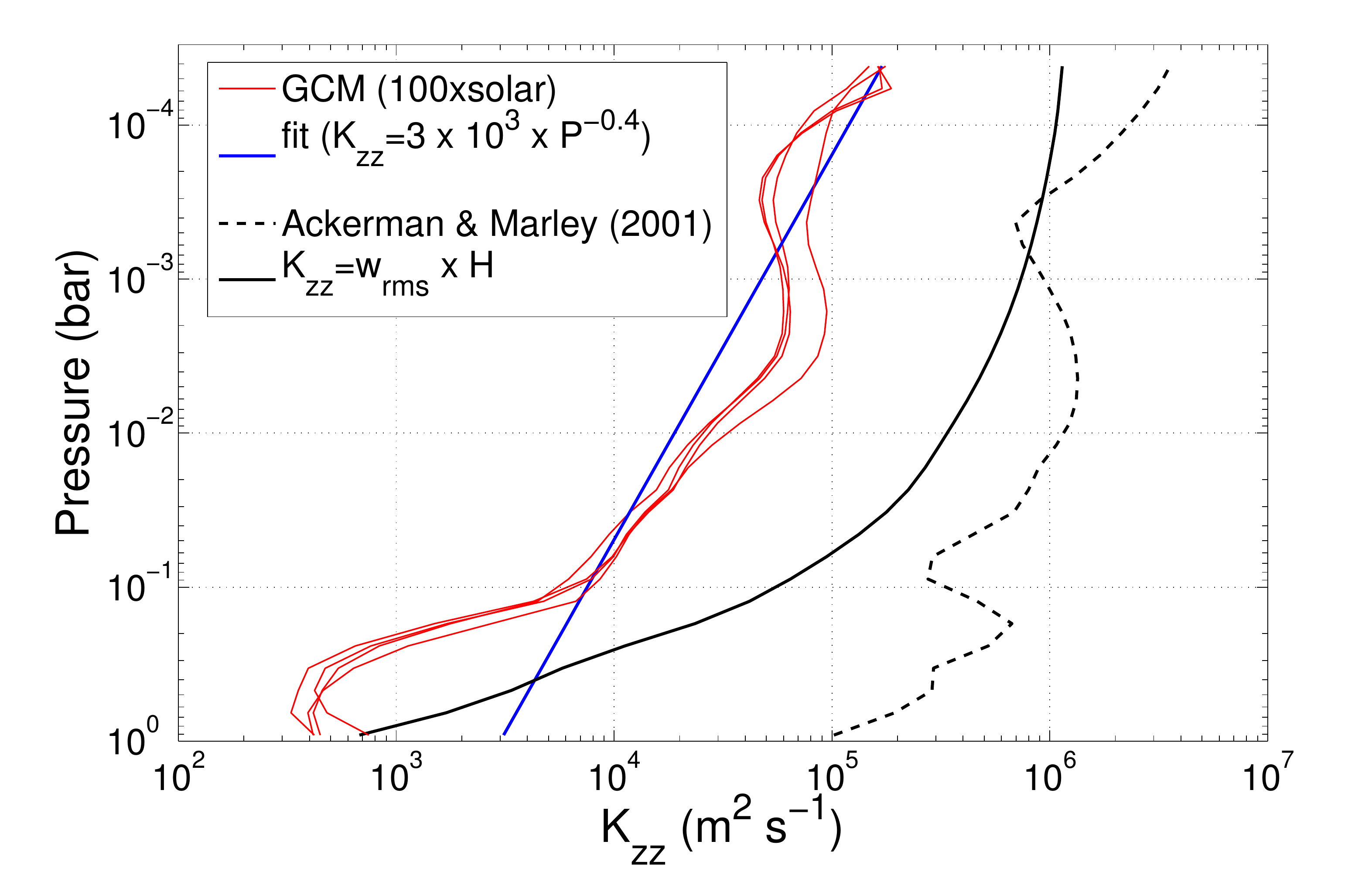}
	\includegraphics[width=8cm]{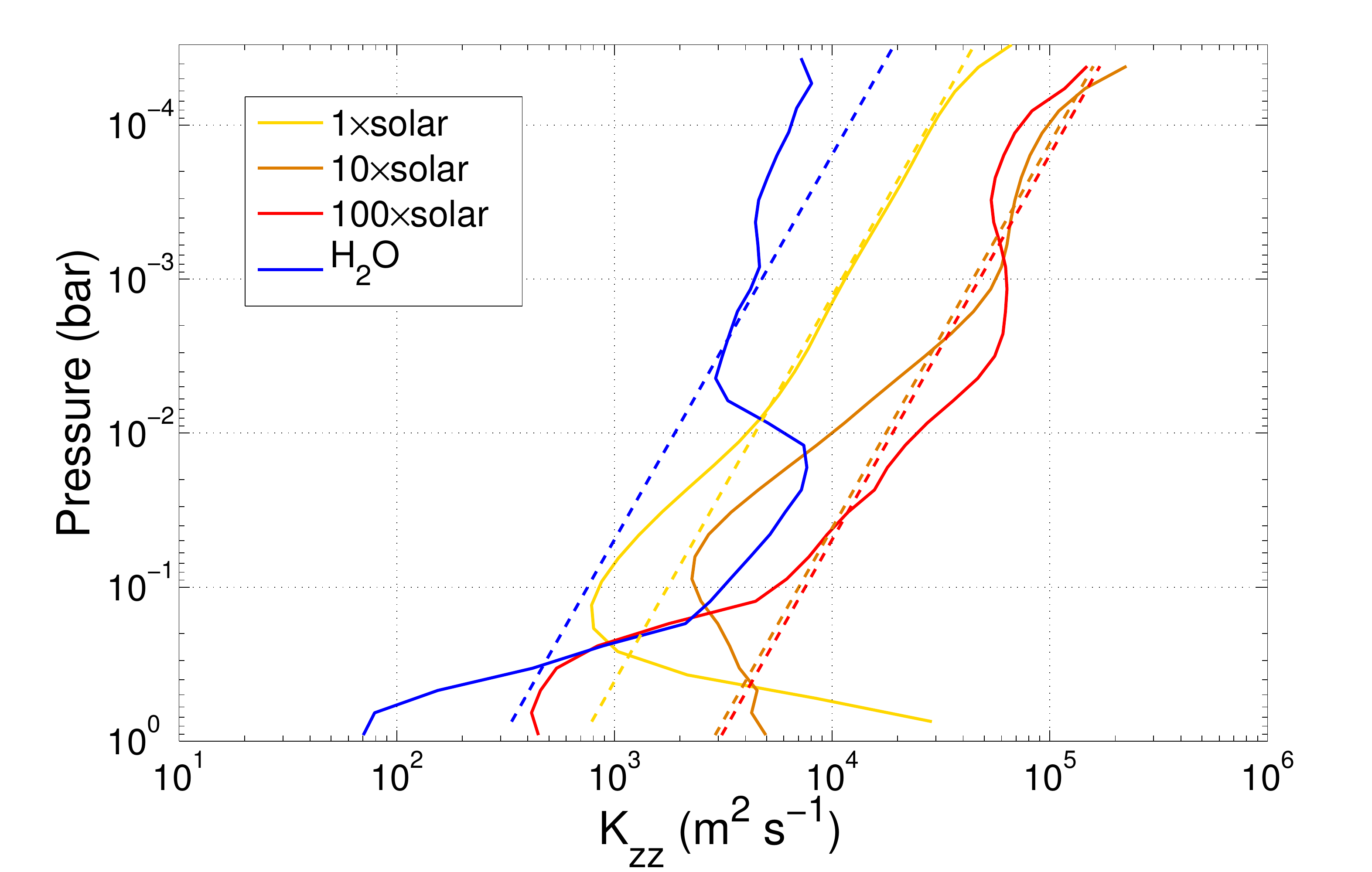}
\end{center}  
\caption{Profiles of equivalent 1D eddy diffusion coefficients. 
The left panel shows K$_{zz}$ for the 100$\times$solar metallicity case. Red lines are from the GCM for particle radii of 0.1, 0.3, 1, 3 microns. The blue line is the simple fit ($K_{zz}=3\times 10^3$ $\times$P$_{bar}$$^{-0.4}$ m$^2$/s). Solid and dashed black lines are estimations from simple formula \citep{gierasch85, ackerman01,lewis10, moses11}. 
The right panel shows K$_{zz}$ derived from the GCM for the different atmospheric compositions with particle radius of 1 micron. The dashed lines are the simple fits ($K_{zz}$=$K_{zz0}\times P_{bar}^{-0.4}$  with $K_{zz0}$=$7\times 10^2$, $2.8\times 10^3$, $3\times 10^3$, $3\times 10^2$ m$^2$/s for the 1, 10, 100$\times$solar metallicity and pure water case respectively).
}
\label{figure_13}
\end{figure}

Figure \ref{figure_13} shows $K_{zz}$ obtained from the 3D simulations using equation (12). The left panel corresponds to the 100$\times$solar metallicity case. $K_{zz}$ was computed for particle radii from 0.1 to 3 microns (red lines) and is similar for these different radii. For the dependence on pressure, a good parametrization is $K_{zz0}=3\times 10^3 \times P^{-0.4}$ m$^2$/s. We also compared $K_{zz}$ obtained from the GCM to previous estimates using 1) the simple formula $K_{zz}= w_{rms} L$ where $w_{rms}$ is the root mean square of the vertical wind and $L$ is the mixing length assumed to be equal to the scale height $H$ \citep{moses11, lewis10}, and 2) the following formula for thermal vertical mixing from \cite{gierasch85} and used in the 1D cloud model of \cite{ackerman01}:

\begin{equation} 
K_{zz}=\frac{H}{3}\left(\frac{L}{H}\right)^{4/3}\left(\frac{RF_c}{\mu \rho_a c_p}  \right)^{1/3}
\end{equation} 

where $R$ is the perfect gas constant, $\mu$ is the mean molecular weight, $\rho_a$ is the atmospheric density, $c_p$ is the specific heat and $F_c$ is the convective heat flux. The latter is computed from our GCM by subtracting the mean net thermal flux by the mean net stellar flux (figure \ref{figure_3}). The factor 1/3 on the left is empirical, derived from observations of giant planets in our own Solar System. This formula was also used by \cite{morley13} for clouds on GJ1214b.

The equivalent 1D $K_{zz}$ from our model is generally one order of magnitude lower than these two estimations. The mixing length has to be lower than the scale height, as predicted by \cite{smith98}. A much better estimate is obtained for both formulae using a mixing length 10 times lower than the scale height.
Since 1D simulations are generally at radiative equilibrium ($F_c$=0), \cite{morley13} assumed $F_c=\sigma T_{eff}^4$ for the convective heat flux. While such an estimation can be quite good for the convective region of exoplanets or brown dwarfs, it strongly overestimates the convective heat flux for irradiated planets. According to our GCM, $F_c$ varies between 1 to 100 W/m$^2$. The use of $F_c=\sigma T_{eff}^4$ overestimates $K_{zz}$ from formula (15) by one order of magnitude. 
Therefore, the vertical mixing is likely overestimated by one order of magnitude in \cite{morley13}. This will tend to overestimate the particle abundance and size at high altitude in their model. Consequently lofting larger particles to higher altitudes to explain GJ1214b's flat spectrum may be more challenging than their results suggest.

Such simple parametrizations of $K_{zz}$ should therefore be used with caution since they give a very rough estimate of the mixing. However, they can give a good trend of the evolution of $K_{zz}$ with pressure. In particular, the variations with pressure of $K_{zz}$ computed with our GCM follow quite well those from formula (15). Even if the atmosphere is never convective in our GCM, we can expect that the global circulation, driven by the horizontal temperature gradient, globally acts as convection for the vertical mixing. The equivalent convective flux $F_c$ in our simulations slowly decreases from the altitude where most of the stellar flux is absorbed. For the 100$\times$solar case, it is more or less constant from 0.1 bar to 0.1 mbar. Since $\rho_a \propto P$, equation (15) gives $K_{zz} \propto P^{-1/3}$. \cite{parmentier13} found $K_{zz} \propto P^{-1/2}$, while in our model the best fit is with -0.4 as the exponent. In fact, an exponent of -0.4 or -1/3 can also provide a good fit to their values. 

The vertical mixing in planetary atmospheres is mostly done by molecular diffusion above the homopause and by waves and convection below. On Earth, the vertical mixing is primarily done by convection (dry and moist) in the troposphere and by gravity waves in the stratosphere. The amplitude of the latter grows as $P^{-1/2}$ until they break in the mesosphere \citep{lindzen81}. In warm exoplanets, the mixing should mostly be done by both largescale upwelling/downwelling (acting as convection) and planetary waves, which are well simulated by GCMs and generally grow as $P^{-1/2}$. We can therefore expect for such atmospheres an eddy diffusion coefficient with an exponent between -1/3 and -1/2 for the pressure dependence, justifying the intermediate exponent used in our fit.

The right panel in figure \ref{figure_13} shows the equivalent 1D $K_{zz}$ for the different atmospheric compositions. It decreases with metallicity in the deeper atmosphere (i.e. at around 1 bar) and increases with metallicity above 0.1 bar. This is explained by the difference in altitude where most of the stellar energy is deposited (i.e. below 100 mbar for the solar metallicity case and at around 10 mbar for the 100$\times$solar metallicity case). 
The equivalent 1D eddy diffusion coefficients can be parametrized as $K_{zz}$=$K_{zz0}\times P^{-0.4}$  (P in bar) with $K_{zz0}$=$7\times 10^2$, $2.8\times 10^3$, $3\times 10^3$, $3\times 10^2$ m$^2$/s for the 1, 10, 100$\times$solar metallicity and pure water case respectively. The parameters in our fits were chosen to primarily match $K_{zz}$ and the mean tracer abundance (Fig. \ref{figure_10}) for pressures lower than 10 mbar. At higher pressures, there are quite strong deviations from this simple power-law dependence, in particular for the pure water case.
For the H-dominated atmosphere, we find that $K_{zz}$ increases with metallicity. 
A higher metallicity leads to stronger dayside/nightside temperature contrasts and stronger vertical motions producing a stronger vertical mixing.
$K_{zz}$ for the pure water case is lower than for H-rich atmospheres  except between 0.1-0.01 bar. However, the pure water case always leads to a higher particle abundance because of its smaller scale height and the weaker sedimentation velocity.

\section{Discussion}

The atmospheric circulation and the vertical mixing we described in the last section have strong implications for the formation of high clouds/haze on GJ1214b and other tidally locked exoplanets. The circulation patterns we obtained, and in particular the anti-Hadley cell, are very general and can be applied to a wide range of warm synchronously rotating exoplanets, from mini-Neptunes to Jupiter-mass exoplanets.

\subsection{Implications for cloud formation} 

The temperature profile in the atmosphere of GJ1214b allows the condensation of potassium chloride (KCl) and zinc sulfide (ZnS) \citep{miller-ricci12, morley13}. According to our simulations for a H-rich atmosphere, the condensation should occur between 40 mbar and 0.4 bar (see section 3.1.). If the atmospheric circulation can transport cloud particles to the upper atmosphere (i.e. 0.1 mbar and less), they should remain condensed (in solid phase) and could produce the observed flat transit spectrum \citep{kreidberg14a}. Micrometric-size particles (or bigger) are required to produce a flat spectrum between 1.1 and 1.7 micron. Transporting such particles from around 0.1 bar to 0.01 mbar requires a strong circulation that can be accurately quantified only with GCMs.
Our 3D simulations with simple tracers can represent quite well clouds that form deeper in the atmosphere and do not evaporate at higher altitudes on the dayside. They reveal that micrometer particles can be lofted to the upper atmosphere quite efficiently. For instance, the abundance is reduced by around a factor 5 at 0.1 mbar for the 100$\times$solar metallicity with a particle radius of 1 $\mu$m. The vertical mixing tends to increase with metallicity. Therefore, metallicity impacts the formation of clouds not only by fixing the amount of condensable species but also by controlling the vertical mixing. A high metallicity should therefore favor the formation of high clouds on GJ1214b. 
Assuming the solar nebula atomic abundances from \cite{lodders03}, the abundance of KCl vapor in the deep atmosphere would be around 8.3$\times$10$^{-6}$kg/kg for the solar metallicity and around 4.3$\times$10$^{-4}$kg/kg for the 100$\times$solar metallicity. Using the mean profiles of tracer relative abundances for particle radii of 1 $\mu$m (Fig. \ref{figure_10}), we find that KCl clouds would be optically thick in transit at visible and near-infrared wavelength at pressures higher than 3 mbar for the solar metallicity and at pressures higher than 0.1 mbar for the 100$\times$solar metallicity. Therefore the latter case is more likely to produce the observed flat transit spectrum than the former case.

Clouds could also modify the circulation and the vertical mixing by absorbing and scattering stellar radiation. If clouds are strongly scattering, we can expect a high planetary albedo and in particular a cooling of the atmosphere below the cloud deck (i.e. below 0.1-1 bar) impacting the atmospheric circulation.
The impact of this cloud-climate feedback on the possibility of high clouds on GJ1214b is best understood using a GCM that can take into account the radiative effect of clouds.
Finally, our GCM predicts a minimum of tracers at the equator for pressures higher than 0.1 mbar for all cases considered here, and we expect the cloud distribution to have the same behavior. A minimum of cloud at the equator could have observable impacts on transmission spectra and also on reflection or emission spectra of the planet.

\subsection{Implications for photochemical haze formation} 
Another possible explanation for the flat transit spectrum of GJ1214b is the formation of photochemical haze, possibly produced by Titan-like methane photolysis. Such a process can be more efficient than cloud condensation at producing high aerosols because particles are produced directly in the upper atmosphere. However, sedimentation may be too strong to maintain a particle density large enough for blocking stellar light. Even on Titan, which is very hazy and has a low gravity (with correspondingly small sedimentation velocities), the haze is not opaque enough to block all methane absorption bands in near-infrared transit spectra \citep{robinson14b}. Therefore GJ1214b's flat spectrum can only be produced by photochemical haze if there is a large methane flux counter-balancing the haze sedimentation flux. Based on the atmospheric circulation simulations, we predict that methane would be supplied at mid-latitudes (see figure \ref{figure_7}). Once produced, haze would be removed mostly at the poles and at the equator. This would impact the hazes' latitudinal distribution. Haze particles should fall until the temperature is high enough to pyrolyze them, mostly into methane. The Huygens Probe Aerosol Collector and Pyrolyser (ACP) experiment showed that Titan's hydrocarbon haze was completely pyrolized at temperatures close to 900K \citep{israel05}. In our simulations of GJ1214b, that  temperature corresponds to pressure of around 0.1 bar for high metallicity and 1 bar for solar metallicity (see Fig. \ref{figure_2}). Therefore, there could be a complete methane cycle for pressures lower than these values, that could recycle haze into methane. 
\cite{miller-ricci12} and \cite{morley13} explored the photochemistry in GJ1214b with a 1D model using eddy diffusion coefficient values from 10$^6$ to 10$^9$ cm$^2$/s and predicted that more methane and more haze would be present at high altitude with a high $K_{zz}$ (i.e. 10$^9$ cm$^2$/s). 
With our model, we predict eddy diffusion coefficient values of around 3$\times$10$^7$ cm$^2$/s at 1 bar to 8$\times$10$^9$ cm$^2$/s at 10$^{-6}$ bar for the 100$\times$solar metallicity case. The photolysis of methane, which occurs mostly above 10$^{-5}$ bar, is thus better simulated with $K_{zz}$=10$^9$ cm$^2$/s in \cite{miller-ricci12} and \cite{morley13}. Moreover the radiative effect of haze, which can absorb stellar radiation, could increase the vertical mixing in the upper atmosphere and support the lofting of aerosol particles, increasing the lifetime of the haze, and this feedback is best studied with a GCM.
We conclude that the formation of high haze on GJ1214b is likely and it is also possible that both high condensate clouds and photochemical haze are formed.

\section{Conclusions}

In this study, we analyzed the atmospheric dynamics of GJ1214b for a cloud-free atmosphere with different metallicities as a first application of the Generic LMDZ GCM to a non-rocky exoplanet. 
We obtained results for radiative transfer, temperatures and winds very similar to those from the other state-of-the-art GCM \citep{kataria14}, validating our model for this kind of planet.

In addition to this intercomparison, we showed that the zonal mean circulation of GJ1214 is characterized by the presence of an anti-Hadley circulation on the nightside, which leads to zonally-averaged subsidence at the equator. This particular regime likely occurs on a wide range of synchronously rotating  exoplanets, from warm mini-Neptunes to hot Jupiters.

Using simple tracers, we analyzed the vertical mixing in the atmosphere of GJ1214b. The tracers showed that atmospheric transport is primarily driven by the zonal mean circulation, leading to an upwelling at mid-latitudes and a downwelling at the equator. In particular, a minimum of tracer abundance appears at the equator, strengthened in the upper atmosphere by a poleward meridional transport. This should be a fundamental feature of the cloud/haze distribution on many synchronously rotating warm exoplanets.
We also found that the vertical mixing increases with metallicity for H-rich atmospheres. For a solar metallicity, the circulation cannot loft micrometer particles into the upper atmosphere. Therefore, if the upper atmosphere of GJ1214b is opaque because of condensate clouds, the atmospheric metallicity must be higher than solar and likely more than 10$\times$solar. 
From these simulations, we derived equivalent 1D  eddy coefficients. We found that:
$K_{zz}=7\times 10^2 \times P_{bar}^{-0.4}$ m$^2$/s is a good fit for 1$\times$solar metallicity and $K_{zz}=3\times 10^3 \times P_{bar}^{-0.4}$ m$^2$/s for 100$\times$solar metallicity. We compared these values to classical formulae for $K_{zz}$ used for instance in \cite{morley13}, showing how these simple formulae can overestimate the mixing by one order of magnitude.
Our parametrizations of $K_{zz}$ can be used in 1D cloud or photochemical models \citep{miller-ricci12, morley13} for which they are expected to lead to an efficient formation of photochemical haze.

Finally, we showed that the atmospheric circulation of GJ1214b could be strong enough to loft micrometric particles of KCl or ZnS from 1 bar to 0.1-0.01 mbar. However, the radiative effects of clouds, by absorbing or diffusing stellar radiation, could modify the strength of the circulation. The next step will be to simulate the atmosphere of GJ1214b with realistic KCl and ZnS clouds, taking into account the latent heat release and the radiative effects. For this next study, the Generic LMDZ GCM benefits from all the developments done for previous studies of cloudy rocky planets. Such 3D simulations would give strong indications of the possibility of high condensate clouds producing the observed flat transit spectrum. Such simulations could also help identify the best observations for probing cloudy atmospheres, with, for example, emission/reflection spectra and phase curves.

\acknowledgments

B.C. acknowledges support from an appointment to the NASA Postdoctoral Program at NAI Virtual Planetary Laboratory, administered by Oak Ridge Affiliated Universities.
VSM is supported by the NASA Astrobiology Institute's Virtual Planetary Laboratory Lead Team, under the National Aeronautics and Space Administration solicitation NNH12ZDA002C and Cooperative Agreement Number NNA13AA93A.  
This work was facilitated though the use of advanced computational, storage, and networking infrastructure provided by the Hyak supercomputer system at the University of Washington.

\clearpage

\begin{table}[!h] 
\begin{tabular}{lc}
\hline
\hline 
   Parameters &    \\ \hline
   $\Omega_p$ (planetary rotation rate, rad s$^{-1}$) & 4.615$\times$10$^{-5}$   \\
   $P_{rev}$ (revolution period, s) & 136512   \\
   $e$ (eccentricity) & 0   \\
   $R_p$ (planetary radius, m) & 1.7$\times$10$^{7}$   \\
   $g$ (gravitational acceleration, m s$^{-2}$) & 8.93   \\
   $F_{star}$ (stellar flux at top of the atmosphere, W m$^{-2}$) & 23600  \\
   $F_{int}$ (internal thermal flux, W m$^{-2}$) & 0.73   \\ \hline

\end{tabular}
\label{table1}
\caption{Orbital and physical parameters used in the model.}
\end{table}

\begin{table}[!h] 
\begin{tabular}{lccc}
\hline
\hline 
   Atmospheric composition & c$_p$ (J kg$^{-1}$ K$^{-1}$) & $H$(km) & Mean molecular weight (g mol$^{-1}$) \\ \hline
   1$\times$solar & 12800 & 220 & 2.3  \\
   10$\times$solar & 12000 & 200 & 2.5  \\
   100$\times$solar & 6474 & 115& 4.38  \\
   pure water & 1864 & 28 & 18.0  \\\hline
\end{tabular}
\label{table1}
\caption{Values of specific heat (c$_p$), scale height ($H$) and mean molecular weight used for the different atmopheric compositions.}
\end{table}



\clearpage


\bibliographystyle{apj}

\end{document}